\newif\ifdouble
\newif\ifsingle
\newif\ifchange
\doubletrue

\ifdouble
  \documentclass[sigconf, usenames, dvipsnames]{acmart}
\else
  \documentclass[sigconf]{acmart}
  \singletrue
\fi

\usepackage{dblfloatfix} 
\usepackage{booktabs} 
\usepackage{arydshln}
\usepackage{subfig}
\newcommand{\remove}[1]{{\color{red}{\sout{#1}}}}

\renewcommand{\remove}[1]{}

\AtBeginDocument{%
  \providecommand\BibTeX{{%
    \normalfont B\kern-0.5em{\scshape i\kern-0.25em b}\kern-0.8em\TeX}}}

\copyrightyear{2023}
\acmYear{2023}
\setcopyright{acmlicensed}\acmConference[UIST '23]{The 36th Annual ACM Symposium on User Interface Software and Technology}{October 29-November 1, 2023}{San Francisco, CA, USA}
\acmBooktitle{The 36th Annual ACM Symposium on User Interface Software and Technology (UIST '23), October 29-November 1, 2023, San Francisco, CA, USA}
\acmPrice{15.00}
\acmDOI{10.1145/3586183.3606727}
\acmISBN{979-8-4007-0132-0/23/10}

\begin{document}
\pagenumbering{arabic}
\pagestyle{plain}

\newcommand{\system}{HoloBots}
\title{\system{}: Augmenting Holographic Telepresence with Mobile Robots for Tangible Remote Collaboration in Mixed Reality}

\author{Keiichi Ihara}
\affiliation{%
  \institution{University of Tsukuba}
  \city{Tsukuba}
  \country{Japan}}
\affiliation{%
  \institution{University of Calgary}
  \city{Calgary}
  \country{Canada}}
\email{kihara@iplab.cs.tsukuba.ac.jp}

\author{Mehrad Faridan}
\affiliation{%
  \institution{University of Calgary}
  \city{Calgary}
  \country{Canada}}
\email{mehrad.faridan1@ucalgary.ca}

\author{Ayumi Ichikawa}
\affiliation{%
  \institution{University of Tsukuba}
  \city{Tsukuba}
  \country{Japan}}
\email{aichikawa@iplab.cs.tsukuba.ac.jp}

\author{Ikkaku Kawaguchi}
\affiliation{%
  \institution{University of Tsukuba}
  \city{Tsukuba}
  \country{Japan}}
\email{kawaguchi@cs.tsukuba.ac.jp}

\author{Ryo Suzuki}
\affiliation{%
  \institution{University of Calgary}
  \city{Calgary}
  \country{Canada}}
\email{ryo.suzuki@ucalgary.ca}

\renewcommand{\shortauthors}{Ihara, et al.}
\begin{abstract}
This paper introduces HoloBots, a mixed reality remote collaboration system that augments holographic telepresence with synchronized mobile robots. 
Beyond existing mixed reality telepresence, HoloBots lets remote users not only be visually and spatially present, but also \textit{physically} engage with local users and their environment. HoloBots allows the users to touch, grasp, manipulate, and interact with the remote physical environment as if they were co-located in the same shared space. 
We achieve this by synchronizing holographic user motion (Hololens 2 and Azure Kinect) with tabletop mobile robots (Sony Toio).
Beyond the existing physical telepresence, HoloBots contributes to an exploration of broader design space, such as object actuation, virtual hand physicalization, world-in-miniature exploration, shared tangible interfaces, embodied guidance, and haptic communication. 
We evaluate our system with twelve participants by comparing it with hologram-only and robot-only conditions. Both quantitative and qualitative results confirm that our system significantly enhances the level of co-presence and shared experience, compared to the other conditions.
\end{abstract}



\begin{teaserfigure}
\centering
\includegraphics[width=0.245\textwidth]{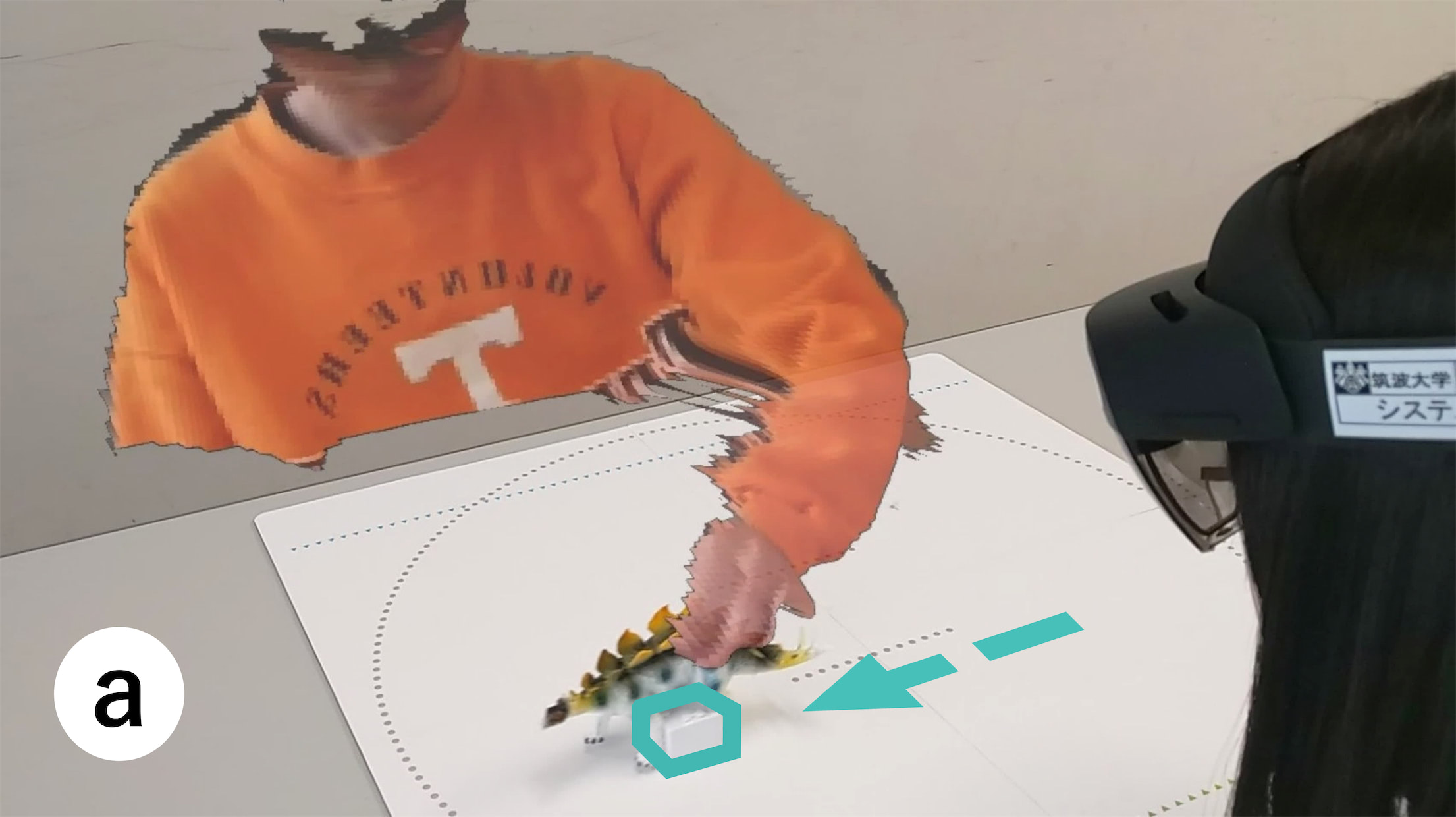}
\includegraphics[width=0.245\textwidth]{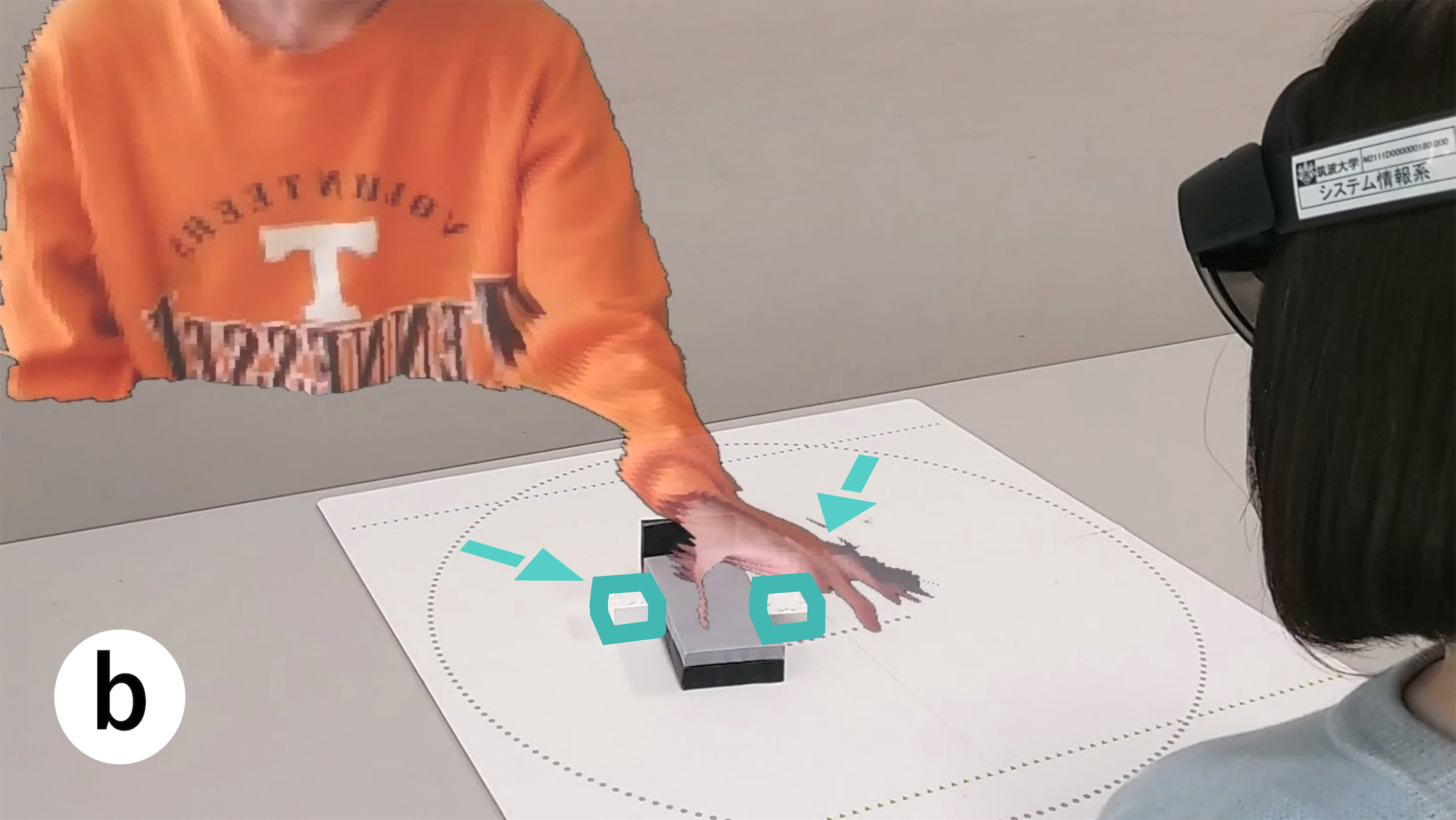}
\includegraphics[width=0.245\textwidth]{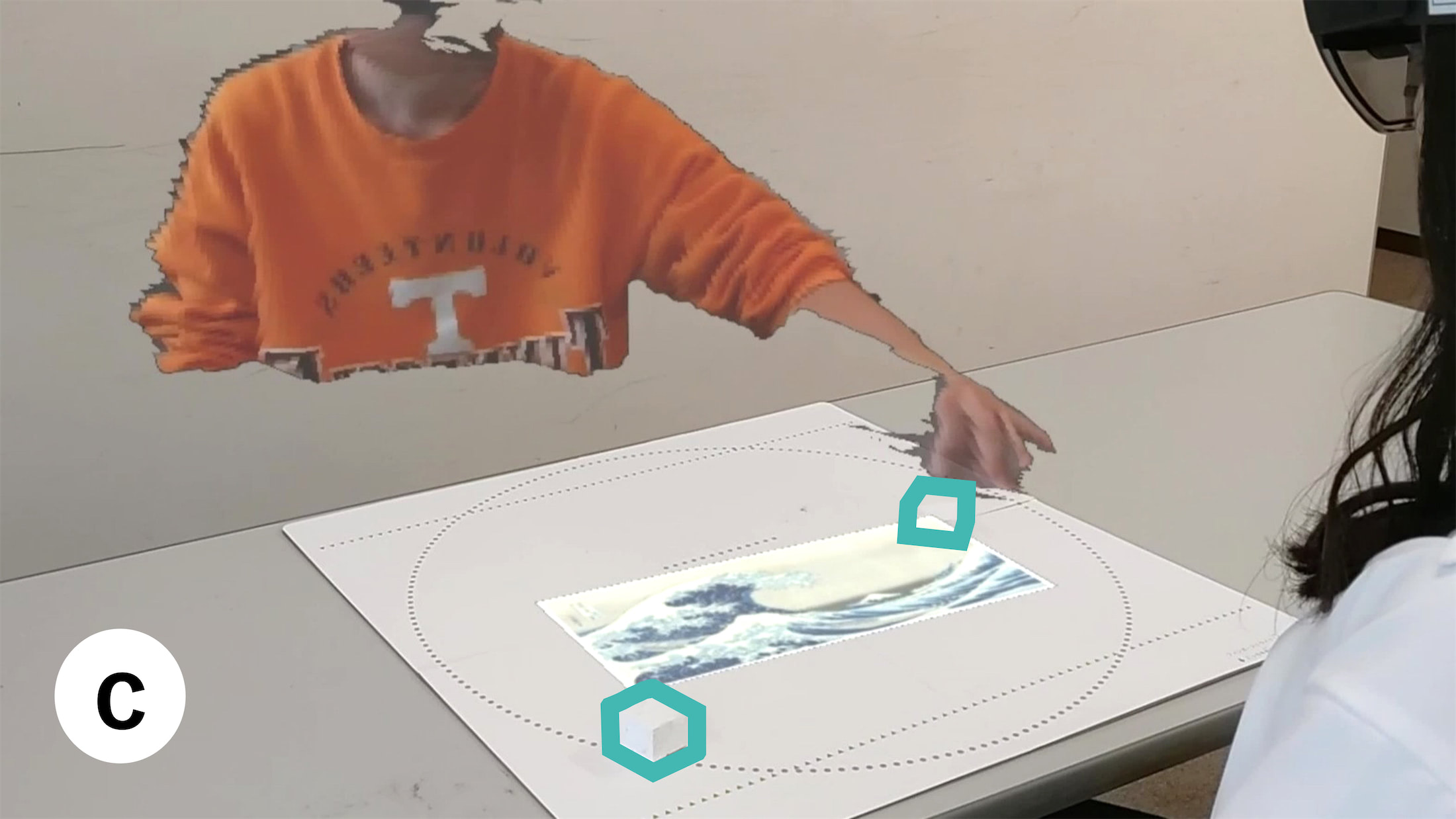}
\includegraphics[width=0.245\textwidth]{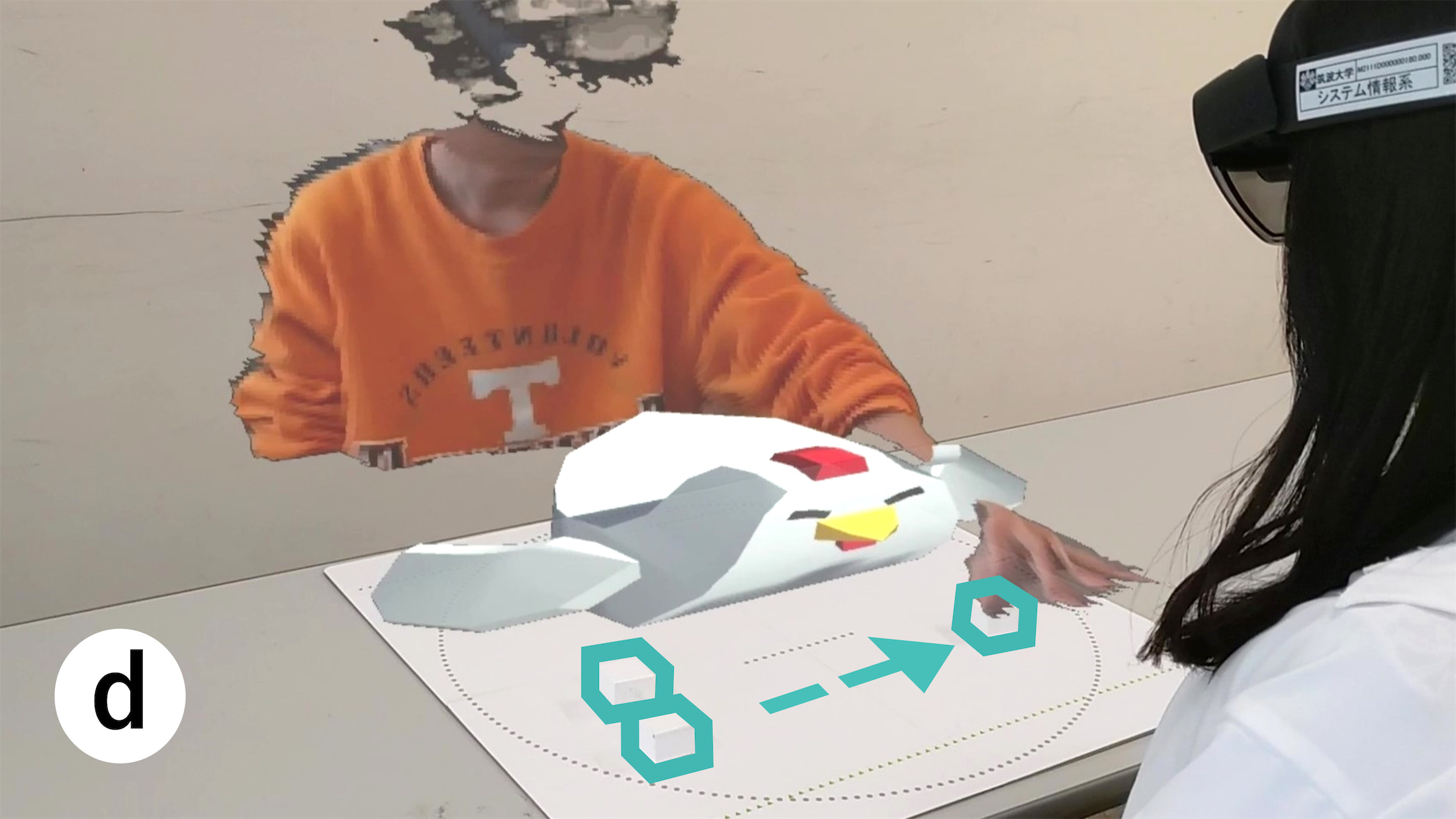}
\includegraphics[width=0.245\textwidth]{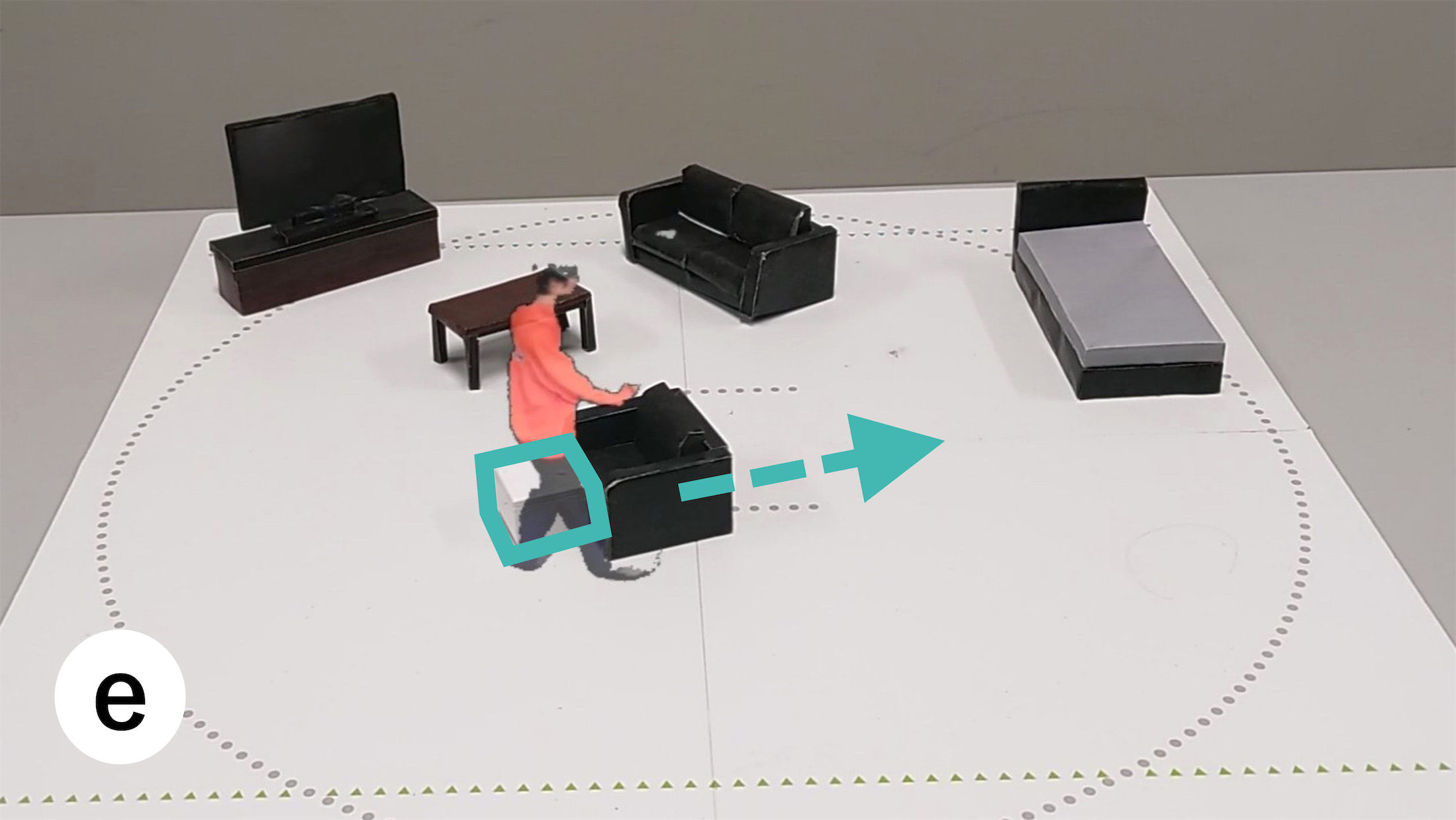}
\includegraphics[width=0.245\textwidth]{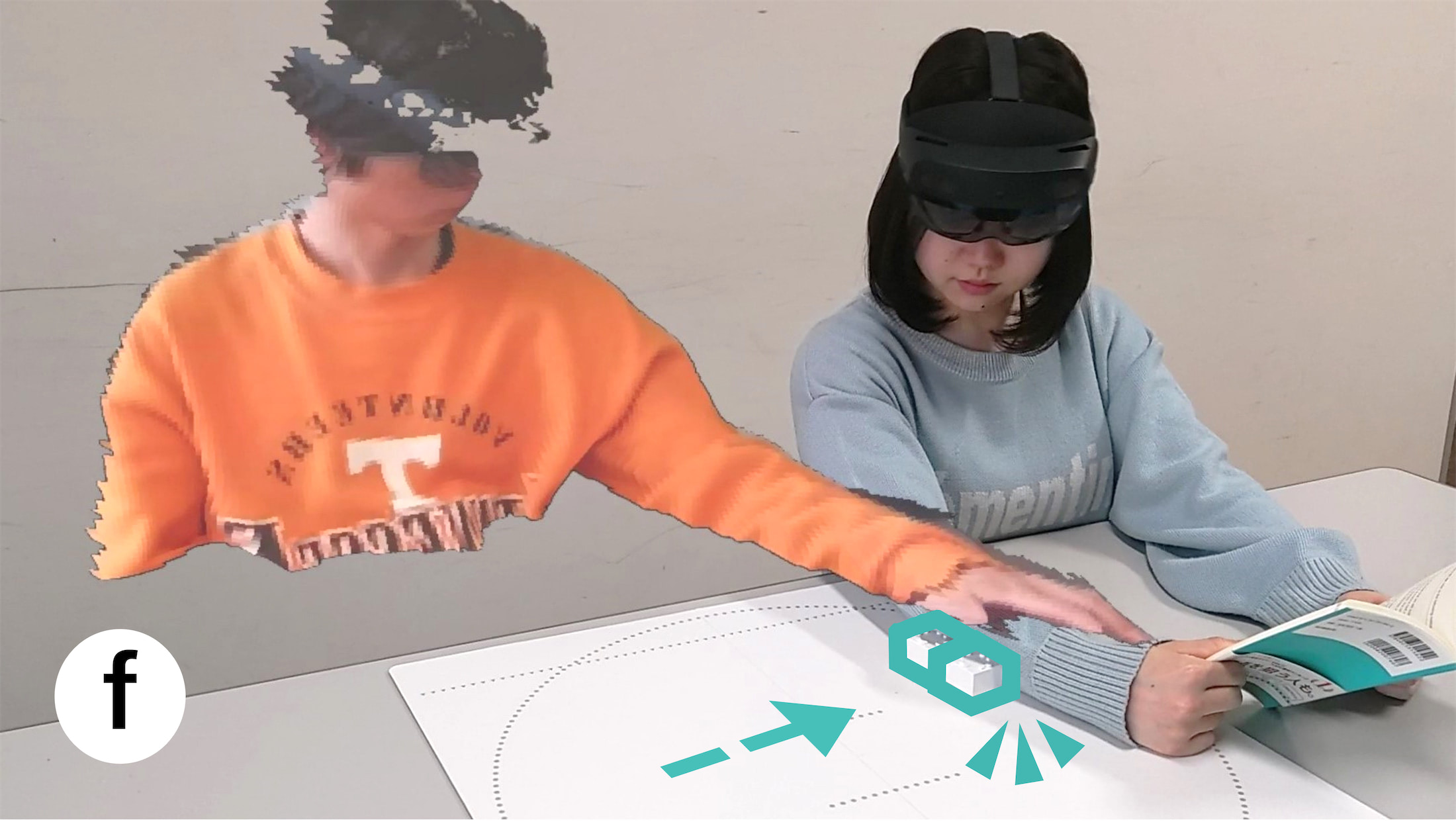}
\includegraphics[width=0.245\textwidth]{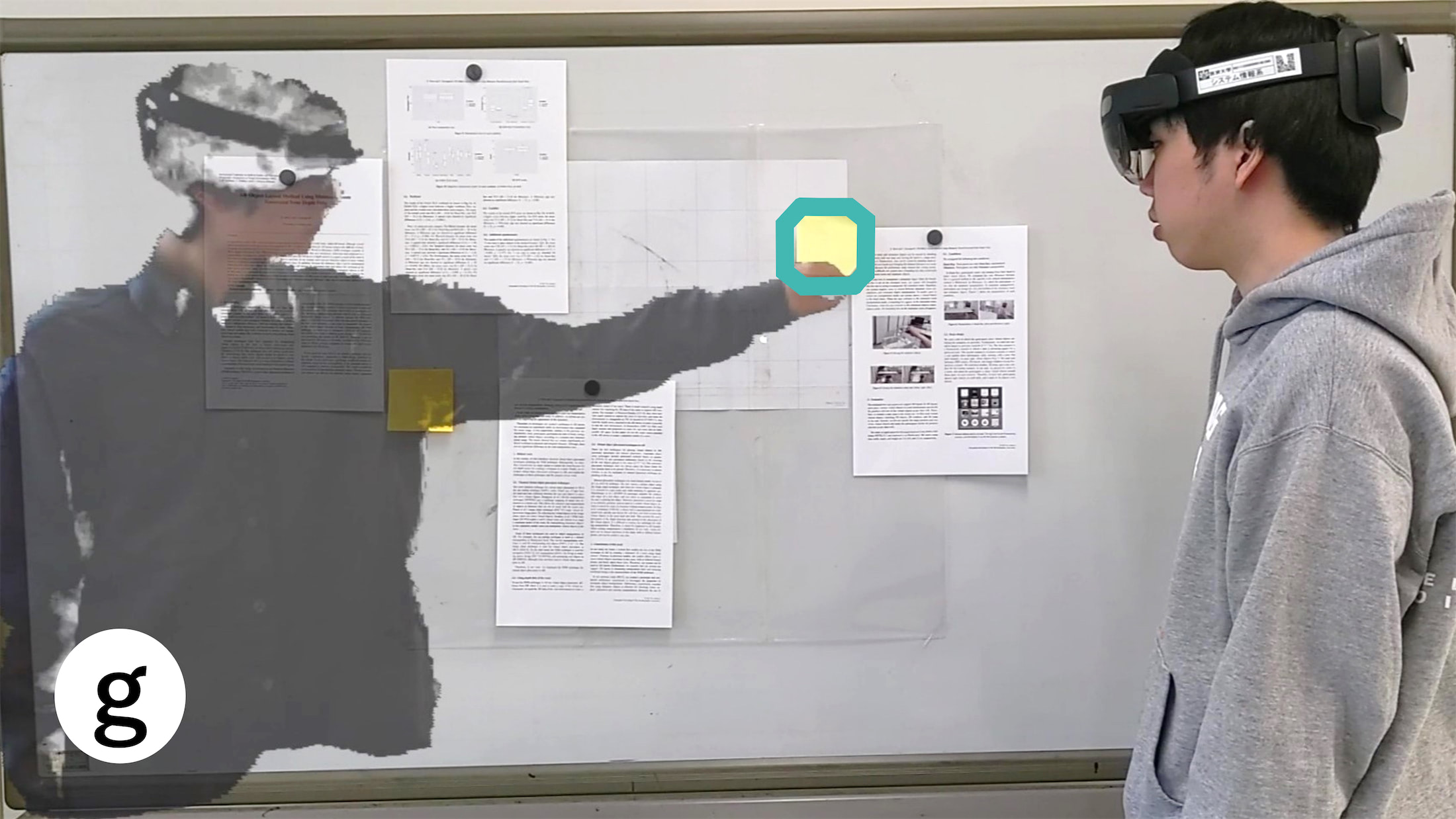}
\includegraphics[width=0.245\textwidth]{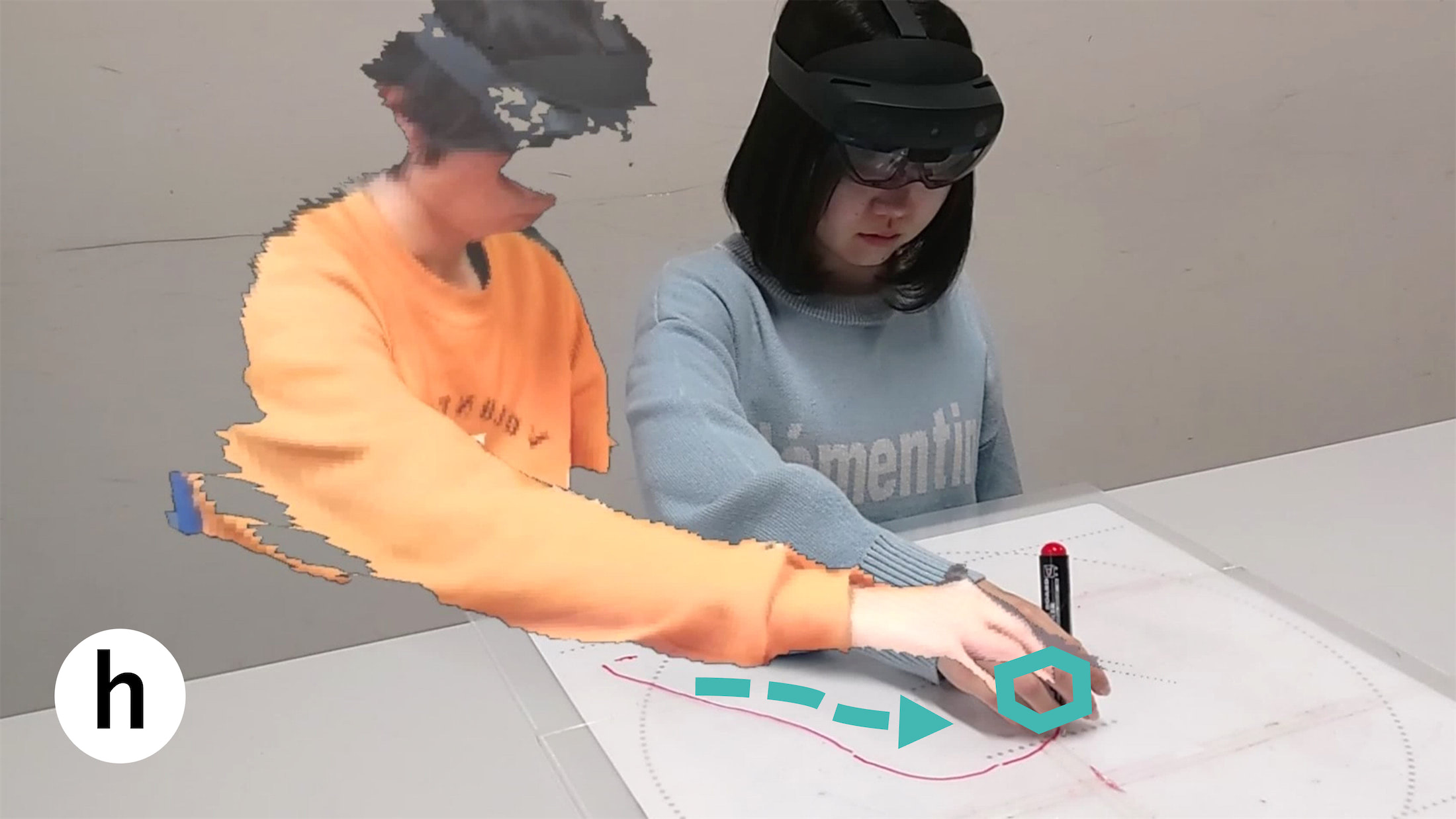}
\caption{\system{} explores augmenting holographic telepresence with tabletop mobile robots for remote collaboration.
The remote user can interact with the local user through various methods, such as (a, b) actuating objects, (c, d) sharing tangible user interfaces, (e) representing the body, and (f) providing haptic feedback.
By using attachments, \system{} is adaptable in situations that involve (g) vertical surface and (h) drawing scenarios.
}
\label{fig:teaser}
\end{teaserfigure}

\maketitle

\section{Introduction}
Today's mixed reality telepresence still falls short of replicating the rich tangible experiences that we naturally enjoy in our everyday lives.
In real-world collaboration, for example, we casually grasp and manipulate physical objects to facilitate discussions, employ touch for social interactions, spatially arrange physical notes for brainstorming, and guide others by holding their hands.
However, such tangible interactions are not possible with current mixed reality telepresence systems, as the virtual remote user has no way to \textit{physically} engage with the local user and environment. 

In this paper, we introduce \system{}, a mixed reality interface to achieve \textit{tangible remote collaboration} by synchronously coupling holographic telepresence with an actuated physical environment. Beyond existing holographic telepresence like \textit{Holoportation}~\cite{orts2016holoportation}, \system{} lets remote users not only visually and spatially co-present but also physically touch, grasp, manipulate, and interact with remote tangible objects, as if they were co-located in the same shared space. 
We achieve this by synchronizing the remote user's motion rendered in a mixed reality headset (Hololens 2 and Azure Kinect) with physical actuation enabled by multiple tabletop mobile robots (Sony Toio). 

Our idea builds up on the existing \textit{physical telepresence}~\cite{leithinger2014physical} and other related approaches~\cite{lee2018physical, siu2018investigating}, but we make two key contributions beyond them.
First, we explore a \textbf{\textit{broader design space}} of tangible remote collaboration with holographic telepresence, which are not fully investigated in the prior work~\cite{siu2018investigating, lee2018physical, he2017physhare}.
For example, we showcase various interactions, such as object actuation, virtual hand physicalization, world-in-miniature exploration, shared tangible interfaces, embodied guidance, and haptic communication. 
We also demonstrate use cases and application scenarios for each interaction, such as physical storytelling, remote tangible gaming, and hands-on instruction.

Second, we contribute to a \textbf{\textit{holistic user evaluation}} to better understand how mobile robots can enhance holographic telepresence in different application scenarios.
To this end, we compare our approach (hologram + robot) with hologram-only and robots-only conditions through a within-subject user study with twelve participants. 
Both quantitative and qualitative results suggest that our system significantly enhances the level of co-presence and shared experience for mixed reality remote collaboration, compared to the other two conditions. Based on the insights, we also discuss the future of tangible remote collaboration that leverages robotic environments.

\remove{Our idea builds up on the existing \textit{physical telepresence}~\cite{leithinger2014physical} and other related approaches~\cite{lee2018physical, siu2018investigating}, but we make two key contributions beyond them. First, by leveraging \textbf{\textit{multiple tabletop robots}}, rather than shape-changing displays~\cite{leithinger2014physical} or single-point actuation of X-Y plotters~\cite{lee2018physical}, our approach enables scalable, deployable, yet generalizable tangible remote collaboration.
Second, we contribute to the \textbf{\textit{demonstration and evaluation of a broader design space}} that is not fully investigated in the prior work~\cite{siu2018investigating, lee2018physical, he2017physhare}. For example, we demonstrate and evaluate various interactions, such as object actuation, virtual hand physicalization, world-in-miniature exploration, shared tangible interfaces, embodied guidance, and haptic communication. 
We also showcase various use cases and application scenarios, such as physical storytelling, remote tangible gaming, and hands-on instruction.
Collectively, our paper contributes to establishing a new approach to holographic tangible telepresence beyond the existing works.}

\remove{To evaluate our approach, we conduct a user study with twelve participants, where we compare our approach (hologram + robot) with hologram-only and robots-only conditions. Both quantitative and qualitative results suggest that our system significantly enhances the level of co-presence and shared experience for mixed reality remote collaboration, compared to the other two conditions. Based on the insights, we also discuss the future of tangible remote collaboration that leverages robotic environments.
}

Finally, this paper contributes the following:

\begin{enumerate}
\remove{\item \system{}, a system to augment holographic telepresence with multiple tabletop mobile robots that enables scalable, deployable, and generalizable tangible remote collaboration.}
\item A design space exploration and application demonstrations that showcase a set of possible interactions and use cases enabled by our system.
\item Results and insights from our user study that confirm the benefits of our approach over hologram-only and robots-only conditions. 
\end{enumerate}
\newpage
\section{Related Work}

\subsection{Remote Collaboration}
\subsubsection{Mixed Reality Remote Collaboration}
Recent advances in mixed reality technologies have enabled immersive remote collaboration that was not possible with traditional desktop interfaces. 
Prior research has explored various approaches for immersive telepresence, such as holographic teleportation (e.g., \textit{Holoportation}~\cite{orts2016holoportation}, \textit{Virtual Makerspaces}~\cite{radu2021virtual}, \textit{Loki}~\cite{thoravi2019loki}), virtual avatars (e.g., \textit{CollaboVR}~\cite{he2020collabovr}, \textit{Mini-Me}~\cite{piumsomboon2018mini}, \textit{Shoulder of Giant}~\cite{piumsomboon2019shoulder}, \textit{ARTEMIS}~\cite{gasques2021artemis}), and projected video stream (e.g., \textit{Room2Room}~\cite{pejsa2016room2room}, \textit{3D-Board}~\cite{zillner20143d}).
These systems allow remote users to be spatially co-located in the same shared space, which greatly enhances collaborative experiences~\cite{bai2020user, cao2020exploratory}. 
For example, by showing virtual hands and bodies in 3D space, the local users can more easily understand the intention of the remote users for various physical tasks such as block assembly~\cite{zhang2022real}, origami~\cite{kim2019evaluating, kim2020combination}, mechanical tasks~\cite{oyama2021augmented, oyama2021integrating}, and physiotherapy education~\cite{faridan2023chameleoncontrol}.
However, current holographic telepresence lacks the physical embodiment of the remote user, which significantly reduces the sense of co-presence~\cite{lee2018physical}. This limitation also constraints rich physical affordances which we naturally employ in co-located physical collaboration~\cite{leithinger2014physical, siu2018investigating}.

\subsubsection{Robotic Telepresence}
To address this limitation, past research has explored robotic telepresence that aims to physically embody remote users by adding a robotic body to a 2D video screen (e.g., \textit{MeBot}~\cite{adalgeirsson2010mebot},  \textit{RemoteCode}~\cite{sakashita2022remotecode}) or by replicating the remote user with a humanoid or non-humanoid robot (e.g., \textit{TELESAR V}~\cite{fernando2012design}, \textit{Telenoid}~\cite{ogawa2011exploring}, \textit{You as a Puppet}~\cite{sakashita2017you}, \textit{GestureMan}~\cite{kuzuoka2000gestureman}, \textit{Geminoid}~\cite{sakamoto2007android}).
The robotic telepresence can greatly enhance user engagement by enabling physical interactions such as  gestures~\cite{adalgeirsson2010mebot} and body movement~\cite{nakanishi2011zoom, rae2014bodies, lee2011now}.
For example, mobile robots allow remote users to move freely around a table to interact with local users and objects for remote education (e.g., \textit{RobotAR}~\cite{villanueva2021robotar}, \textit{ASTEROIDS}~\cite{li2022asteroids}).
Beyond a screen-based representation, \textit{VROOM}~\cite{jones2020vroom, jones2021belonging} overlays a holographic avatar on a telepresence robot that enriches non-verbal communication such as gestures or eye-contact. 

\subsubsection{Physical Telepresence}
An alternative approach to adding physical embodiment to remote users is using \textit{synchronized distributed physical objects}~\cite{brave1998tangible}, rather than embodying users themselves with robotic telepresence. 
Such an approach was originally explored through \textit{InTouch}~\cite{brave1997intouch}, \textit{ComTouch}~\cite{chang2002comtouch}, and \textit{PsyBench}~\cite{brave1998tangible}, in which synchronized tangible tokens embody the remote user's motion and behavior.
This idea has evolved into a concept of \textit{physical telepresence}~\cite{leithinger2014physical}, which synchronizes physical shape rendering with the remote users' visual appearance.
For instance, Leithinger et al.~\cite{leithinger2014physical} uses a shape-changing display~\cite{follmer2013inform} to physically render a remote user's hand and surrounding objects with screen-based visual feedback. 
Recent works have also expanded this concept by combining a virtual avatar with a motorized X-Y plotter to actuate a single token (e.g., \textit{Physical-Virtual Table}~\cite{lee2018physical}).
However, the existing approach using shape displays lacks deployability due to the dedicated hardware requirement, and X-Y plotters lack scalability and generalizability due to a single point actuation and limited interaction area.
More closely related to our work, a few researchers have explored the use of mobile robots for tangible remote collaboration in VR (e.g., \textit{PhyShare}~\cite{he2017physhare}) and mixed reality environments (e.g., \textit{Siu et al.}~\cite{siu2018investigating}).
However, this approach of using multiple mobile robots has not been fully explored yet, as these prior works do not present the comprehensive design space and have not conducted any user evaluation to understand the benefits and limitations of this approach.
Beyond these prior works, we contribute to 1) an exploration of the broader design space with a demonstration of comprehensive applications, and 2) a holistic user evaluation through condition experiments.

\subsection{Bi-Directional Virtual-Physical Interaction}
Outside the context of remote collaboration, past HCI research has also explored bi-directional virtual-physical interaction by leveraging augmented reality and actuated environments~\cite{suzuki2022augmented}.
For example, systems like \textit{Kobito}~\cite{aoki2005kobito}, \textit{Augmented Coliseum}~\cite{kojima2006augmented}, and \textit{IncreTable}~\cite{leitner2008incretable} explore the synchronous coupling between AR and actuated physical objects, which can enrich visual feedback and affordances of robots and  actuated tangible interfaces.
These interfaces typically employ robot motion (e.g., \textit{exTouch}~\cite{kasahara2013extouch}), actuated tangible tokens (e.g., \textit{PICO}~\cite{patten2007mechanical}, 
\textit{Reactile}~\cite{suzuki2018reactile}, \textit{Actuated Workbench}~\cite{pangaro2002actuated}), IoT devices (e.g., \textit{MechARSpace}~\cite{zhu2022mecharspace}, \textit{WIKA}~\cite{jeong2020wika}, \textit{Kim et al.}~\cite{kim2018does}) to synchronize between virtual and physical outputs in a bi-directional manner.
Similar to our work, \textit{Sketched Reality}~\cite{kaimoto2022sketched} and \textit{Physica}~\cite{li2023physica} explores bi-directional interaction between embedded virtual objects and tabletop robots. 
Our system extends their work in the context of holographic tangible remote collaboration in mixed reality environments. 

\subsection{Actuated Tangible User Interfaces}
Actuated tangible user interfaces were originally developed to address the challenge of digital-physical discrepancies in conventional tangible interfaces~\cite{poupyrev2007actuation}.
Towards this goal, HCI researchers have explored a variety of actuated tangible user interfaces~\cite{poupyrev2007actuation} and shape-changing user interfaces~\cite{rasmussen2012shape, coelho2011shape, alexander2018grand}, using magnetic actuation~\cite{patten2007mechanical}, ultrasonic waves~\cite{marshall2012ultra}, magnetic levitation~\cite{lee2011zeron}, and wheeled and vibrating robots~\cite{nowacka2013touchbugs}.
Rosenfeld et al.~\cite{rosenfeld2004physical} introduced the concept of using physical mobile robots as an actuated tangible user interface.
This concept has been expanded through various systems such as 
\textit{Zooids}~\cite{le2016zooids},
\textit{ShapeBots}~\cite{suzuki2019shapebots}, \textit{HERMITS}~\cite{nakagaki2020hermits}, \textit{Rolling Pixels}~\cite{lee2020rolling}, and \textit{(Dis) Appearables}~\cite{nakagaki2022dis}. 
Swarm user interfaces can also provide haptic sensations~\cite{kim2019swarmhaptics, suzuki2017fluxmarker, suzuki2021hapticbots, zhao2017robotic} and actuate everyday objects~\cite{kim2020user, farajian2022swarm}. 
Inspired by these works, we also leverage multiple tabletop robots for our actuated interfaces.


\section{\system{}: System Design}
\begin{figure}[h]
\centering
\includegraphics[width=\linewidth]{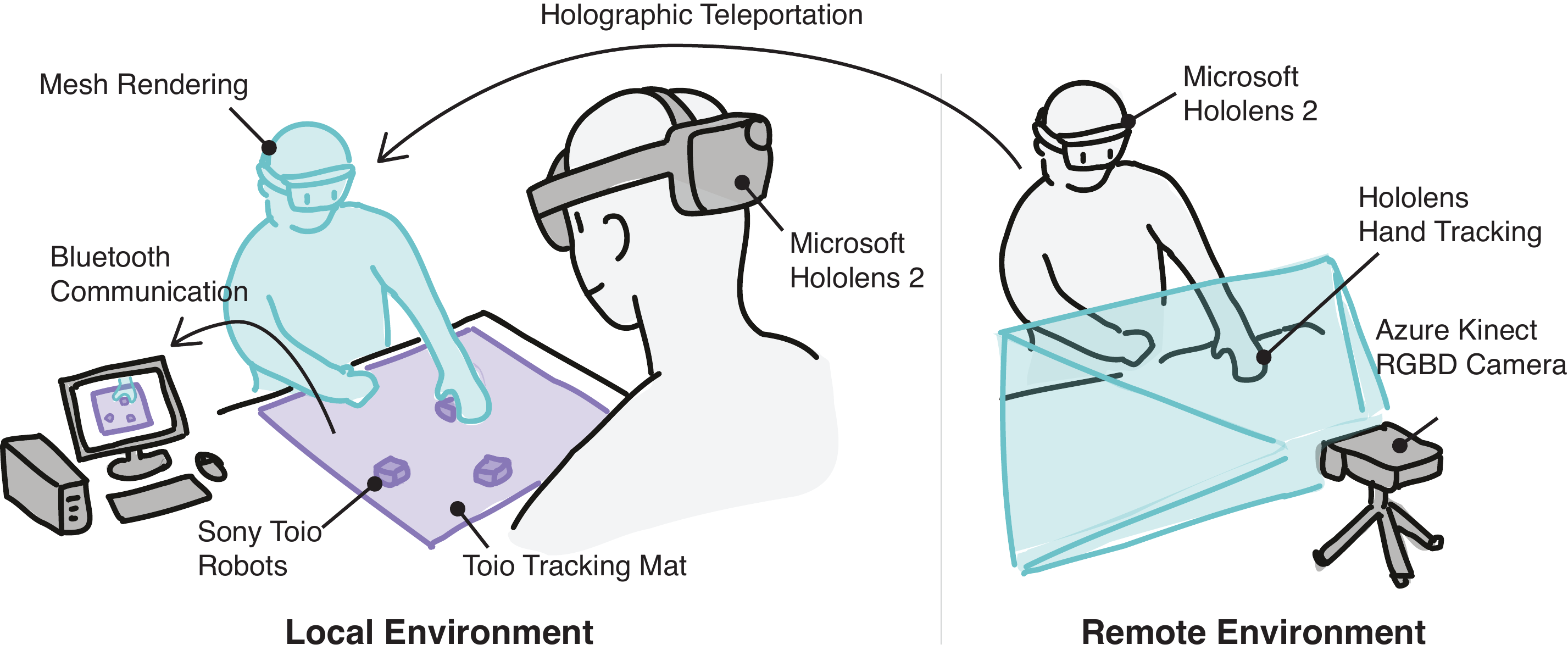}
\caption{System Setup: The local user can see the remote user's avatar and interact Toios or virtual objects with the remote user through Hololens. The remote user's body is tracked by Azure Kinect and the hands are tracked by Hololens.}
\label{fig:system-setup}
\end{figure}

This section introduces \system{}, a system that augments holographic telepresence with multiple tabletop robots. As illustrated in Figure~\ref{fig:system-setup}, \system{} consists of three main components: 1) \textit{capturing a remote user} with the Azure Kinect depth camera, 2) \textit{holographic rendering and hand tracking} with Microsoft Hololens 2 headset, and 3) \textit{synchronized actuation} with Sony Toio tabletop mobile robots.

\subsubsection*{\textbf{Capturing a Remote User with a Depth Camera}}
The Azure Kinect RGB-D camera is used to capture the remote user's body. The camera is positioned in front of the remote user with a tripod stand. 
The Kinect camera is connected to the local PC (G-Tune, Intel Core i7-11800H 2.30GHz CPU, NVIDIA GeForce RTX 3060 GPU, 64GB RAM) via a USB cable.
The depth information is captured through the Azure Kinect Sensor SDK running on the local PC. The depth sensor first generates a point cloud with a resolution of 640 x 576, which is then converted into a real-time colored 3D  mesh using the Azure Kinect Examples for Unity package~\footnote{\url{https://assetstore.unity.com/packages/tools/149700}}. 
Mesh data is captured with 30 FPS and the size of each mesh data is approximately 20 MB.


\subsubsection*{\textbf{Holographic Rendering and Hand Tracking}}
In our setup, both local and remote users wear the Microsoft Hololens 2 mixed reality headset, which has a diagonal field of view of 52 degrees. 
The remote user's holographic mesh generated by the local PC is rendered in Hololens 2 through Holographic 
Remoting~\footnote{\url{https://learn.microsoft.com/en-us/windows/mixed-reality/develop/native/holographic-remoting-player}}, enabling high-quality and low-latency (60 FPS) rendering over an Ethernet connection, allowing the local user to view the mesh. 
Hololens 2 is also used to track the user's hand movements using the MRTK hand-tracking library. 
Tracked hands are used to 1) grab virtual robots to manipulate and synchronize the physical one in the remote environment, or 2) move robots based on the finger position to physicalize the virtual hand. 
These processes are executed on Unity running on the remote PC (G-Tune, Intel Core i7-11800H 2.30GHz CPU, NVIDIA GeForce RTX 3070 GPU, 64GB RAM) and the local PC, respectively, connected with Hololens 2 through Holographic Remoting. 

\begin{figure*}[h]
\centering
\includegraphics[width=\textwidth]{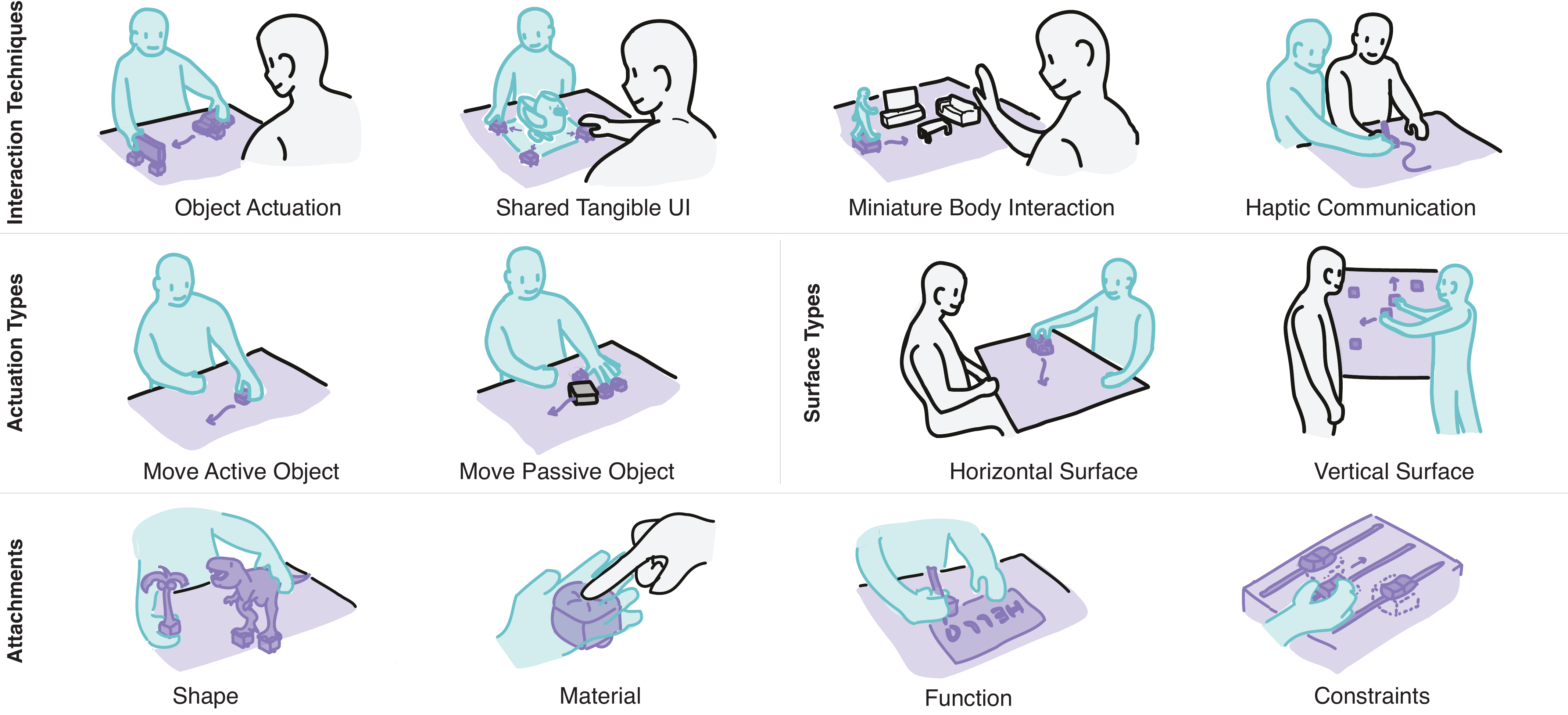}
\caption{Design Space of \system{}.}
\label{fig:design-space}
\end{figure*}

\subsubsection*{\textbf{Synchronized Actuation with Tabletop Mobile Robots}}
Our system uses Sony Toio~\footnote{\url{https://www.sony.com/en/SonyInfo/design/stories/toio/}} as tabletop mobile robots. Each robot measures 3.2 cm × 3.2 cm × 2.5 cm and can move at a speed of up to 35 cm/sec for straight-line movement and 1500 deg/sec for rotation. The robot has a built-in camera that can scan patterns printed on a mat (Toio Tracking Mat) to detect their position and orientation. The size of the tracking mat has 55 cm × 55 cm of covered area, but it can be extended by aligning multiple mats. Each Toio robot is controlled using the Toio SDK for Unity~\footnote{\url{https://github.com/morikatron/toio-sdk-for-unity}} on a PC and continuously sends its position and orientation to the PC via Bluetooth\textregistered \  standard Ver. 4.2 every 10 ms. 
For the controlling algorithm, we adapt to the open-source library~\cite{nakagaki2020hermits} and rewrite the algorithm for our Unity application.
To start using our application, the local user first performs a manual calibration to align the remote user's holographic mesh with the Toio mat.
This alignment can be bypassed in subsequent uses saving the relative position between the Toio mat and the avatar mesh. By placing a QR code on the Toio mat to acquire the mat's position and leveraging the relative position, we can display the avatar mesh in the appropriate position.

After the calibration, each robot's position is controlled through the following three ways: 1) physical Toio movement in the remote environment, 2) virtual object movement in the remote user's Hololens, or 3) finger position movement of the remote user. 
When both users have a physical Toio setup, the system can simply synchronize the position of each environment. 
On the other hand, when only the local user is equipped with the Toio robot, then the remote user can manipulate virtual Toios by grasping and manipulating virtual Toio objects rendered in the Hololens, while the local user manipulates physical Toios. 
Alternatively, the remote user can manipulate these Toio robots with hand and finger tracking. 
For the finger binding, we use the thumb, index, and/or pinky finger positions, depending on the available number of robots
The position data for each robot is sent between the remote and local PCs through UDP communication.
In our implementation, we set the Toio robot's speed up to 17.5 cm/sec taking into account the balance between the speed and accuracy.
Therefore, if the remote user attempts to move the local Toio robot at speed higher than this, it may lead to positional errors.
Considering the tradeoff between precise movements and jittering, we set the default tolerance to 1.1 cm for all interactions, except for the miniature body interaction, where we set it to 0.4 cm since accuracy with the avatar’s body was more crucial than some small jittering.
Finally, to increase the reproducibility, we make our software open source~\footnote{\url{https://github.com/KeiichiIhara/HoloBots}}.

\section{\system{} Design Space}
In this section, we explore the design space of \system{} in the following four dimensions: 1) interaction techniques, 2) actuation types, 3) surface types, and 4) physical attachment (Figure~\ref{fig:design-space}). 


\subsection{Interaction Techniques}
\subsubsection*{\textbf{Object Actuation}}\label{subsec:Storytelling}
\system{} offers various ways for remote users to interact with the local user. The object actuation enables remote users to move and manipulate objects in the local environment. For example, remote users can directly grab the Toio robot to move its location, or the attached object for more expressive engagement. 
\begin{figure}[h]
\centering
\includegraphics[width=0.32\linewidth]{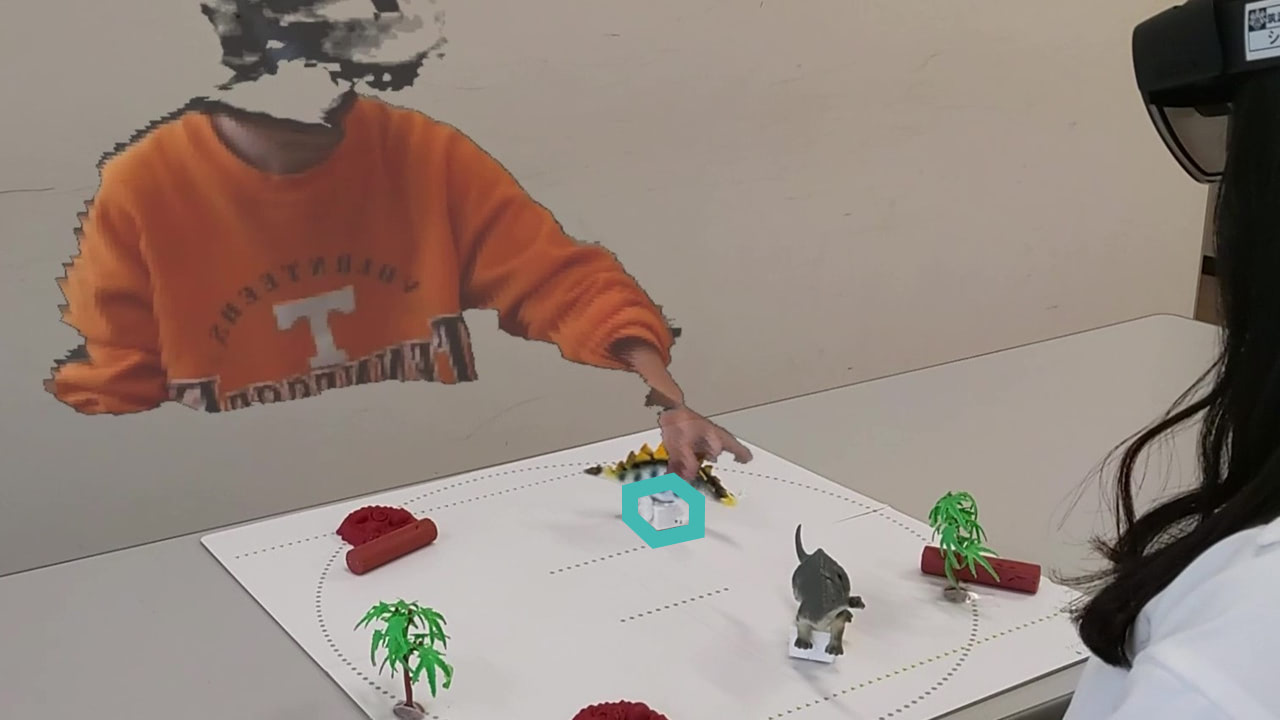}
\includegraphics[width=0.32\linewidth]{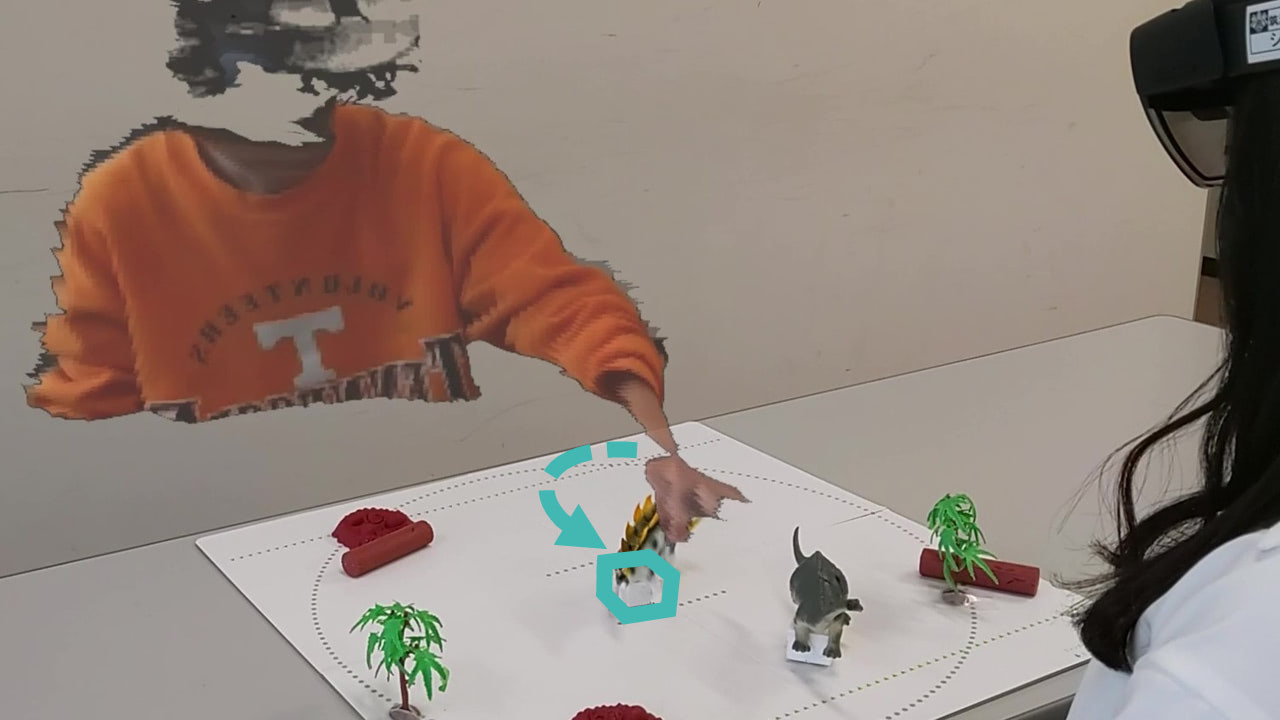}
\includegraphics[width=0.32\linewidth]{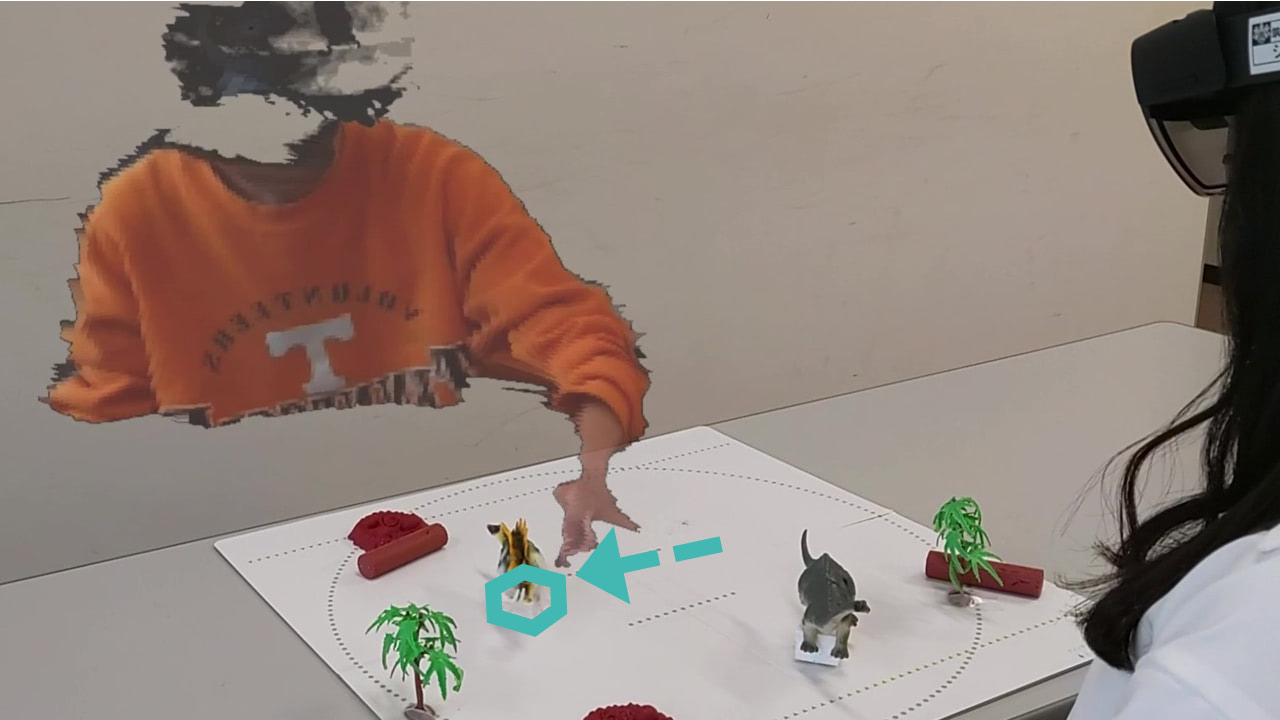}
\caption{Storytelling}
\label{fig:storytelling}
\end{figure}
\remove{
Figure~\ref{fig:object-actuation} illustrates the holographic remote user moving an animal toy.
We demonstrate object actuation for different use cases, such as storytelling, gaming, or even drawing.
\begin{figure}[h]
\centering
\includegraphics[width=0.32\linewidth]{figures/object-actuation-1-c.jpg}
\includegraphics[width=0.32\linewidth]{figures/object-actuation-2-c.jpg}
\includegraphics[width=0.32\linewidth]{figures/object-actuation-3-c.jpg}
\caption{Object Actuation}
\label{fig:object-actuation}
\end{figure}
}

The object actuation can be used for different use cases, such as storytelling, gaming, and drawing.
For storytelling, \system{} allows both local and remote users to participate in creating a story together with tangible objects.
The local user can either observe as an audience member or actively engage with the story-creation process.
Figure \ref{fig:storytelling} illustrates a remote user physically moving a dinosaur toy on a stage to narrate a story to the local user.
This provides, for example, engaging tangible storytelling for children and their remote parents or friends.

\subsubsection*{\textbf{Shared Tangible UI}}
\begin{figure}[h]
\centering
\includegraphics[width=0.32\linewidth]{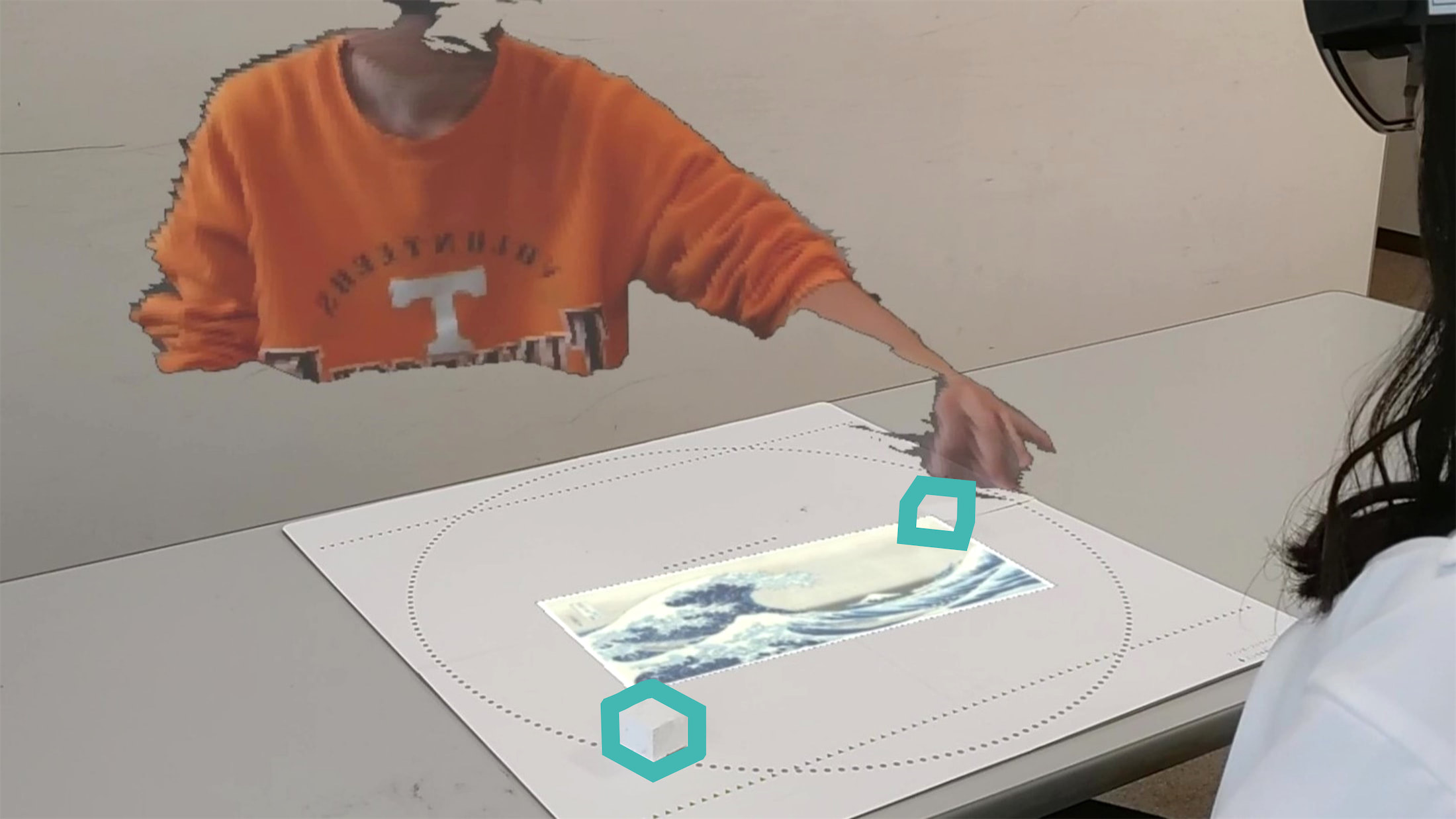}
\includegraphics[width=0.32\linewidth]{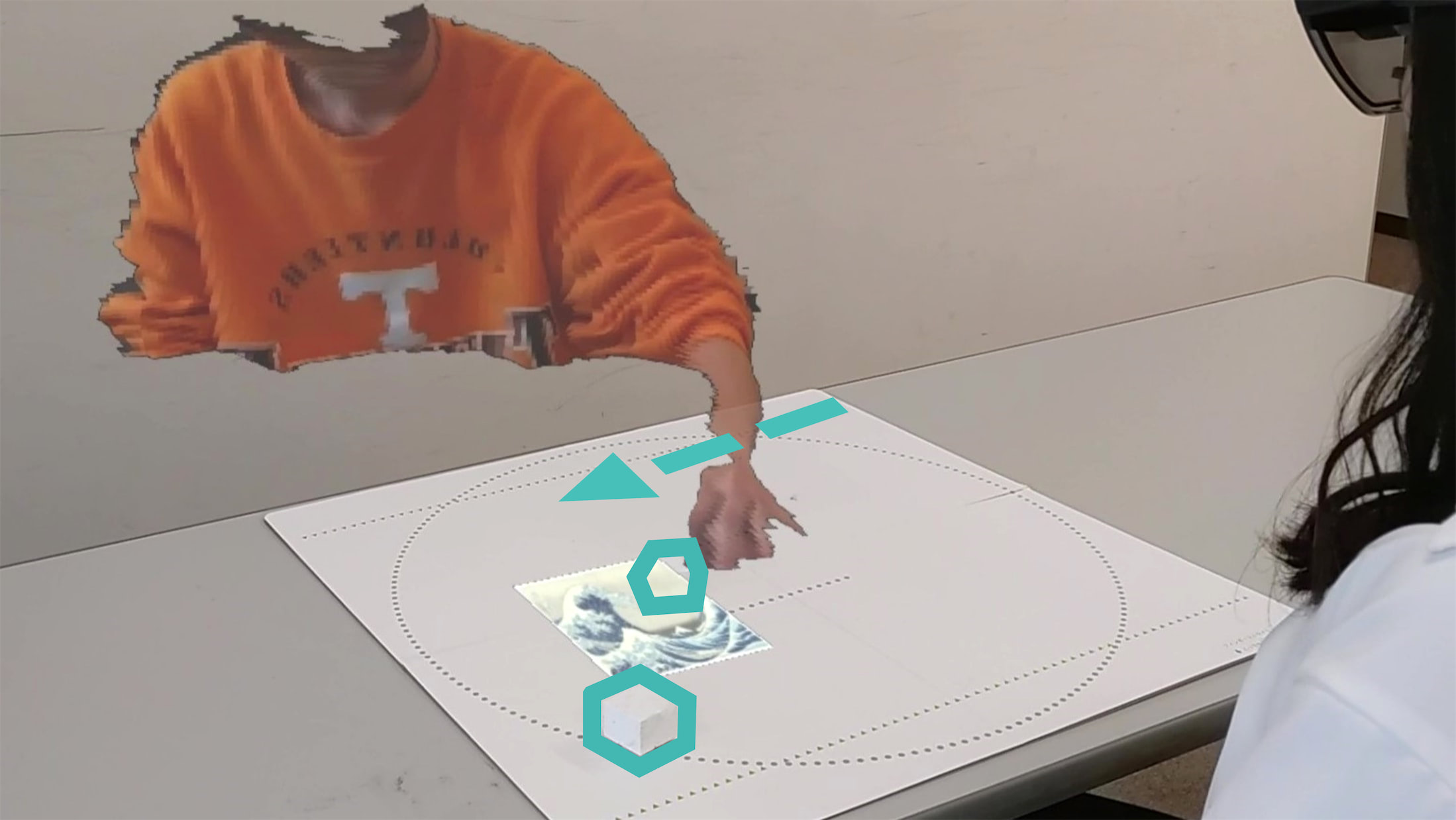}
\includegraphics[width=0.32\linewidth]{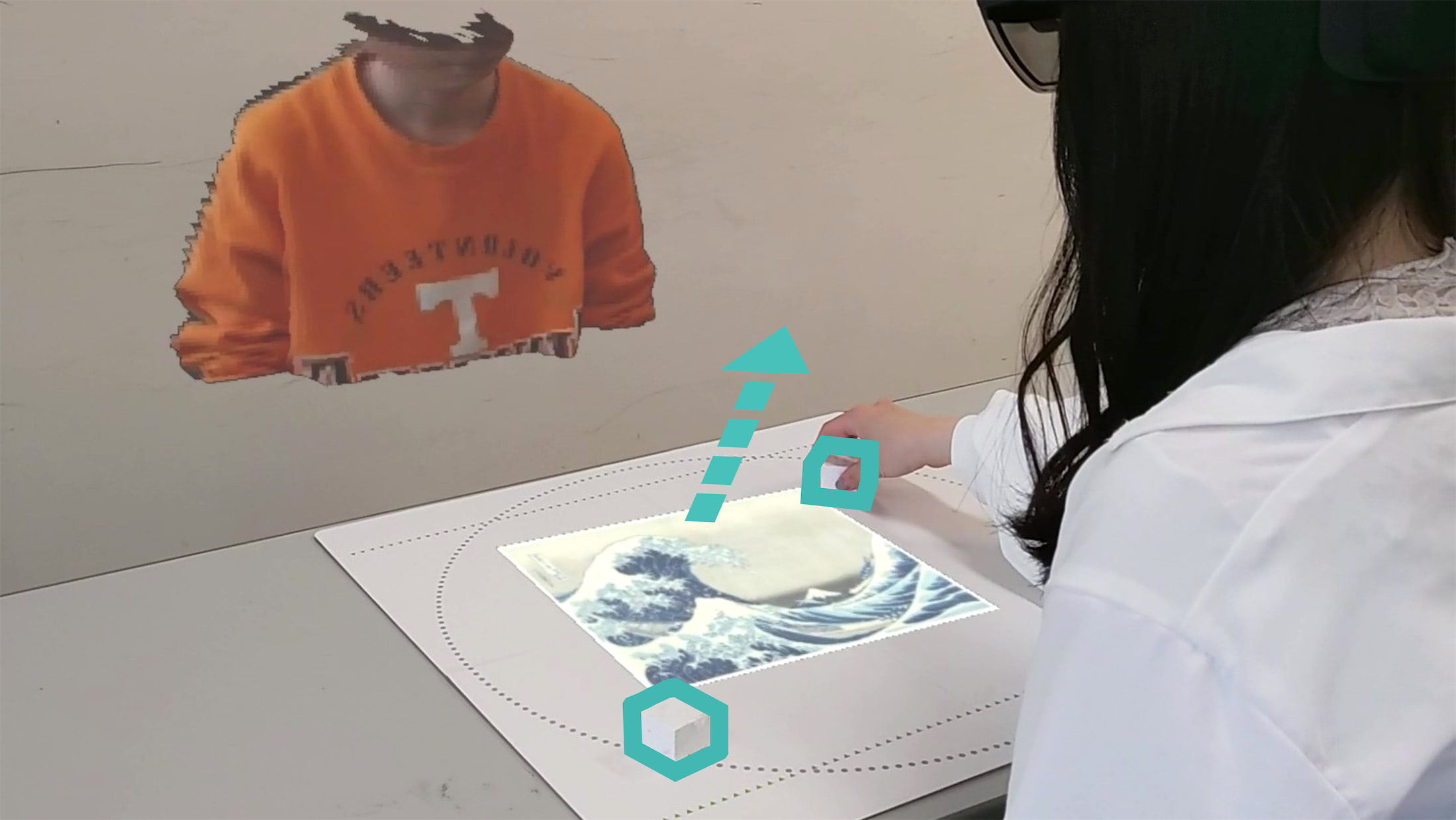}
\caption{Shared Tangible UI}
\label{fig:tangible-ui}
\end{figure}

Another interaction technique is the shared tangible user interface, which allows both local and remote users to manipulate virtual object properties through tangible tokens. 
Toio robots can be represented as various tangible UI elements, such as control points, buttons, sliders, and knobs, so that by controlling the same UI, local and remote users can manipulate the UI together.
For example, Figure~\ref{fig:tangible-ui} illustrates users changing the position and scale of the virtual picture by manipulating robots, which are represented as a control point of the image.

\begin{figure}[h]
\centering
\includegraphics[width=0.32\linewidth]{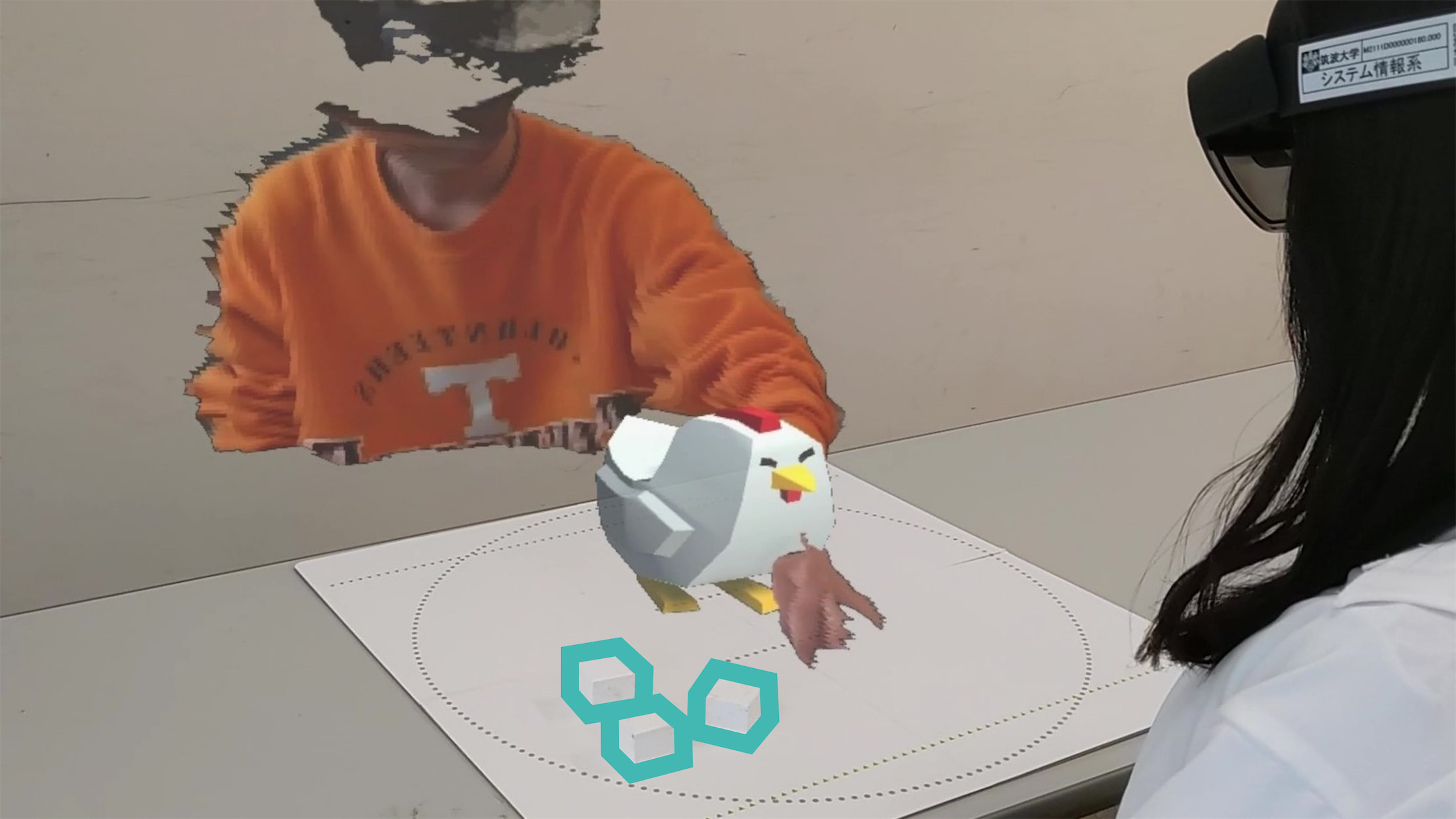}
\includegraphics[width=0.32\linewidth]{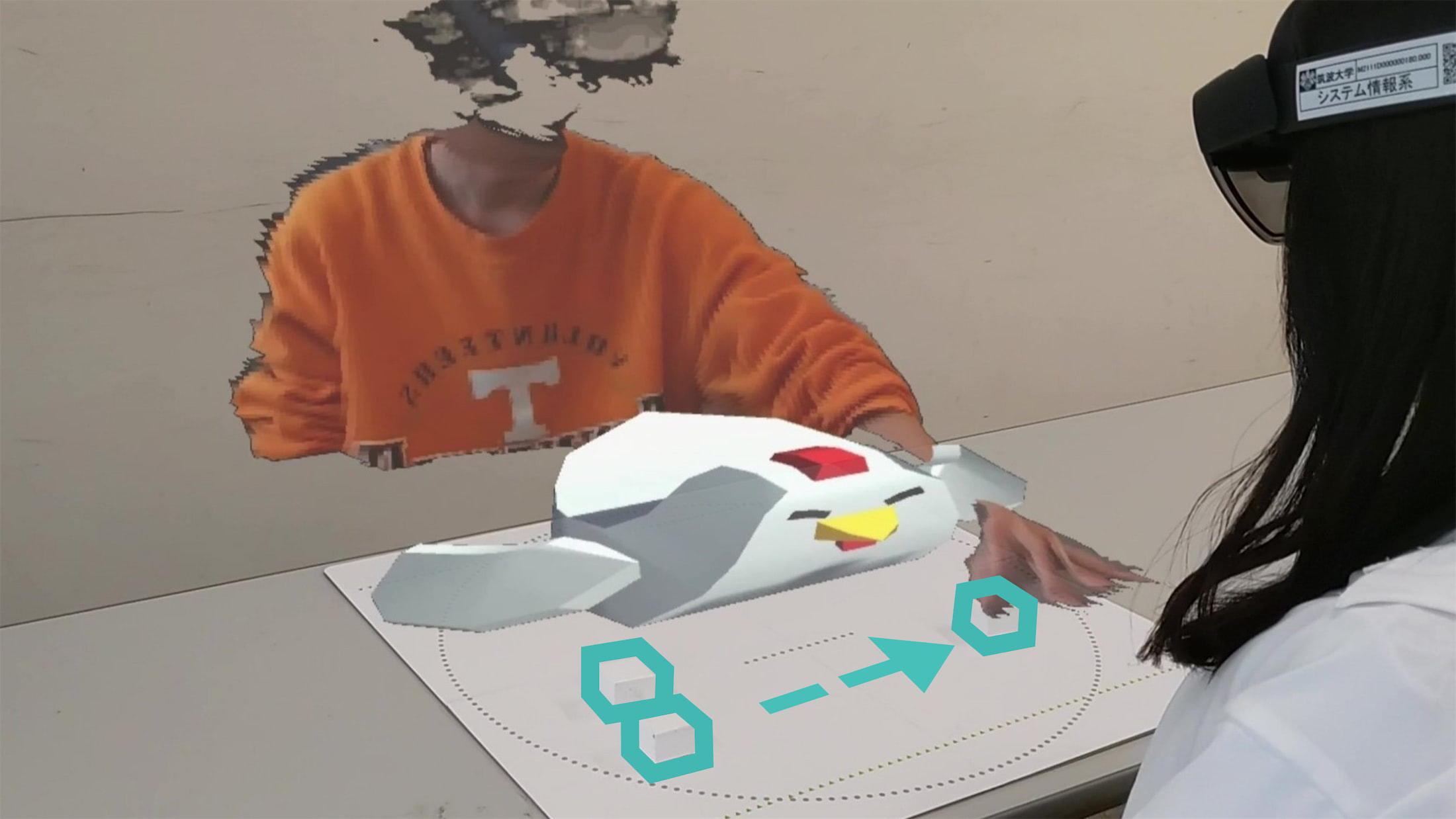}
\includegraphics[width=0.32\linewidth]{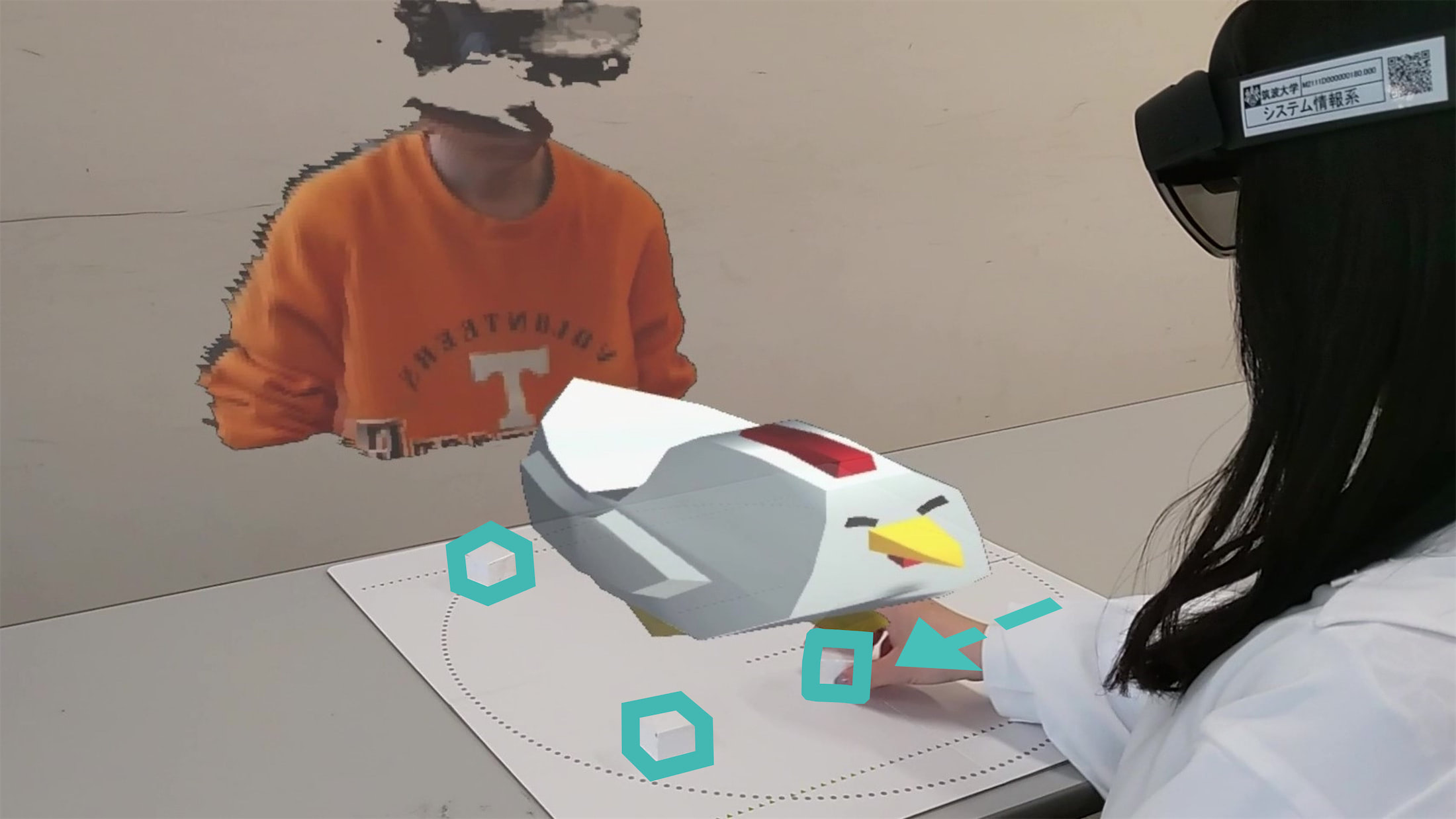}
\caption{Collaborative Design}
\label{fig:collaborative-design}
\end{figure}

In Figure~\ref{fig:collaborative-design}, three Toios are used as a tangible UI for manipulating a 3D object. Two  of the Toios represent sliders and adjust the width and depth of the object, while the third Toio represents a knob and alters the object's height.

\subsubsection*{\textbf{Miniature Body Interaction}}
\begin{figure}[h]
\centering
\includegraphics[width=0.32\linewidth]{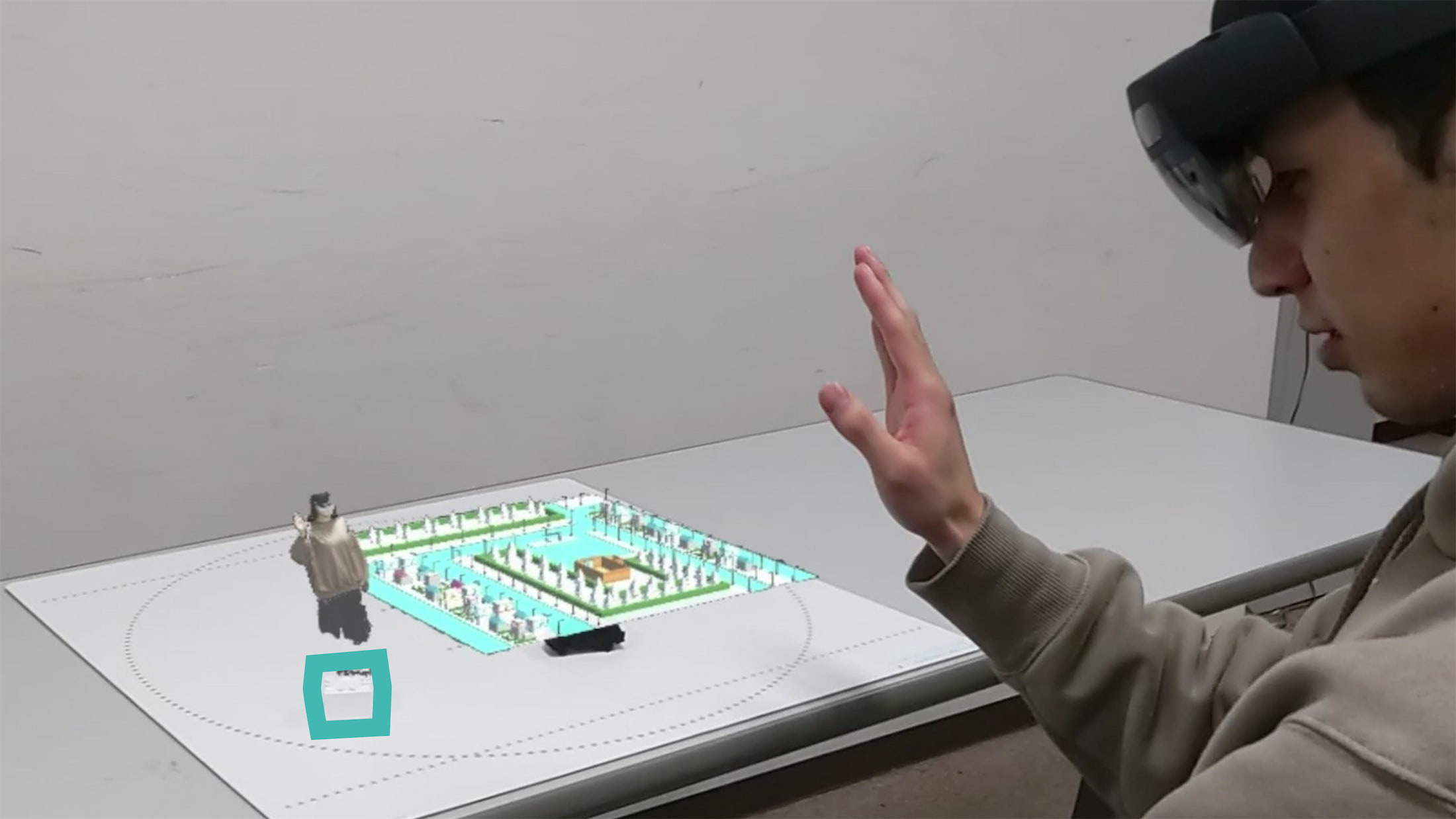}
\includegraphics[width=0.32\linewidth]{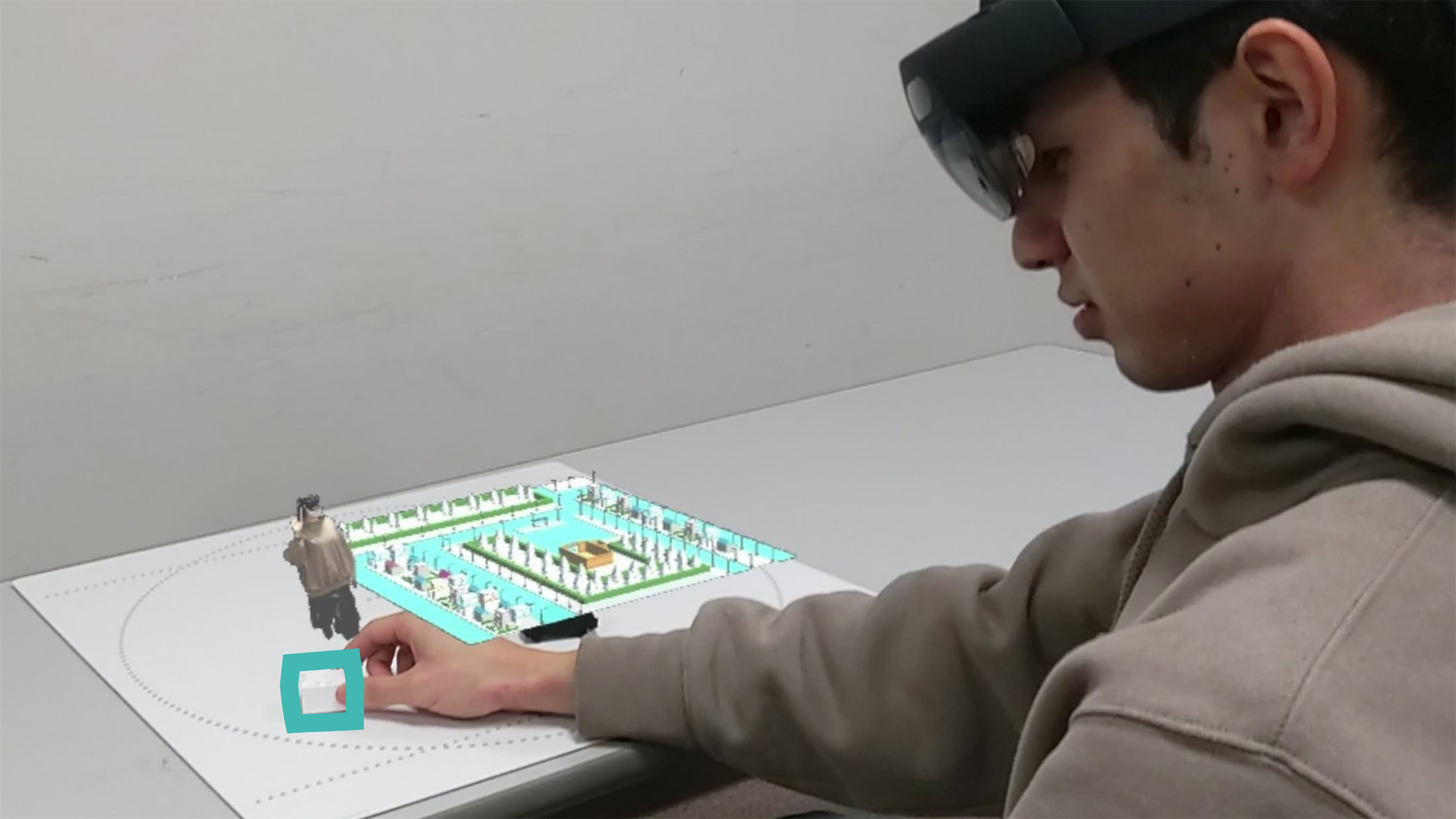}
\includegraphics[width=0.32\linewidth]{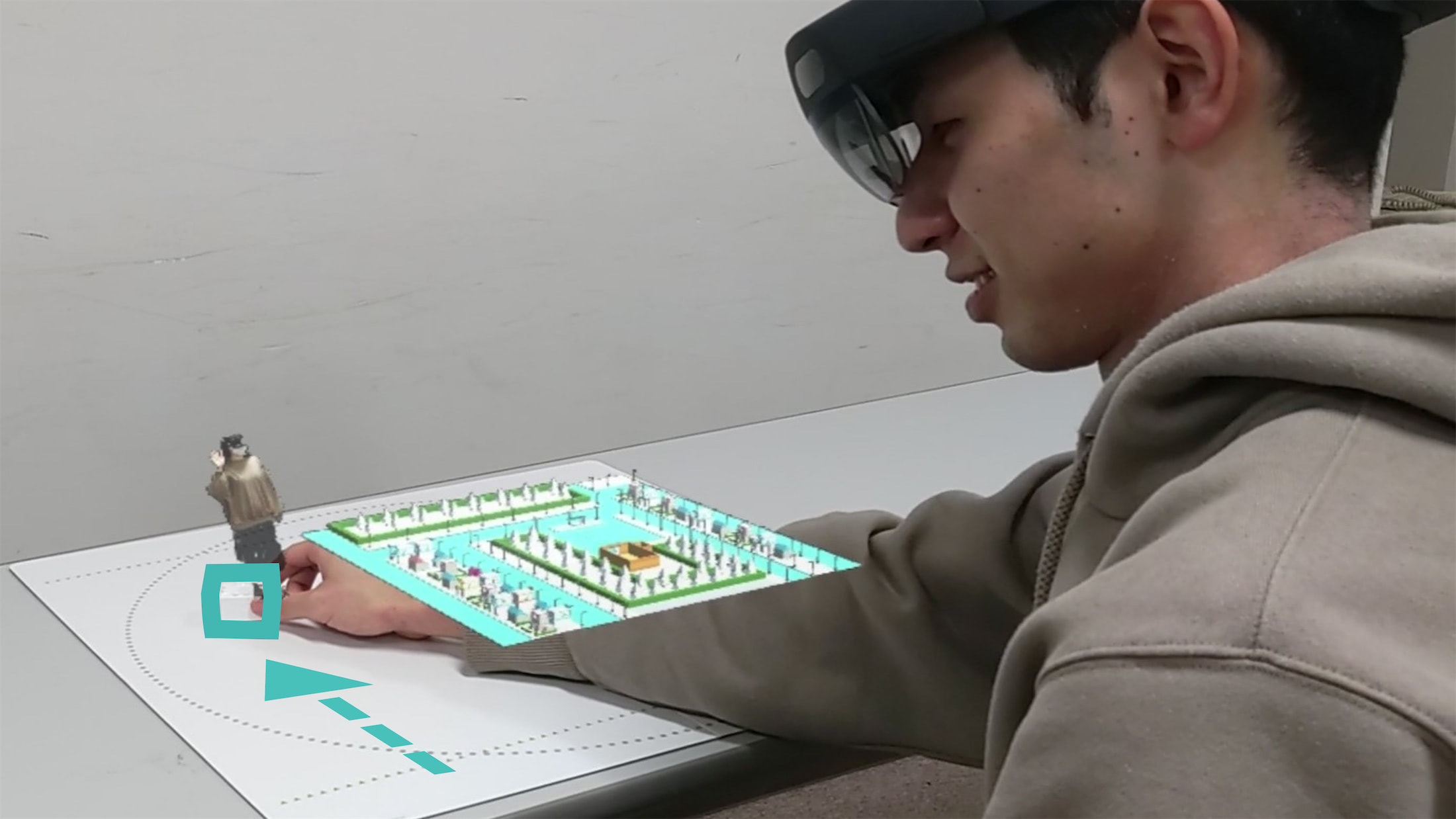}
\caption{Miniature Body Interaction}
\label{fig:miniature-body}
\end{figure}

The robot can also embody the remote user through a miniature body. 
Our system also facilitates the collaborative world-in-miniature exploration, by representing as the miniature user. 
Similar to the prior work that explores tangible world-in-miniature exploration (e.g., \textit{miniStudio}~\cite{kim2016ministudio}, \textit{Does it Feel Real?}~\cite{muender2019does}, \textit{Shoulder of Giants}~\cite{piumsomboon2019shoulder}, \textit{ASTEROIDS}~\cite{li2022asteroids}), the tangible embodiment of the miniature user facilitates rich physical affordances for the world-in-miniature interaction, while providing effective visual feedback through holographic representation.
The remote user can walk around on a real-size environment, which is captured and tracked through Azure Kinect body tracking. 
For example, the remote user can visually instruct the local user using gestures and physically move objects in the local environment by pushing them with Toios (Figure~\ref{fig:miniature-body}). 
\remove{This technique can be used for several applications, such as architecture or interior design.}

\begin{figure}[h]
\centering
\includegraphics[width=0.32\linewidth]{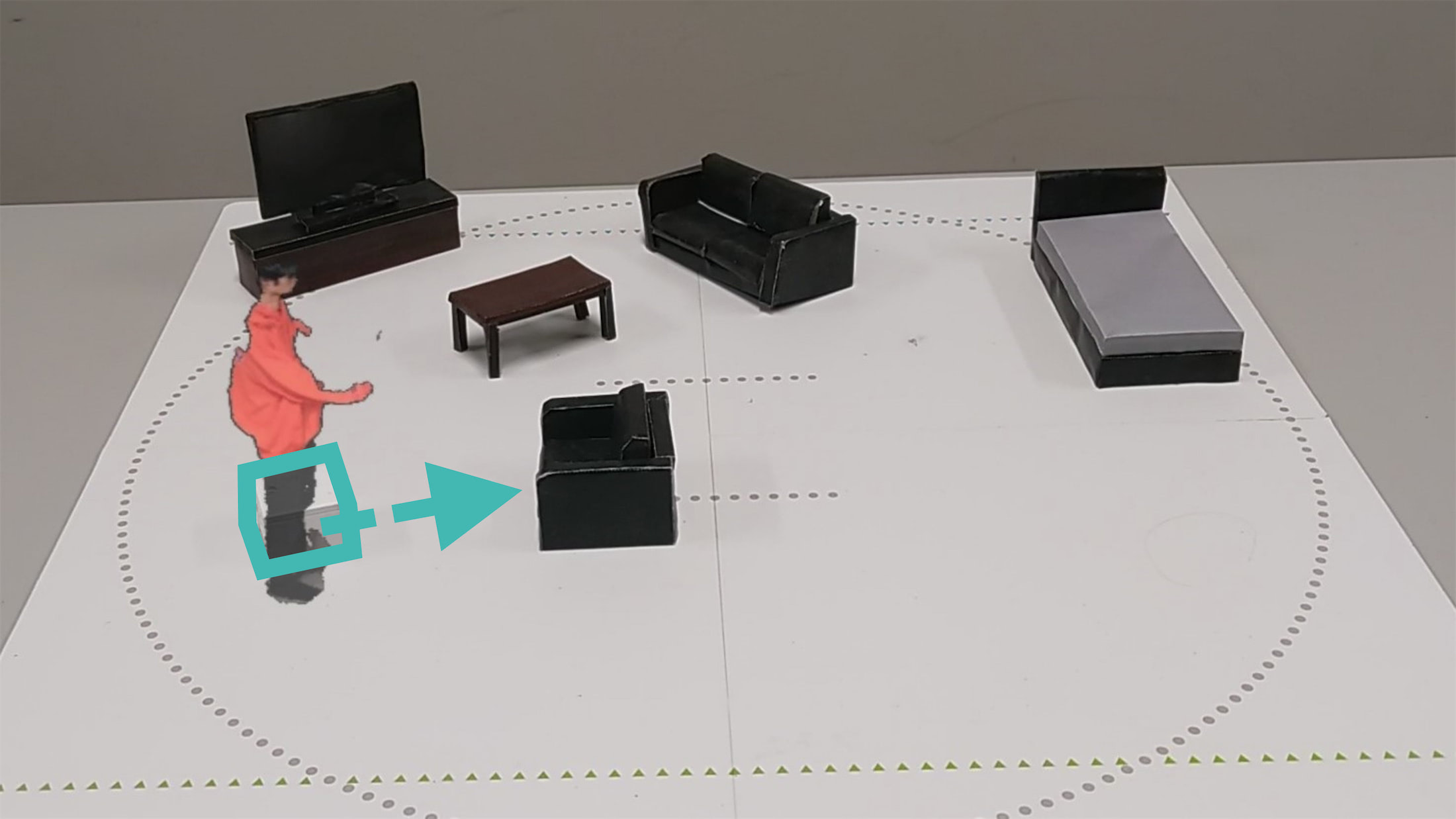}
\includegraphics[width=0.32\linewidth]{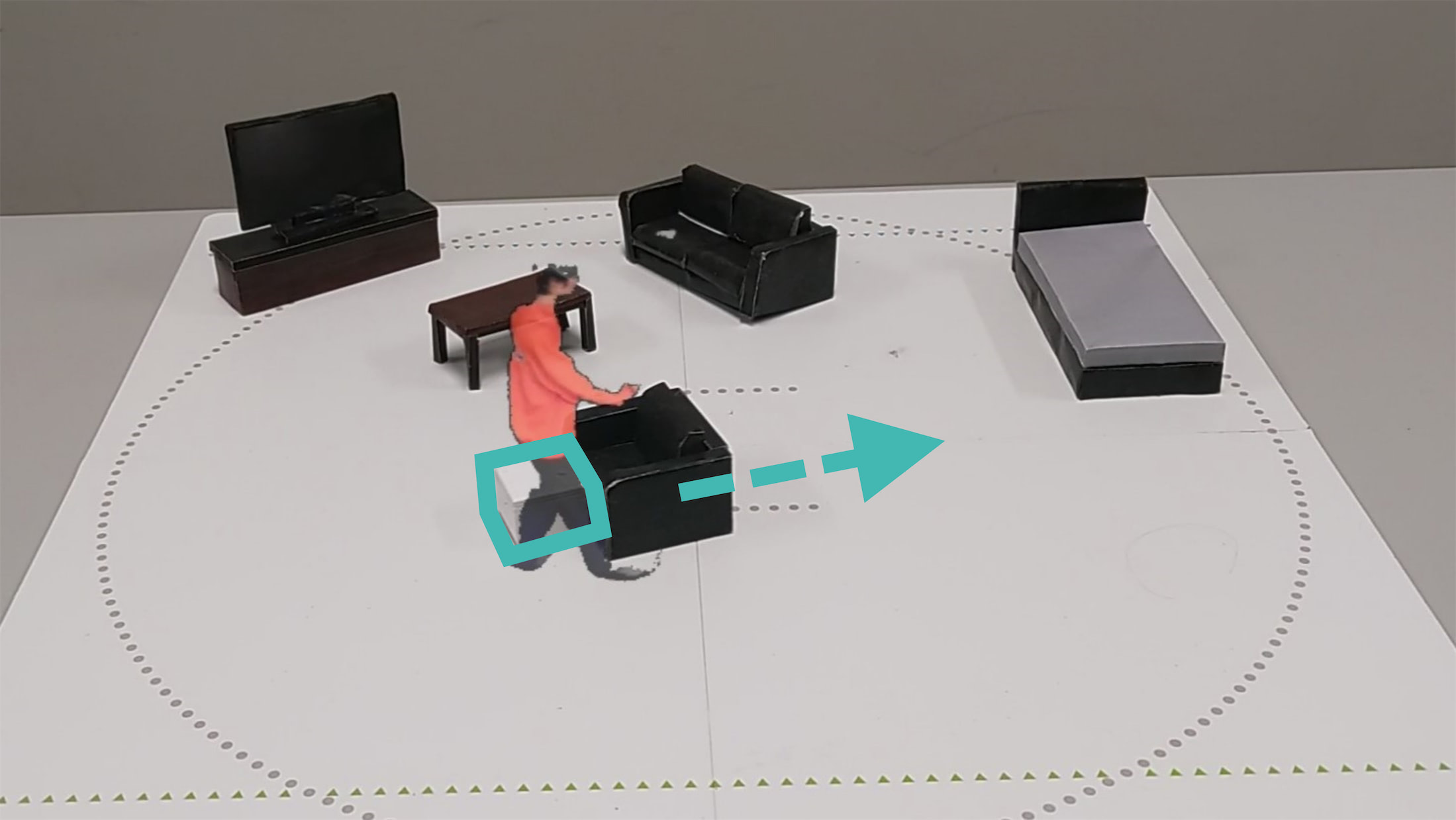}
\includegraphics[width=0.32\linewidth]{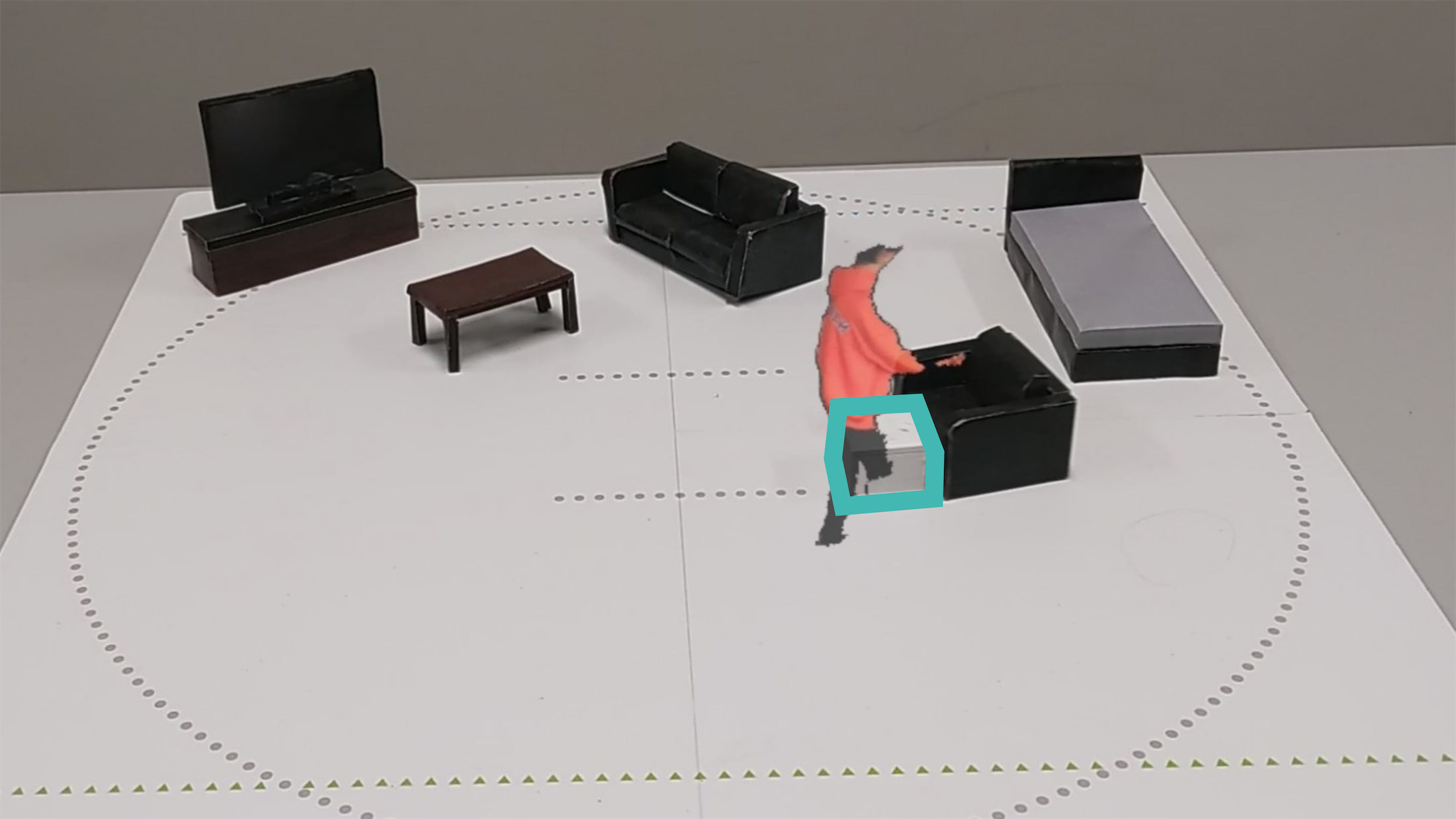}
\caption{Interior Design}
\label{fig:interior-design}
\end{figure}

Taking inspiration from the immersive interior and architectural design (e.g., \textit{DollhouseVR}~\cite{ibayashi2015dollhouse}), this could be used for the collaborative world-in-miniature exploration, in which the robot can embody the physical representation of the miniature user. 
For example, Figure~\ref{fig:interior-design} illustrates an application for collaborative interior design.
This application uses miniature furniture to facilitate discussion and decision-making between remote and local users.
The remote user is visually represented as a miniature avatar, with a Toio representing the remote user's physical body. 
The remote user can visually instruct the local user using gestures and physically push the miniature furniture to arrange the position.

\subsubsection*{\textbf{Haptic Communication}}
Haptic communication is another interaction technique that enables the remote user to provide haptic feedback to the local user. 
There are various ways to provide haptic communication. 
For example, the user can guide the Toio robot to navigate the remote user based on the actuation, similar to \textit{dePend}~\cite{yamaoka2013depend}, as if they were holding their hands. 
This technique can be used for hands-on instruction. 
Alternatively, the remote user can physically touch the local user by moving and touching the local user's body using Toios, similar to \textit{SwarmHaptics}~\cite{kim2019swarmhaptics}.
This can be used for remote social interaction.

\begin{figure}[h]
\centering
\includegraphics[width=0.32\linewidth]{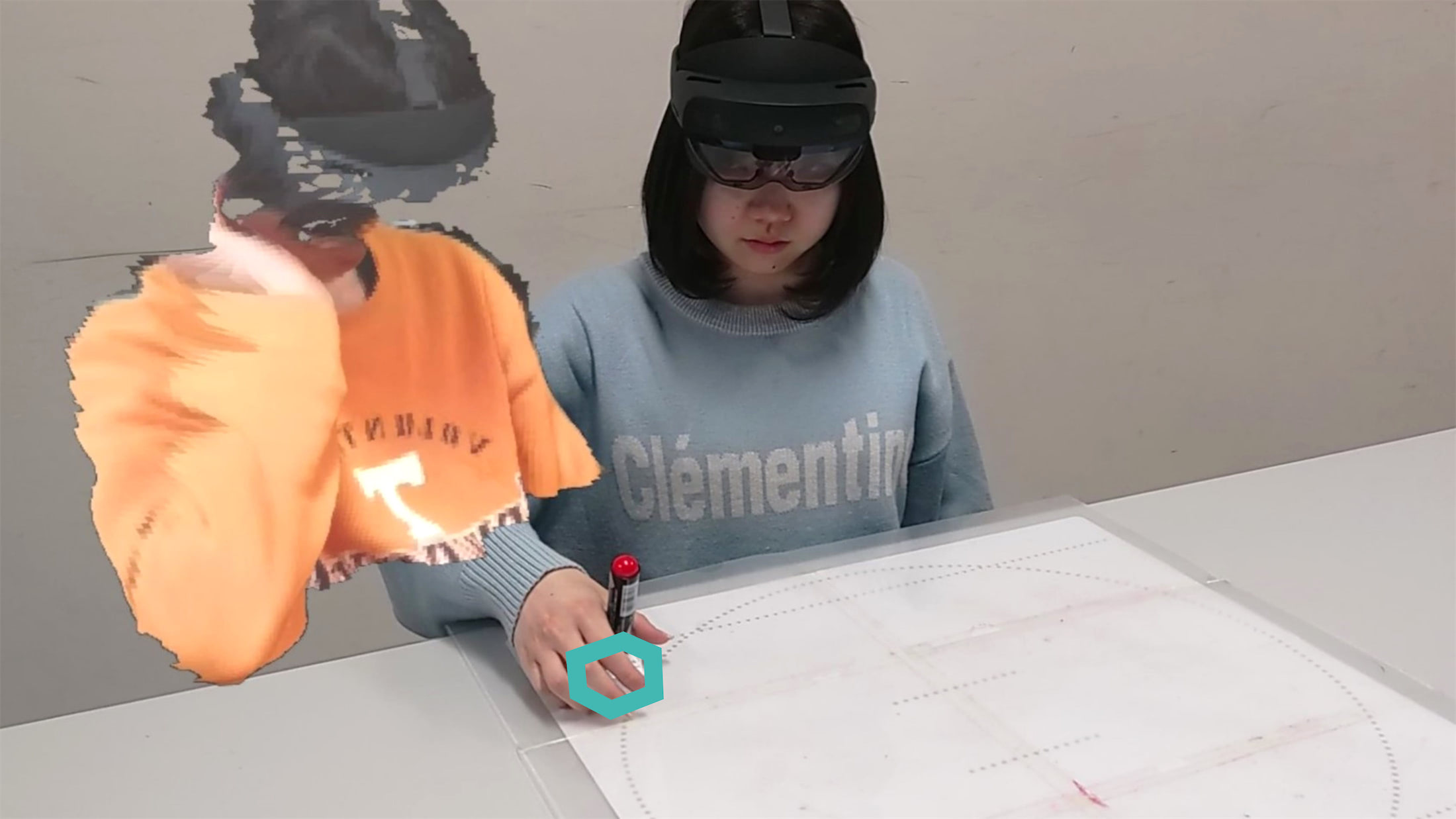}
\includegraphics[width=0.32\linewidth]{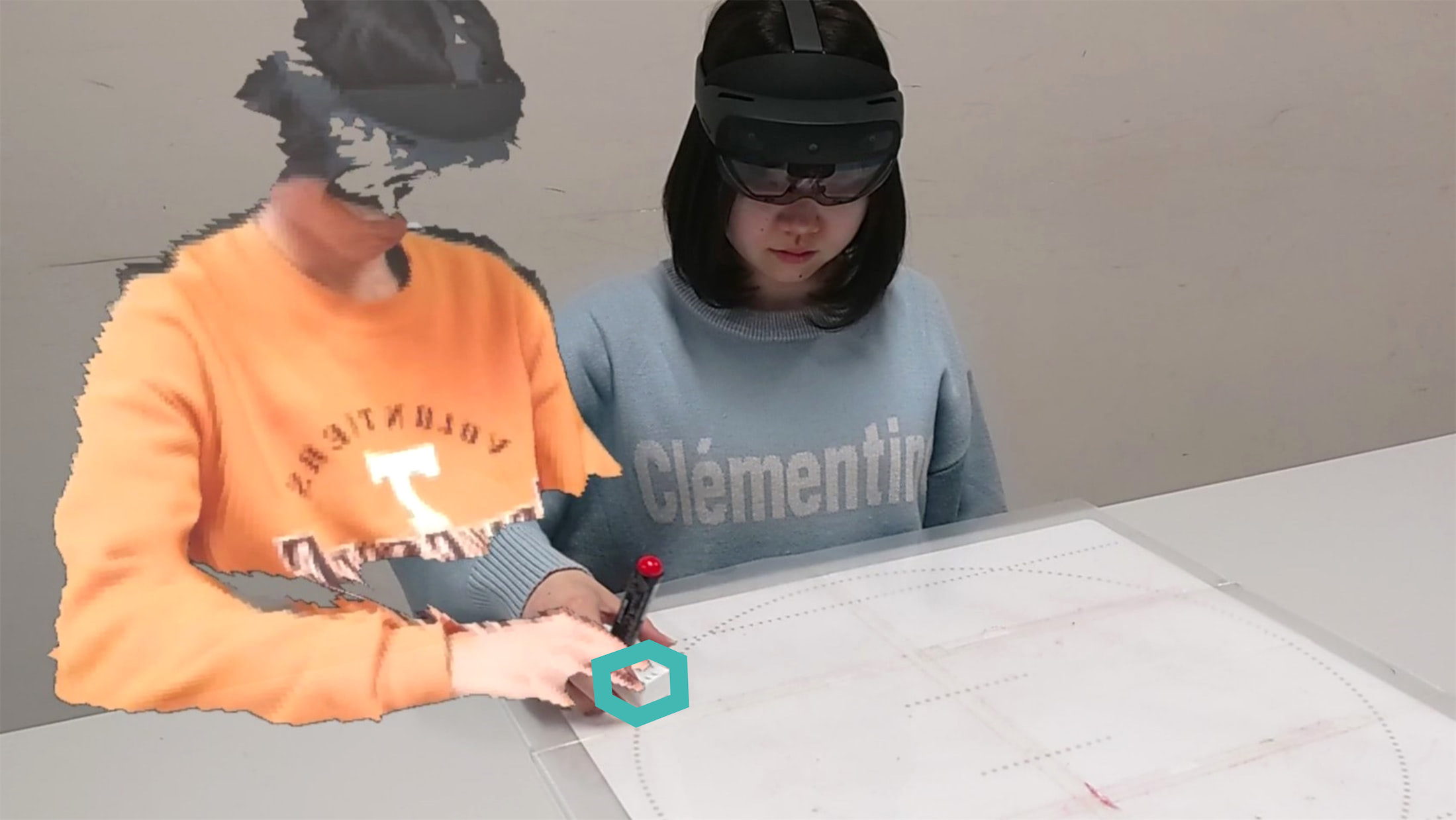}
\includegraphics[width=0.32\linewidth]{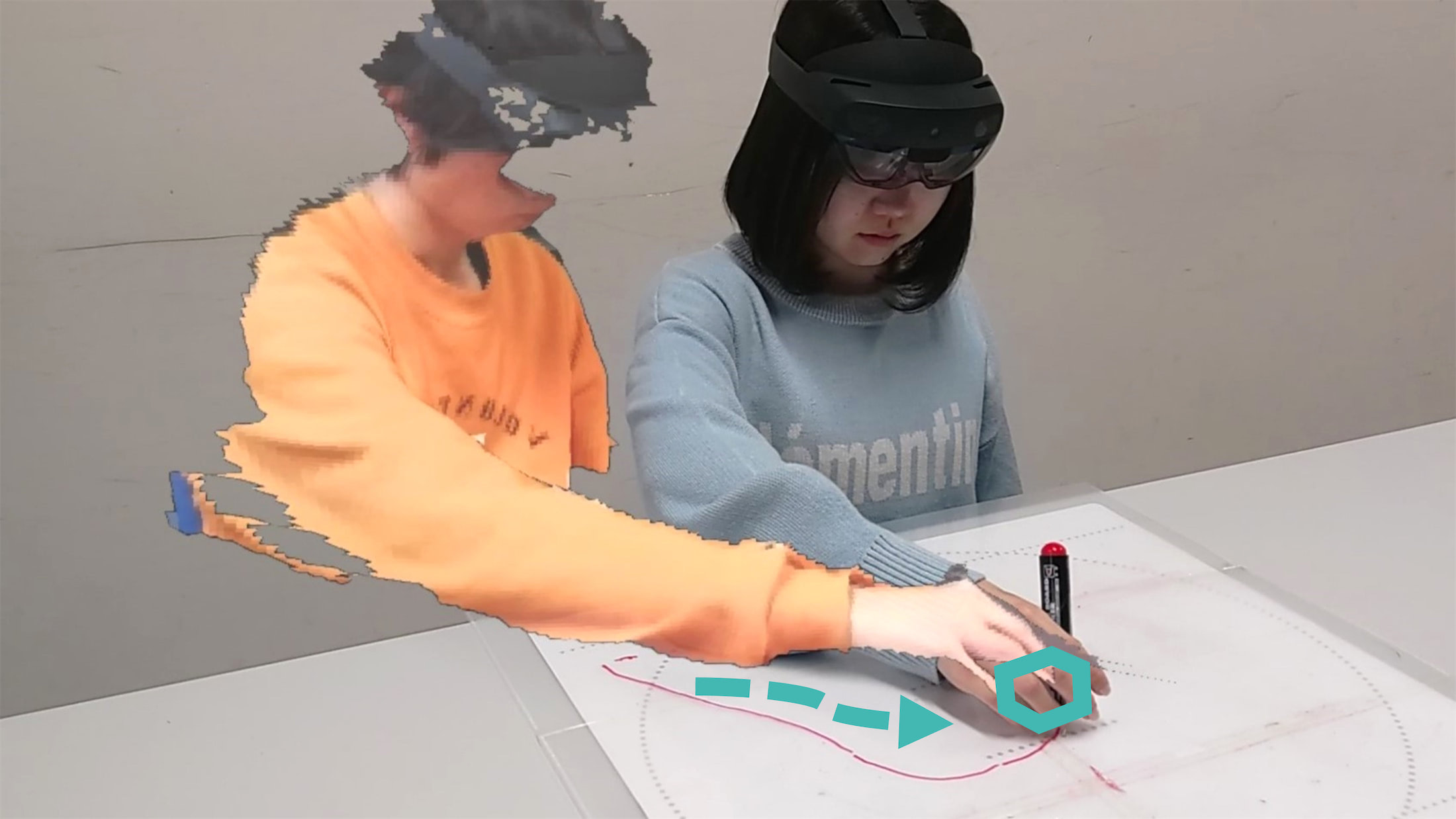}
\caption{Haptic Communication}
\label{fig:guide}
\end{figure}

Figure~\ref{fig:guide} shows a remote user controlling the movement of a red pen to draw on the physical canvas.
By attaching a physical pen to a Toio, the remote user can move the pen and draw on a physical canvas.
Local and remote users can therefore collaborate in real-time to create drawings and illustrations together.


\begin{figure}[h]
\centering
\includegraphics[width=0.32\linewidth]{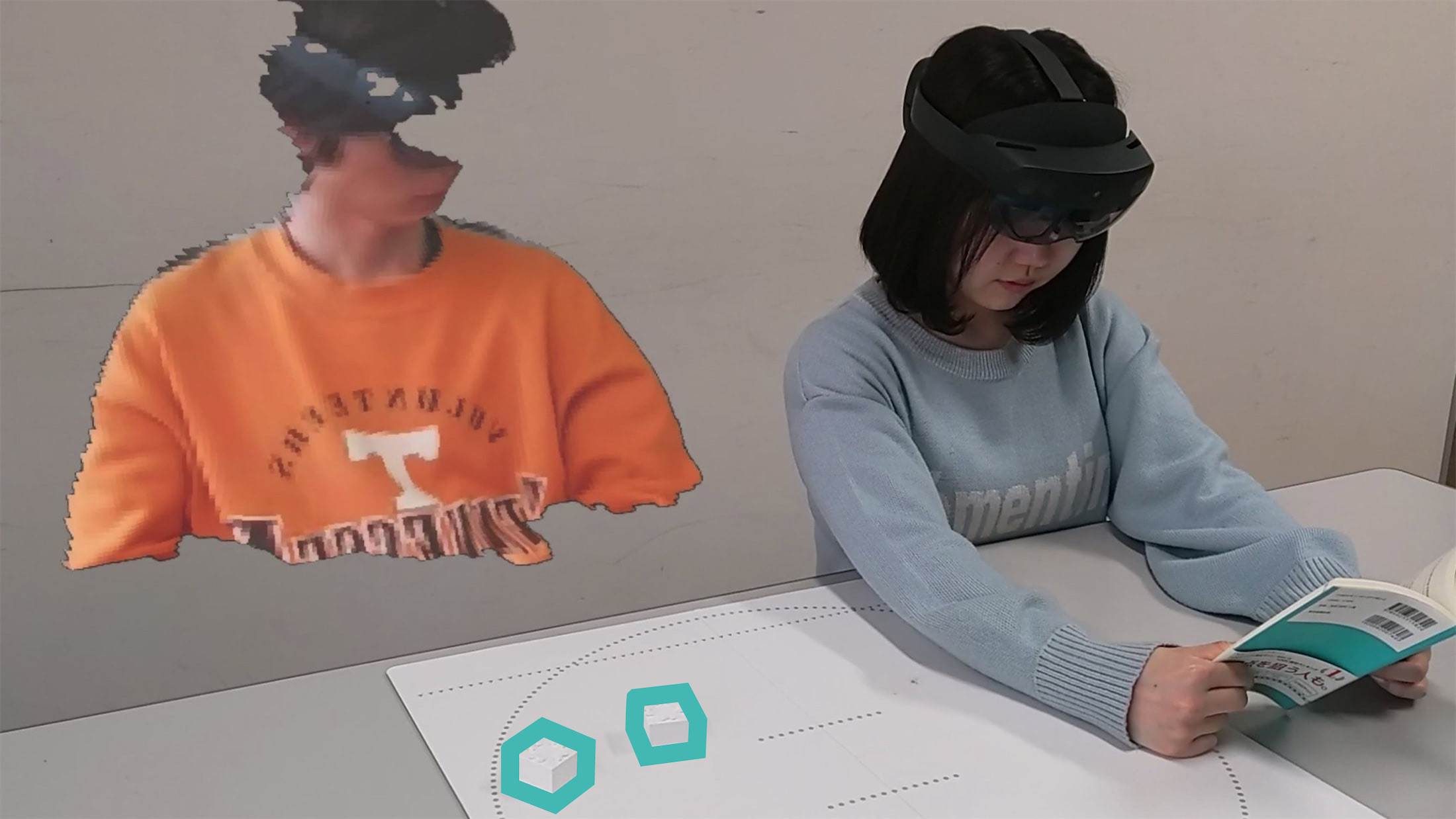}
\includegraphics[width=0.32\linewidth]{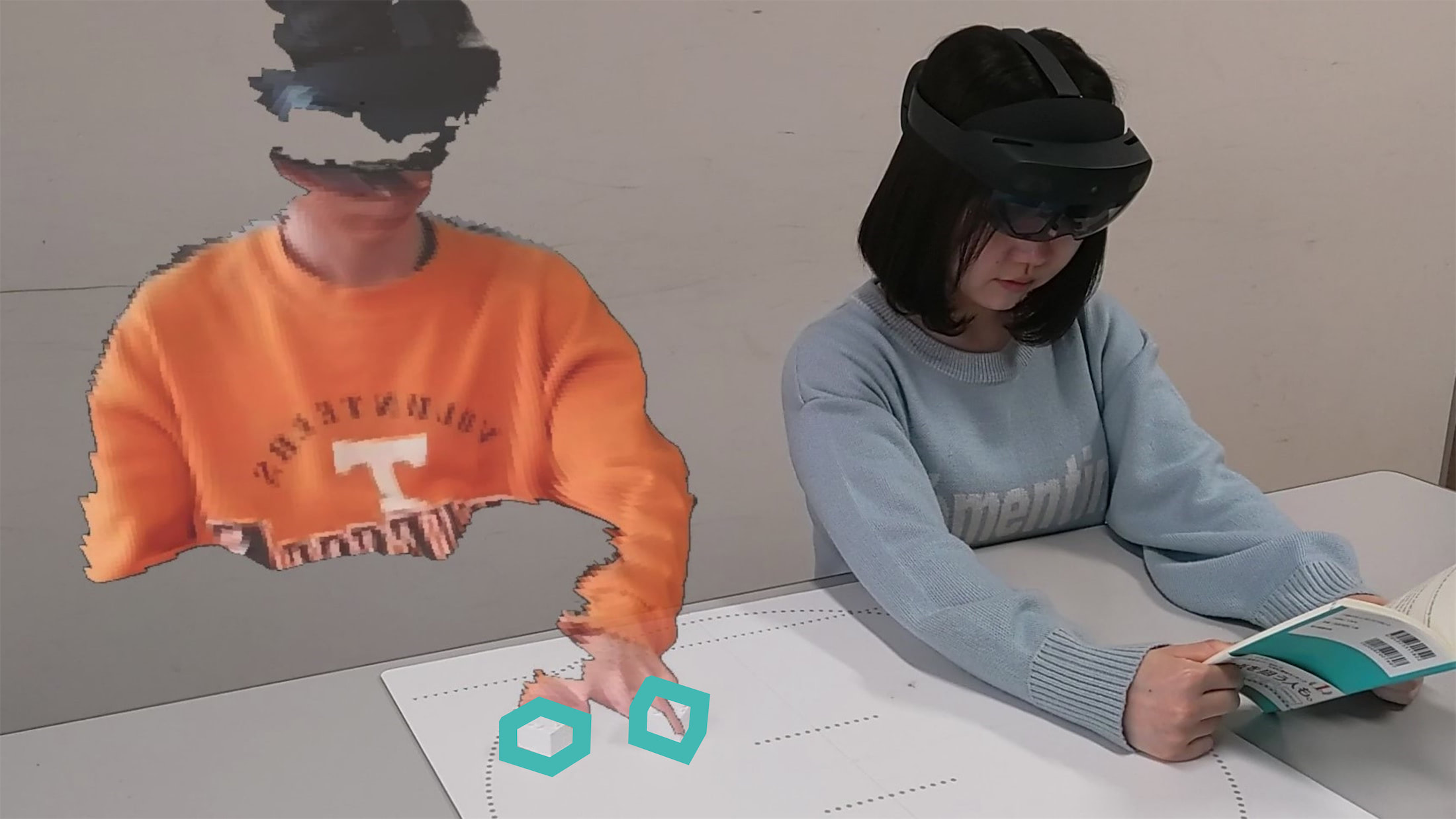}
\includegraphics[width=0.32\linewidth]{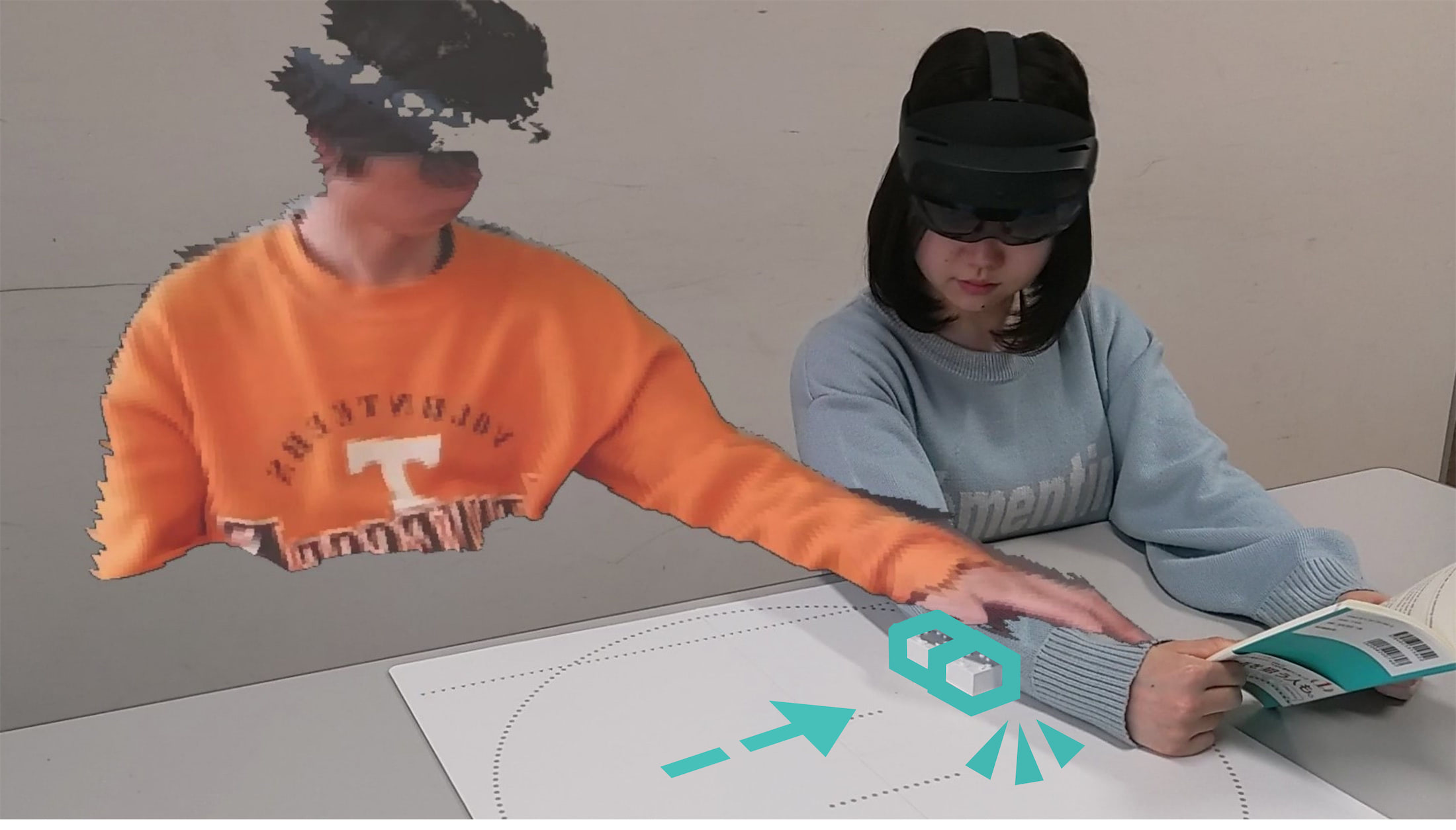}
\caption{Notification}
\label{fig:notification}
\end{figure}

This can also provide haptic notifications, enabling remote users to physically notify local users using Toios.
By attaching Toios to the remote avatar's hand, the remote user can touch the local user and initiate communication.
In Figure ~\ref{fig:notification}, the remote user touches the local user who is reading a book to start a conversation.

\subsection{Actuation Types}
\subsubsection*{\textbf{Move Active Object}}
In \system{}, the remote users can actuate physical objects in two ways. 
First, the user can simply grasp and move the Toio robot itself. By moving the Toio, which is attached to the various object, \system{} enables the remote user to actuate physical objects (Figure~\ref{fig:storytelling}). 

\begin{figure}[H]
\centering
\resizebox{\columnwidth}{!}{
\includegraphics[width=0.32\linewidth]{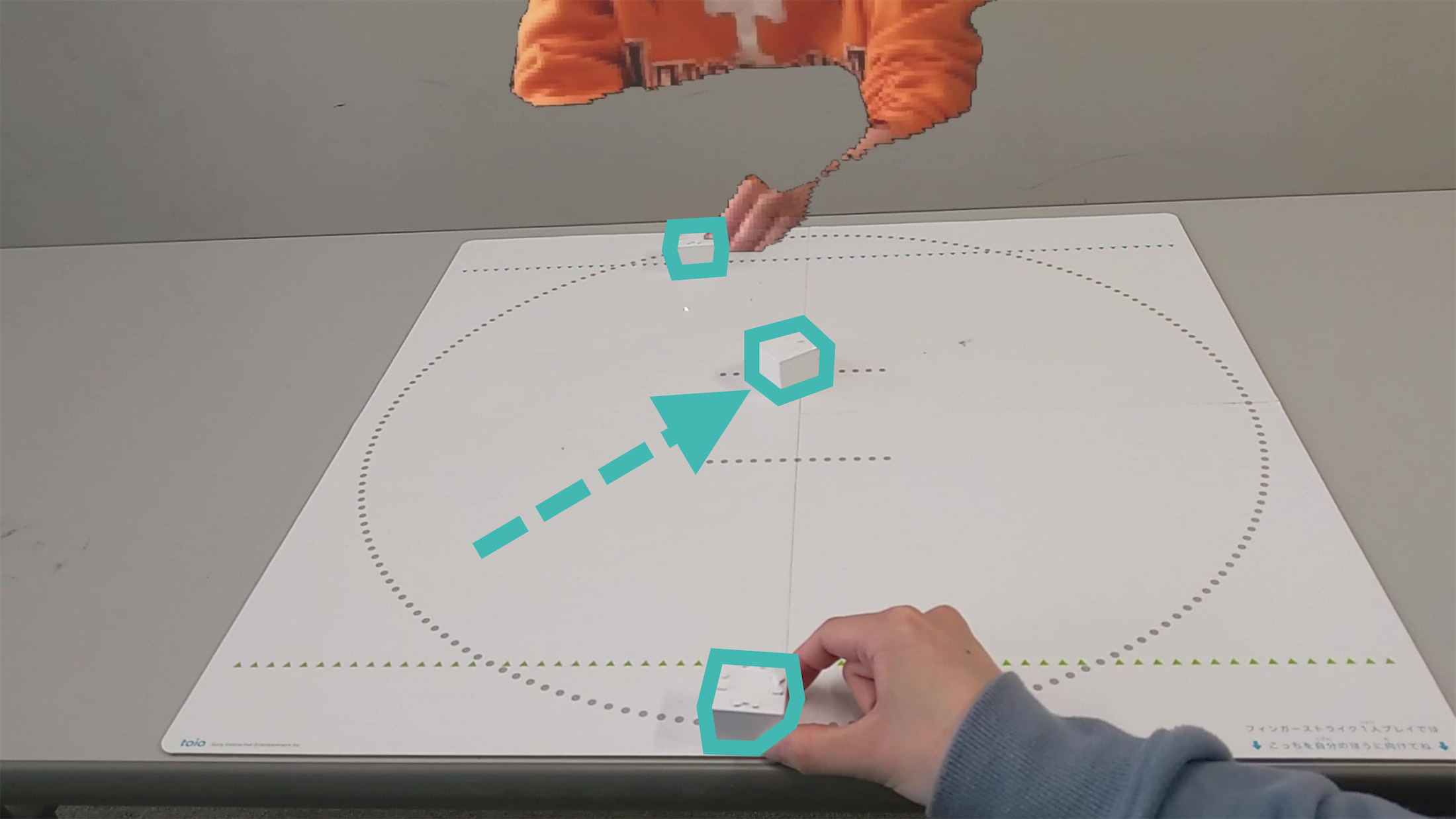}
\includegraphics[width=0.32\linewidth]{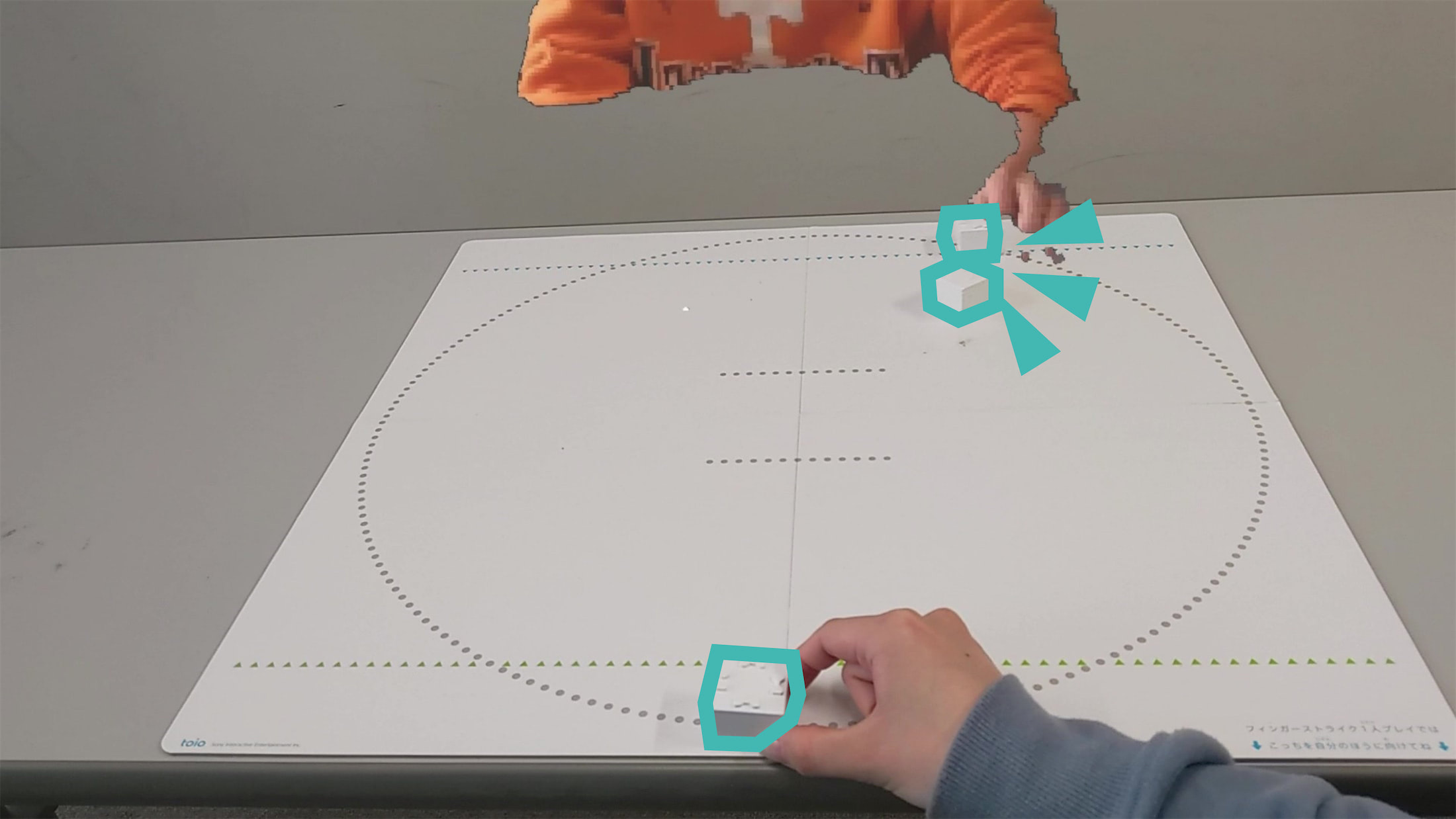}
\includegraphics[width=0.32\linewidth]{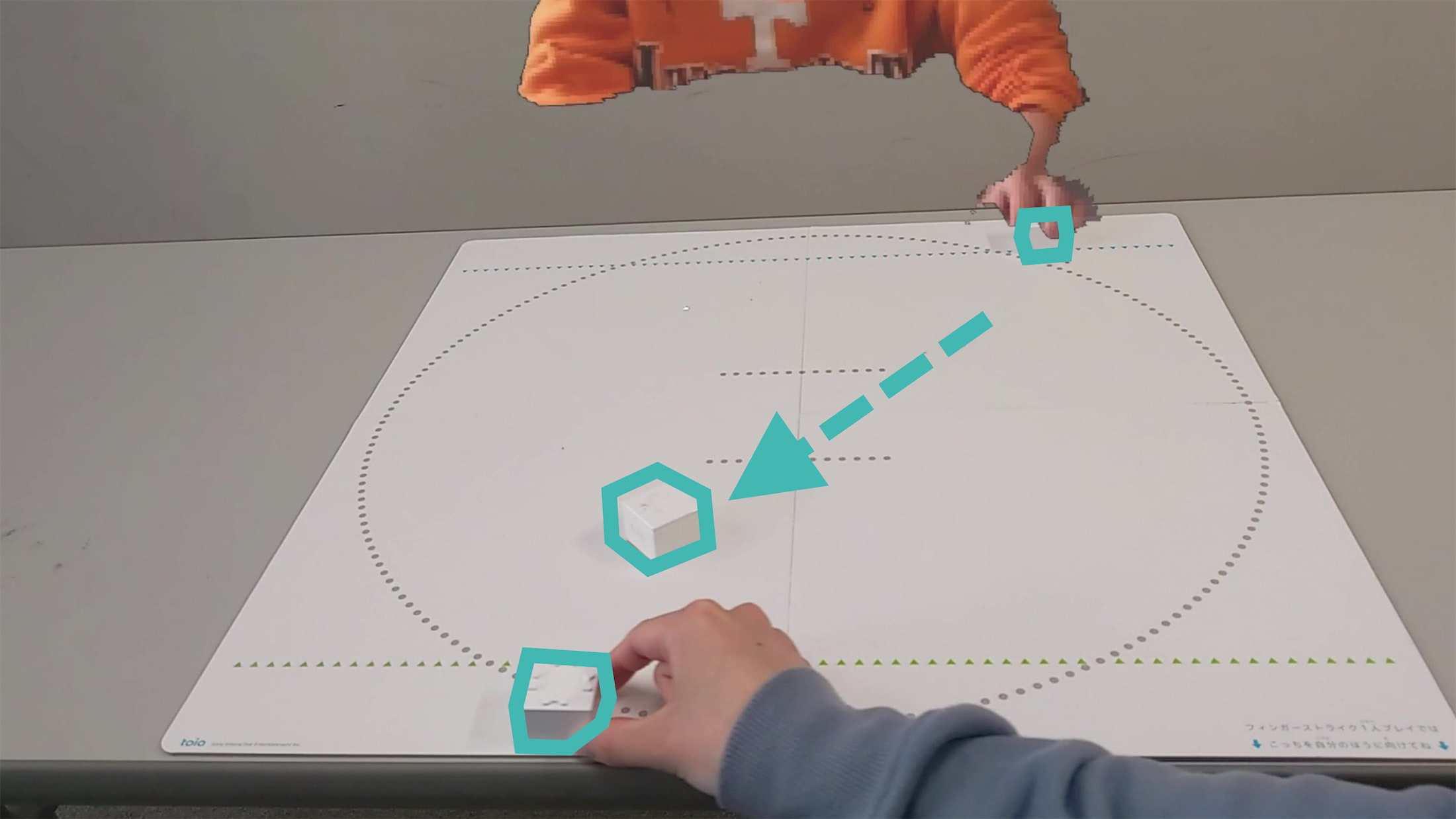}
}
\caption{Table Hockey}
\label{fig:hockey}
\end{figure}

One possible application is the remote gaming experience.
By attaching Toios to game objects, the local user can physically interact with the remote user through the tangible game.
Figure~\ref{fig:hockey} depicts a table hockey game application that utilizes three Toios---two for the mallets and one for the puck, similar to~\cite{nakagaki2022dis, kaimoto2022sketched}. 
This application allows users to play and compete with each other in real-time, creating an engaging and immersive gaming experience.


\subsubsection*{\textbf{Move Passive Object}}
\begin{figure}[h]
\centering
\includegraphics[width=0.32\linewidth]{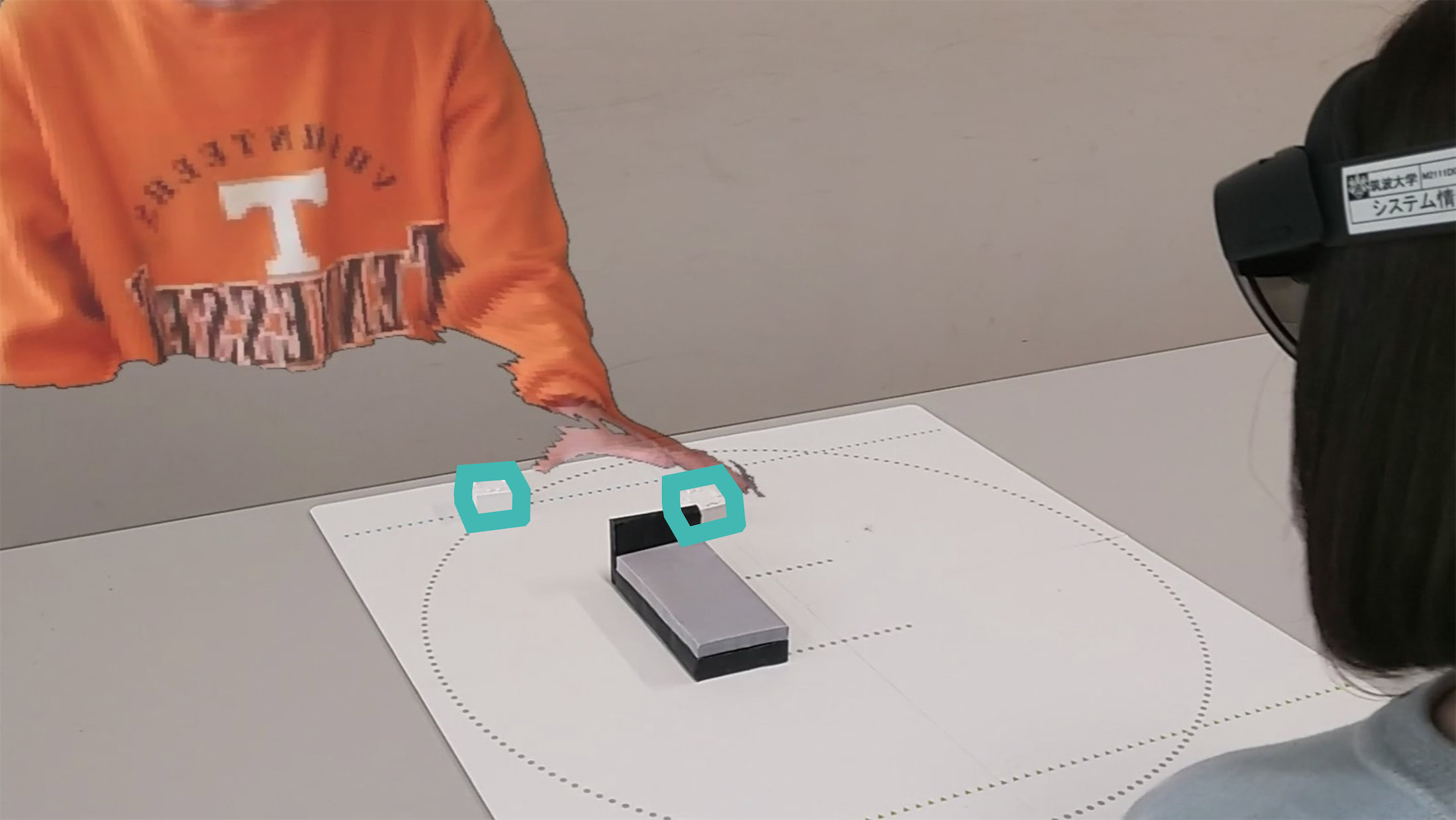}
\includegraphics[width=0.32\linewidth]{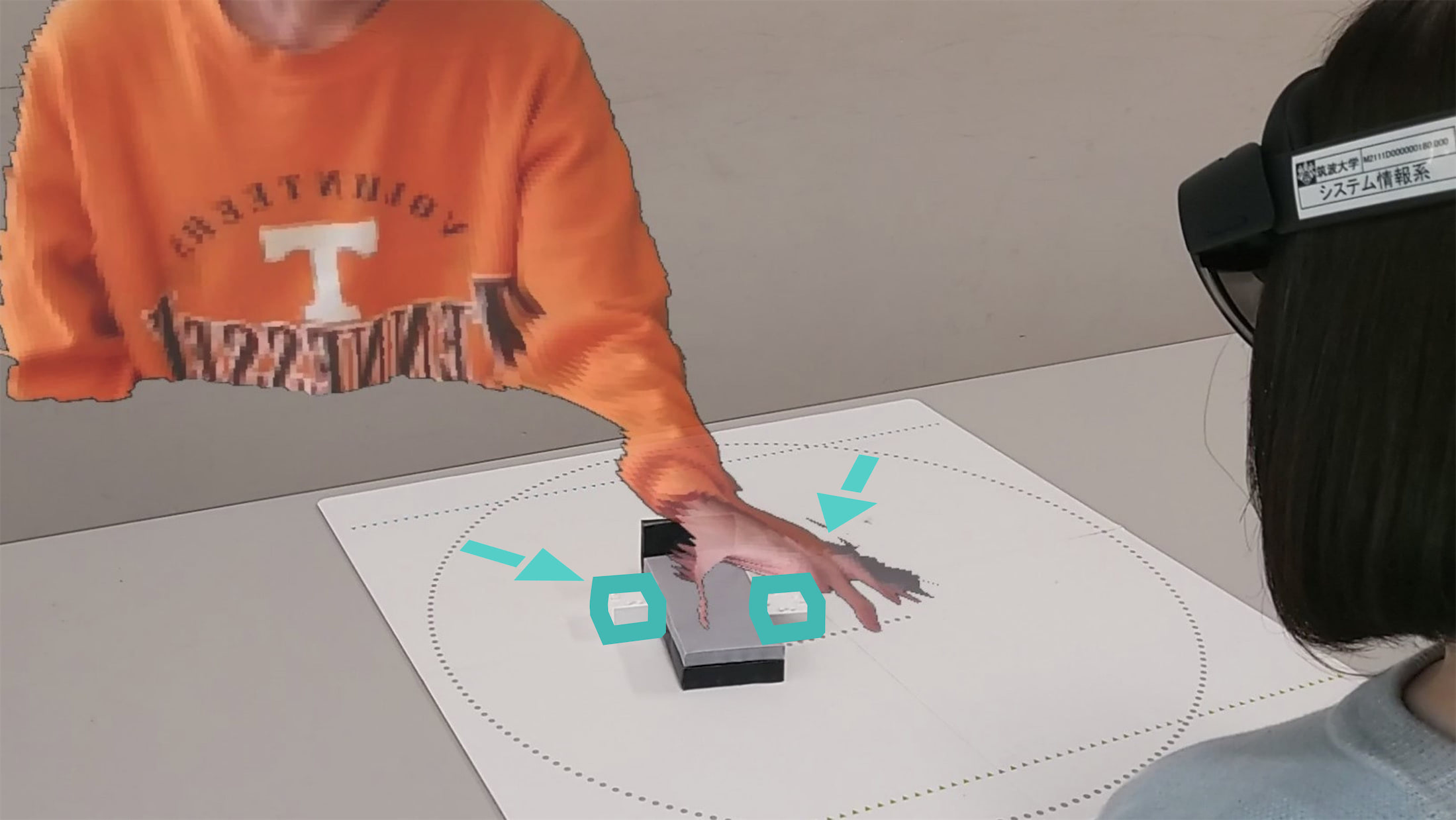}
\includegraphics[width=0.32\linewidth]{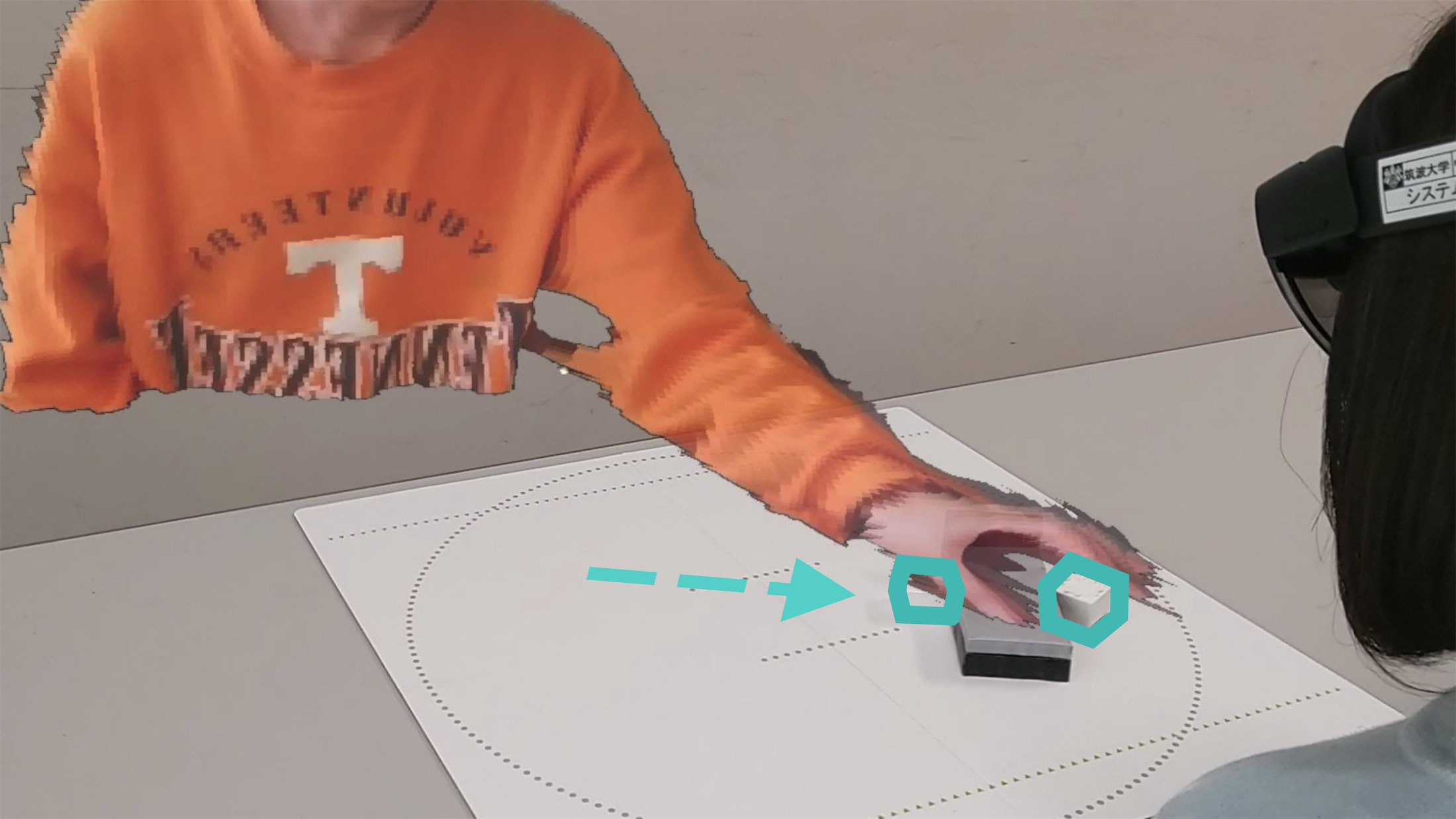}
\caption{Move Passive Object}
\label{fig:passive-object}
\end{figure}

Alternatively, the user can also actuate everyday passive objects by pushing these objects with the Toio.
This allows actuating objects without attaching robots in advance.
Similar to~\cite{kennel2023interacting}, by making the robots follow the user's fingers, the remote user can physicalize their hands and fingers, so that pushing the other passive objects (Figure~\ref{fig:passive-object}).
This method allows an intuitive way of interacting with physical objects, as the remote user can use hand gestures to control objects.
In the current setup, each Toio can push an object up to 32 grams. 


\subsection{Surface Types}
\subsubsection*{\textbf{Horizontal Surface}}
\system{} also supports two different surface types that the robot can move around.
The first one is the horizontal surface, such as a tabletop surface where the users sit down together to manipulate objects on the table. 

\subsubsection*{\textbf{Vertical Surface}}
\begin{figure}[h]
\centering
\includegraphics[width=0.32\linewidth]{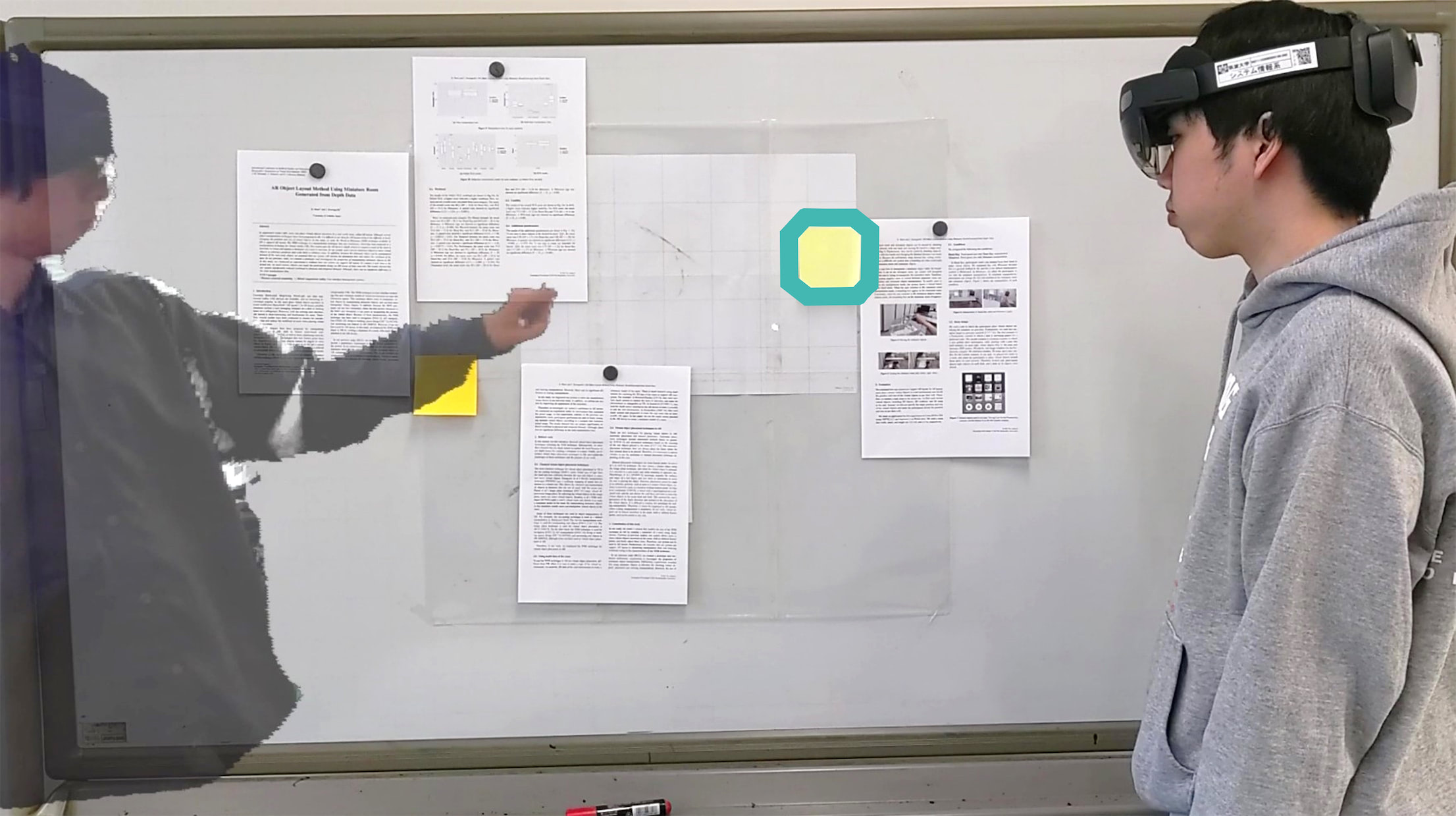}
\includegraphics[width=0.32\linewidth]{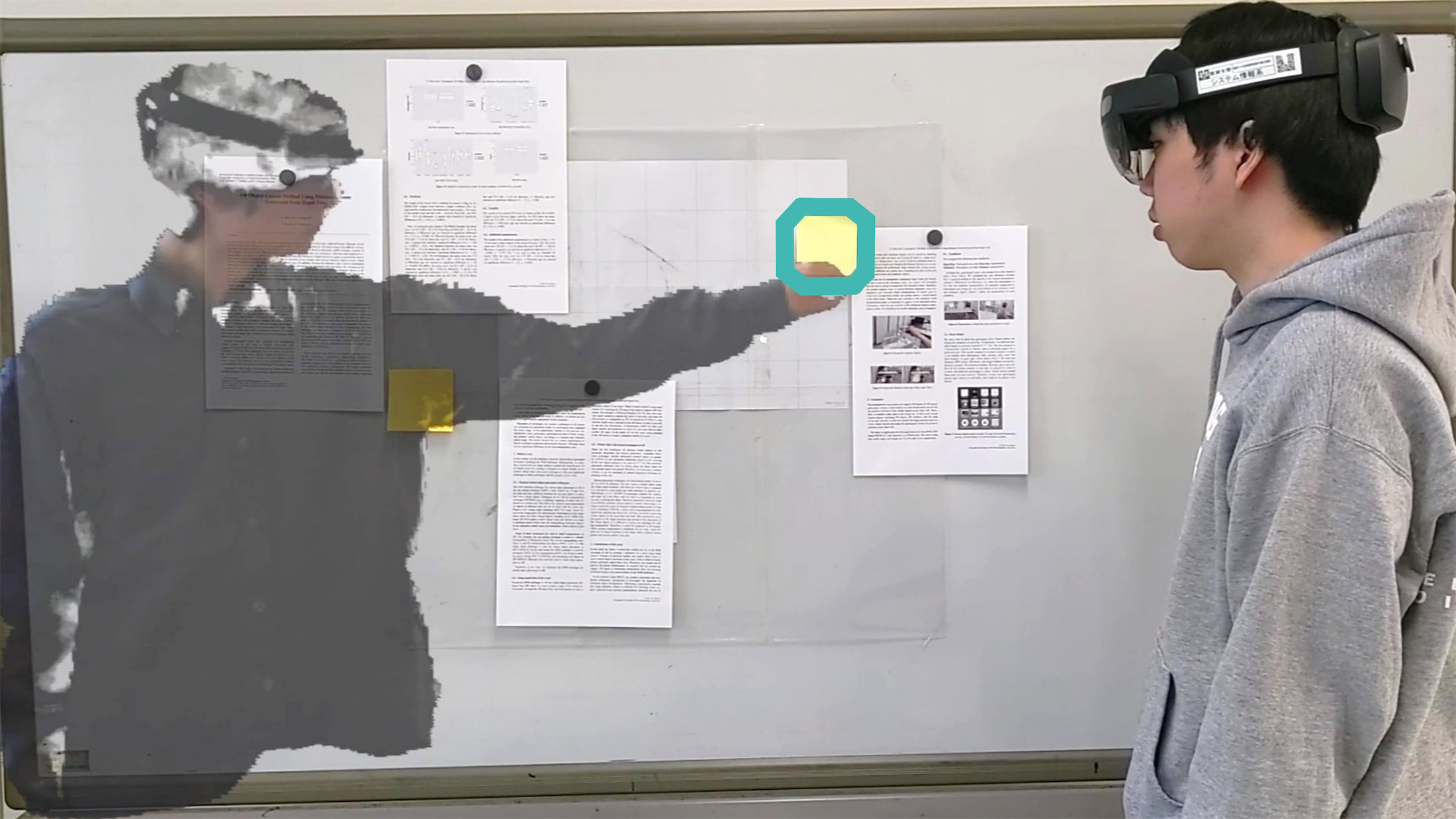}
\includegraphics[width=0.32\linewidth]{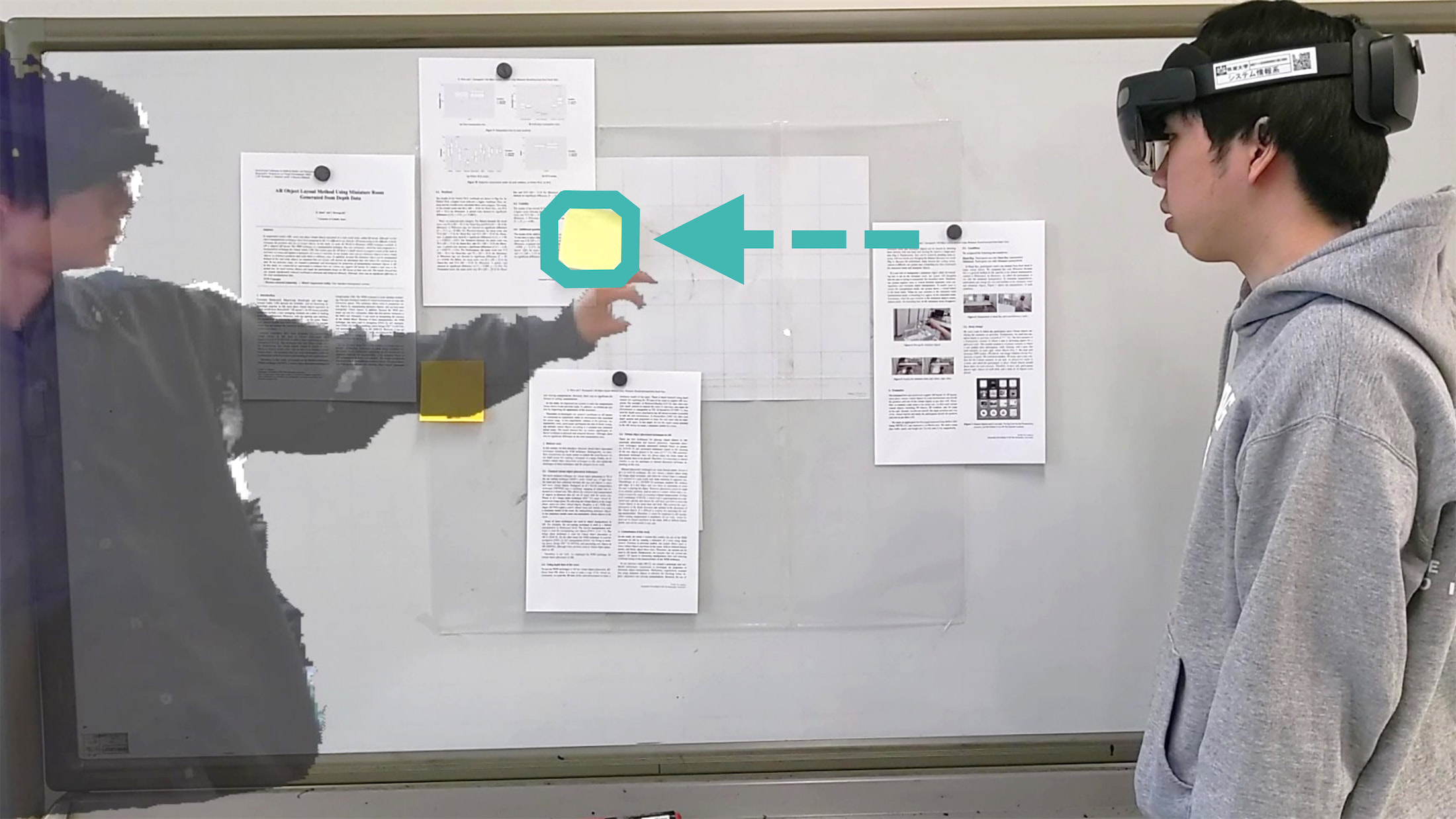}
\caption{Vertical Surface}
\label{fig:vertical-surface}
\end{figure}

Alternatively, by attaching a small magnet at the back of the Toio, Toio can move on a vertical surface such as a whiteboard or a magnetic wall.
By moving Toios on a vertical surface can be useful for applications that require standing up, such as brainstorming or presentations.
In our prototype, we attach an N35 neodymium magnet (8 mm $\times$ 3 mm, 1 mm thickness) to the bottom of the Toio robot with tape, which has a strong attraction force to be attached to the whiteboard, while weak enough to move on a wall. For the tracking of the vertical surface, we use a thinner tracking mat (Toio Developer Mat, 0.1 mm thickness) that can be attached to the whiteboard.
With the vertical surface, we can also expand the application domains, such as collaborative discussion and brainstorming with the post-it notes on a whiteboard (Figure~\ref{fig:vertical-surface}).


\subsection{Attachments of the Robot}
\system{} is also designed to be versatile and adaptable to various applications by allowing the user to attach different components to the robot. These attachments provide additional functionalities and enable the robot to perform a wider range of tasks, making it suitable for a variety of applications. 

\subsubsection*{\textbf{Shape Props}}
Shape props can modify the robot's shape and physical appearance.
As illustrated in Figure~\ref{fig:storytelling}, attaching a dinosaur toy to the robot can be used to represent a dinosaur, expanding its interactive potential.
By attaching Toios to physical objects such as puppets, stuffed animals, toy figures, and LEGO blocks, both local and remote users can move the objects, crafting the story and narrative, as we do in physical space.

\subsubsection*{\textbf{Material Props}}
The addition of material props such as soft materials, fur, and fabric enables the local user to enhance the sensation of remote objects and users.
For example, by attaching soft materials, mobile robots can represent remote users' hands to improve haptic communication.
Also, the use of fabric materials enables the mobile robots to represent portions of the remote user's arm that are clothed.

\subsubsection*{\textbf{Functional Props}}
\begin{figure}[h]
\centering
\includegraphics[width=0.32\linewidth]{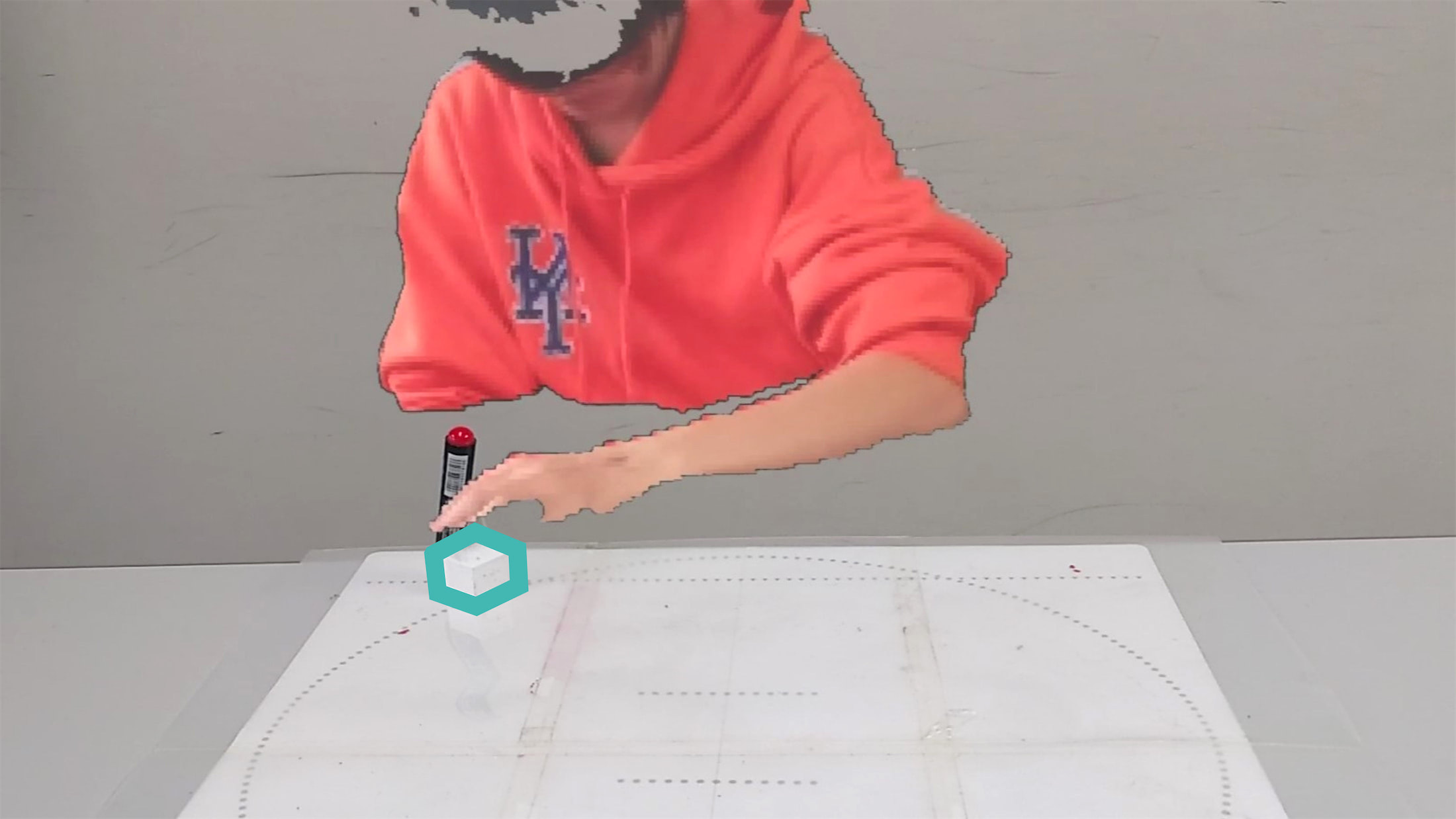}
\includegraphics[width=0.32\linewidth]{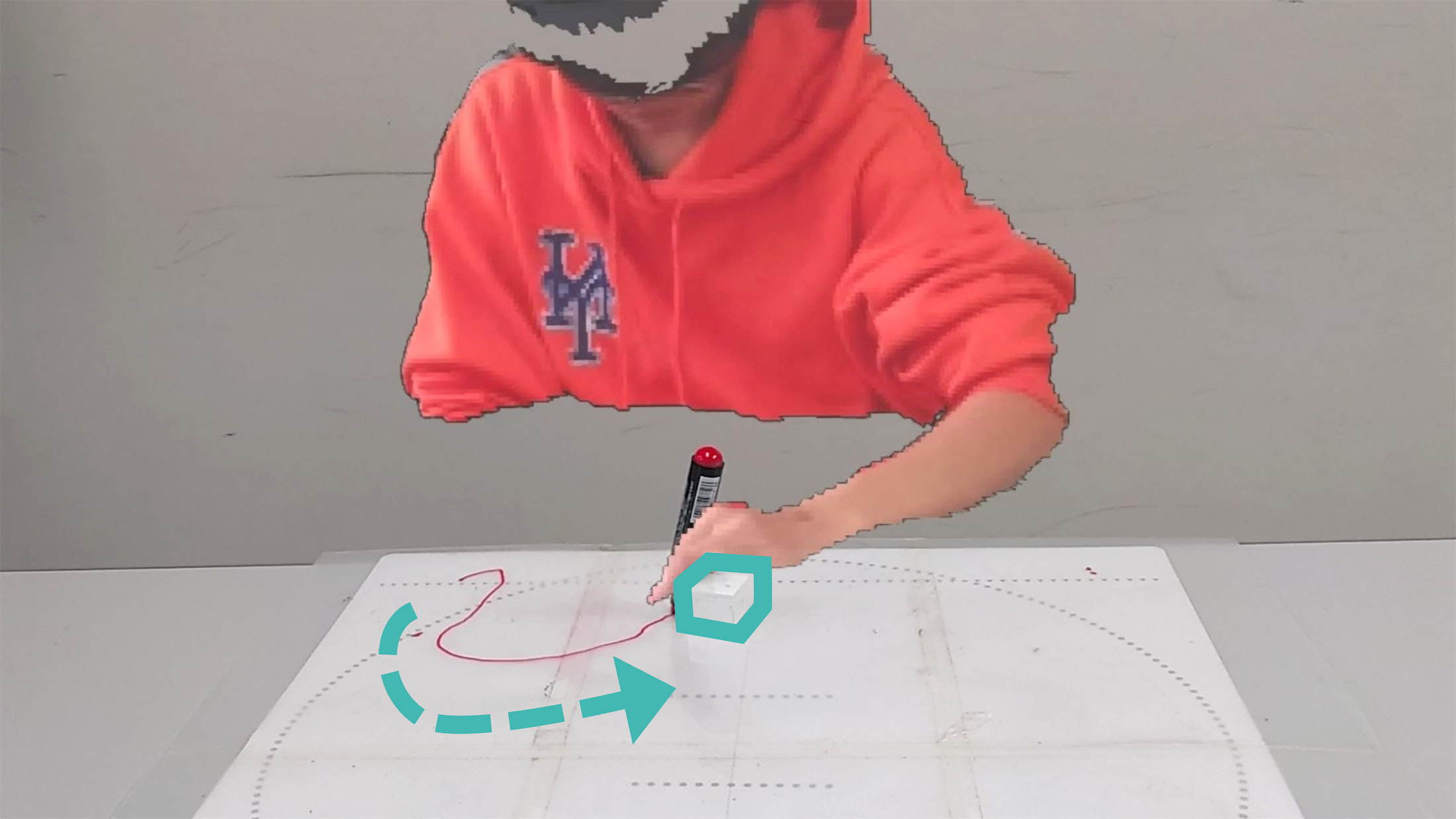}
\includegraphics[width=0.32\linewidth]{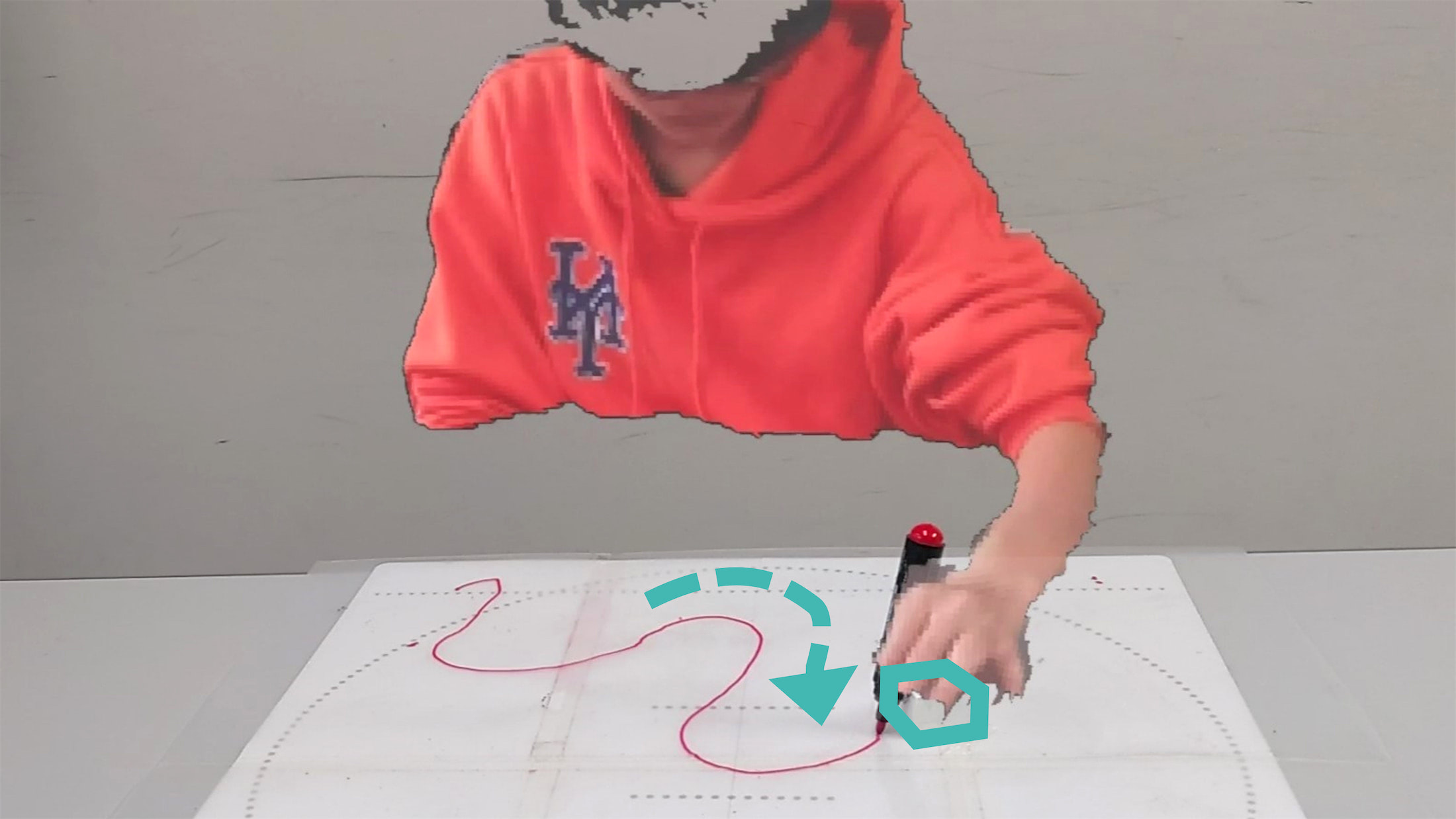}
\caption{Pen Attachment for Drawing}
\label{fig:drawing-attachment}
\end{figure}

Attachments can supplement the robot with added functionalities.
For example, Figure~\ref{fig:drawing-attachment} illustrates remote users drawing on a transparent sheet using a robot equipped with a pen, which facilitates visual communication between users.
As shown in Figure~\ref{fig:vertical-surface}, attaching post-it notes to the mobile robots enables the remote user to highlight specific parts in the local user's environment. 
Also, by attaching magnets to the robots, users can extend their mobility from horizontal to vertical surfaces.

\subsubsection*{\textbf{Constraints}}
Mechanical constraints, such as rings and rubber bands, can be employed to restrict the movements of mobile robots as PICO~\cite{patten2007mechanical}.
This provides both the remote and local users to move the robots within a specific range of movement.
For example, by using a straight ring, the movement of mobile robots can be limited to a straight line, which could help the creation of a precise slider UI.
Also, confining all the mobile robots within a ring can help limit the area in which the remote user can influence the local environment. 





\newcommand{\UIApp}{Shared Tangible UI}
\newcommand{\ObjectApp}{Object Actuation}
\newcommand{\BodyApp}{Miniature Body Interaction}
\newcommand{\HandApp}{Haptic Communication}

\section{User Study}
To evaluate the effectiveness of incorporating both virtual and physical representations in holographic remote collaboration, we conducted a user study comparing our system with Hologram-Only and Robots-Only conditions across four distinct interactions. 
We gathered both quantitative and qualitative measurements for various aspects, such as social presence, system usability, and cognitive workload, with a within-subject user study.

\subsection{Method}
\subsubsection{Participants}
We recruited 12 participants (11 male, 1 female) from our local university, with an age range of 21-24 years (M = 22.1, SD = 1.16).
Participants were surveyed on their familiarity with VR/AR using a 7-point Likert scale from 1 (novice) to 7 (expert), and the average score was 2.75 (SD = 1.71).

\subsubsection{Study Setup}
\begin{figure}[h]
\centering
\includegraphics[width=0.48\linewidth]{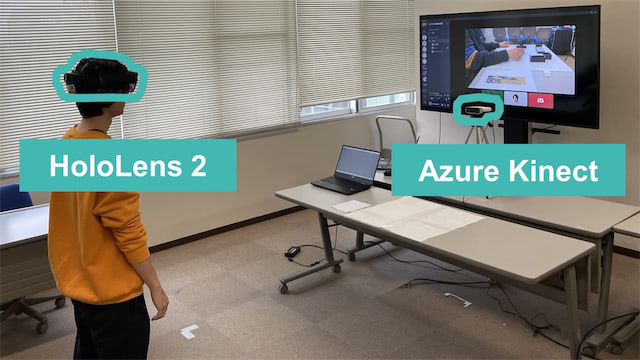}
\includegraphics[width=0.48\linewidth]{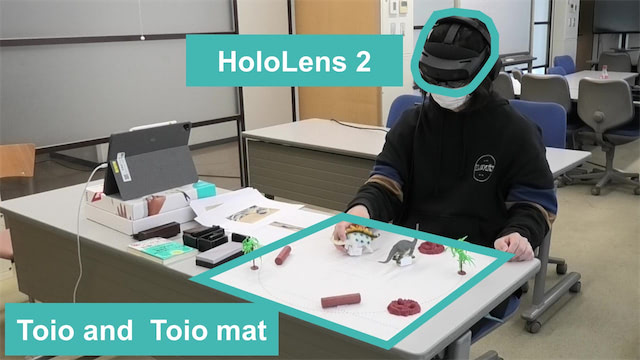}
\caption{Study Setup. Left: Experimenter's room, Right: Participant's room}
\label{fig:study-setup}
\end{figure}

We present the setup used in our study in Figure~\ref{fig:study-setup}.
One of the authors acts as a remote collaborator (referred to as "the experimenter") for each participant to reduce differences in interaction between groups.
The participant and the experimenter are situated in separate rooms and communicate remotely.
The dimensions of the participant's room were approximately 11.3 m by 5.2 m, while the experimenter's room measured approximately 7.5 m by 5.9 m.
Both the participant and experimenter were equipped with Hololens 2 headsets.
On the participant's desk, we placed the Toios and a Toio mat.
To enable the experimenter to view the participant's workspace, we used an iPad to capture the video image and transmitted it to the experimenter's display.

For audio communication, we used Discord\footnote{https://discord.com/}, a voice chat application.
To mitigate any potential interference from the Toio's sound, the participant wore noise-canceling headphones (Sony WH-1000XM4).


\begin{figure*}[t]
\centering
\includegraphics[width=0.24\linewidth]{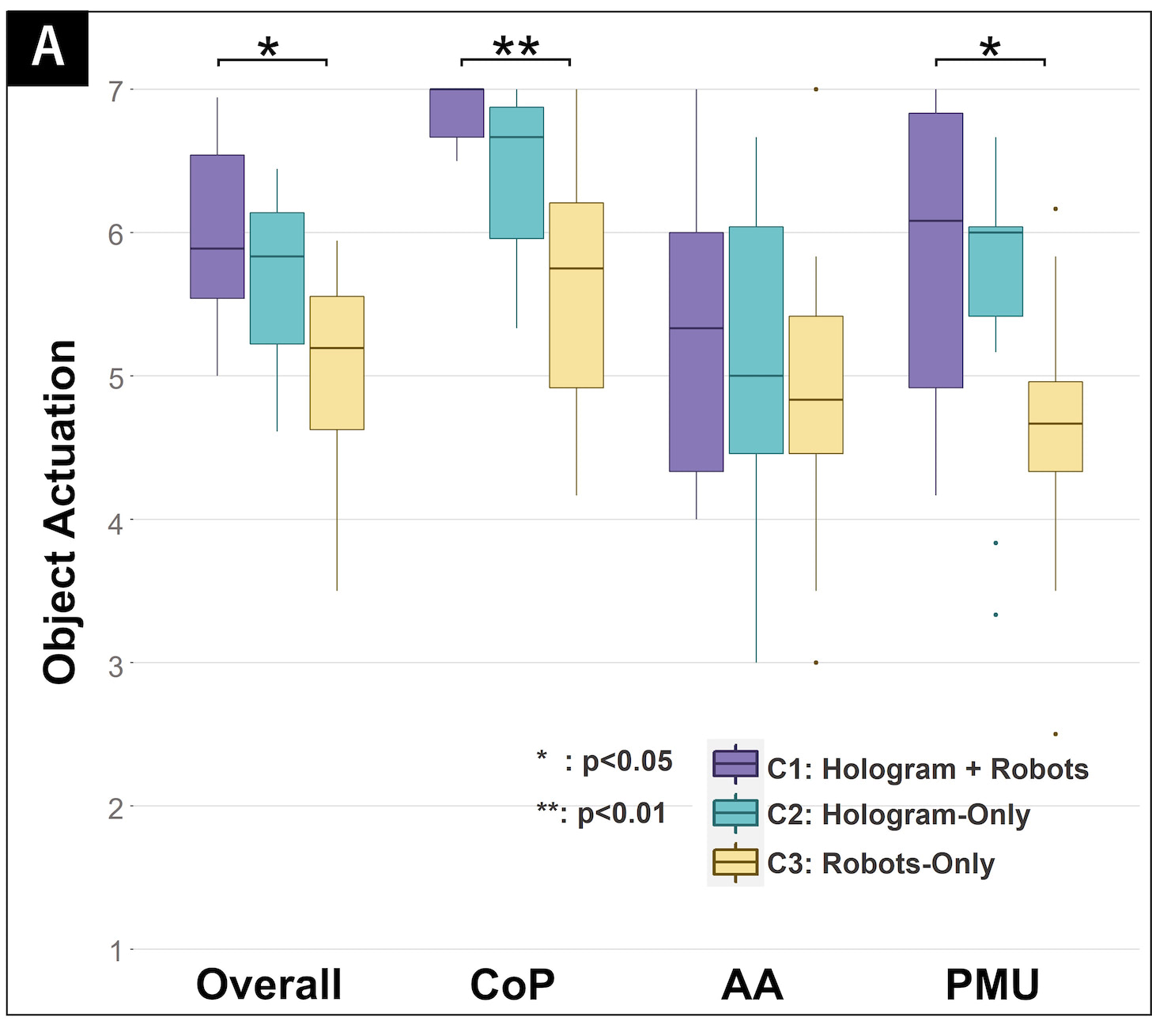}
\hspace{0.010\columnwidth}
\includegraphics[width=0.24\linewidth]{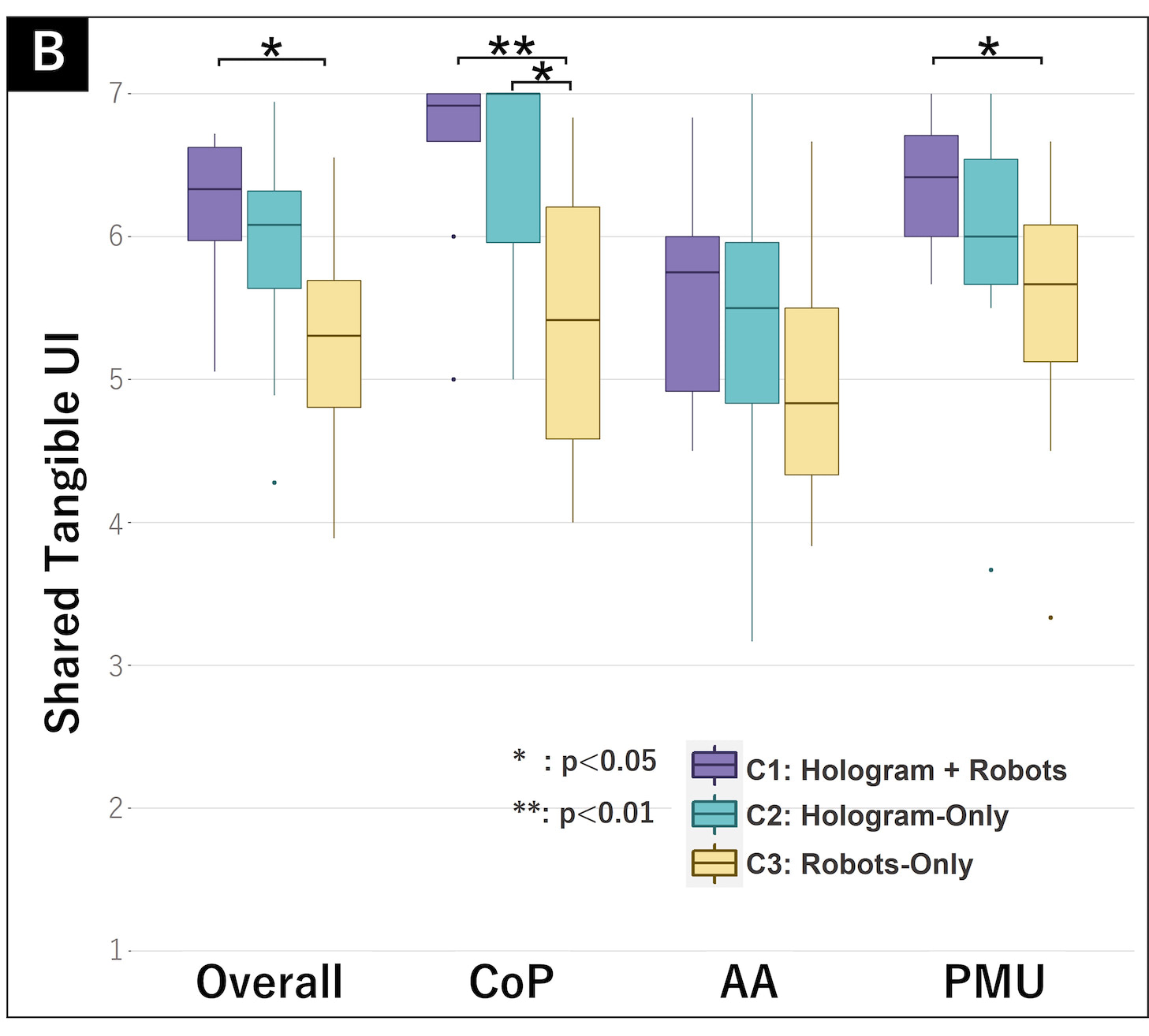}
\hspace{0.010\columnwidth}
\includegraphics[width=0.24\linewidth]{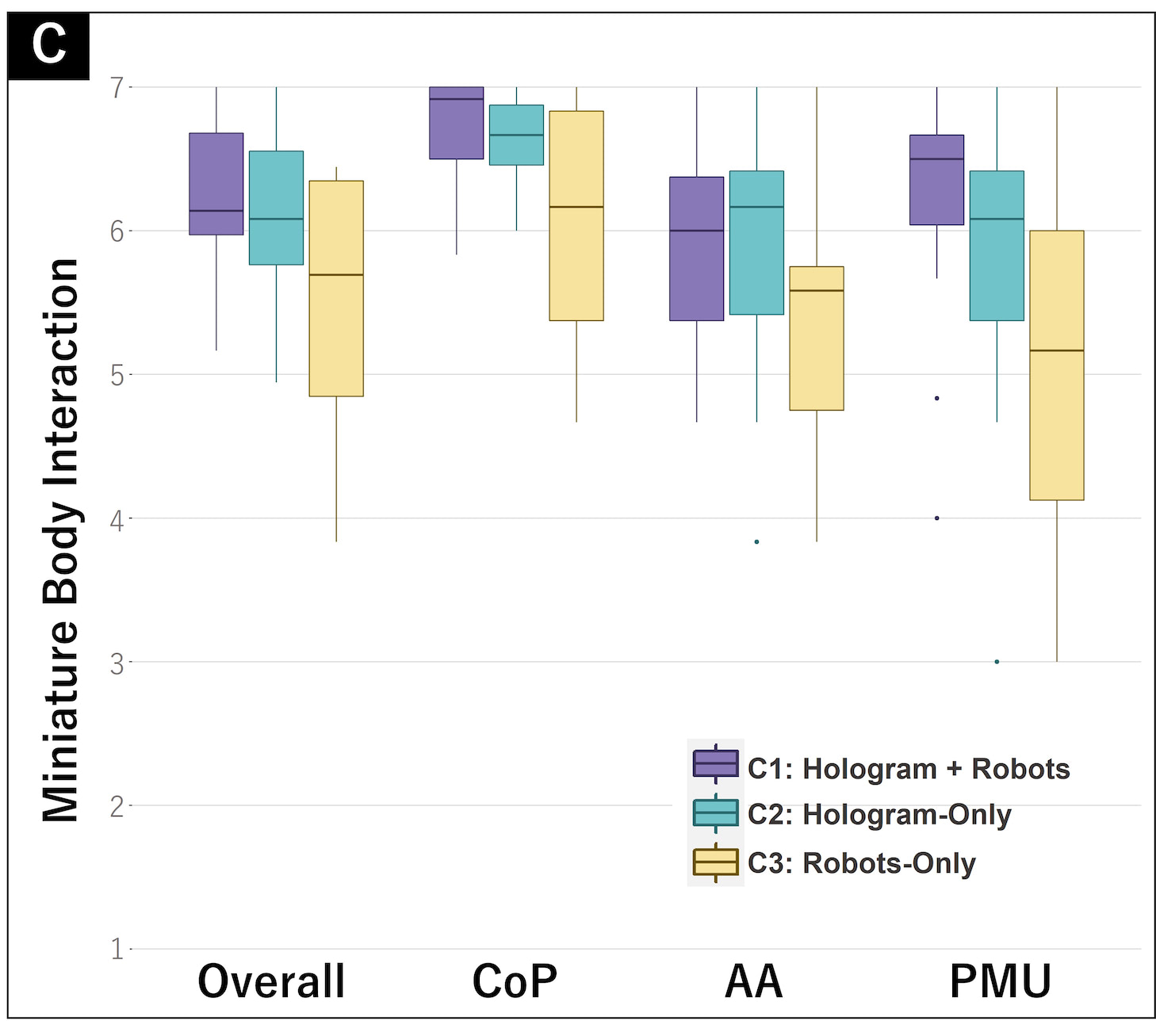}
\hspace{0.010\columnwidth}
\includegraphics[width=0.24\linewidth]{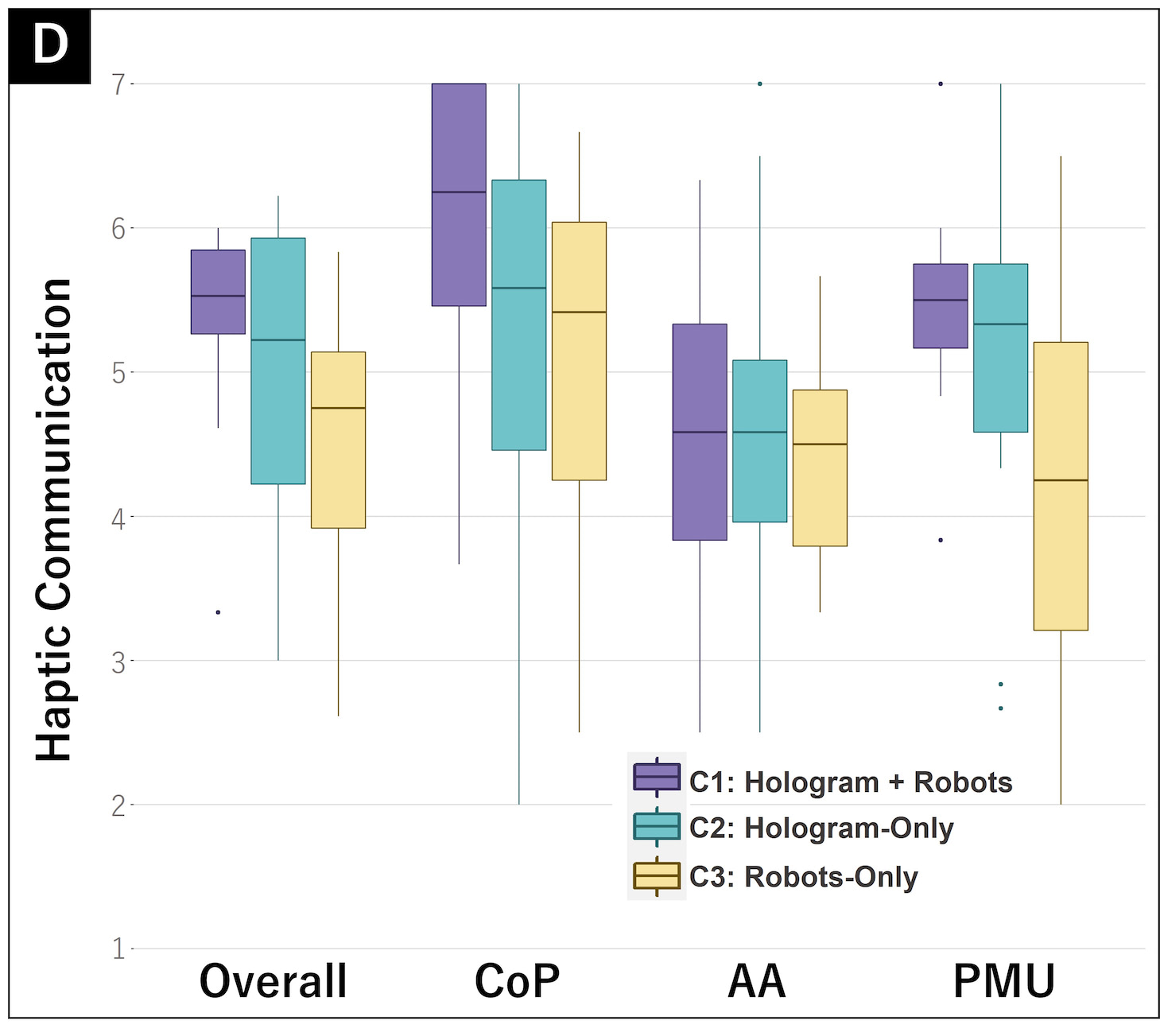}
\caption{Social Presence Questionnaire Results. A: \ObjectApp{}, B: \UIApp{}, C: \BodyApp{}, D: \HandApp{}. CoP: Co-Presence, AA: Attentional Allocation, PMU: Perceived Message Understanding.}
\label{fig:study-social-presence}
\end{figure*}


\begin{figure}[h]
\includegraphics[width=0.27\linewidth]{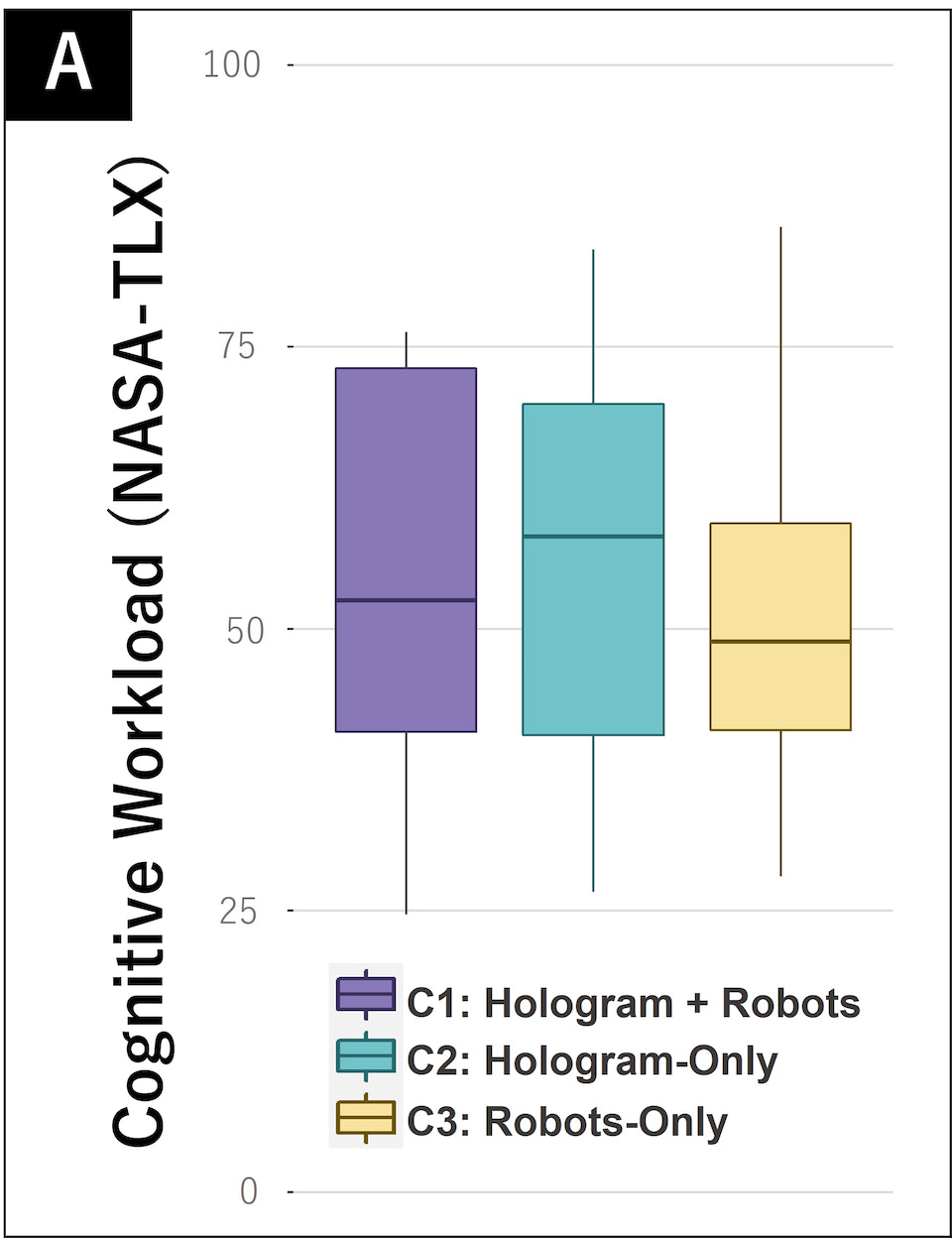}
\hspace{0.035\columnwidth}
\includegraphics[width=0.27\linewidth]{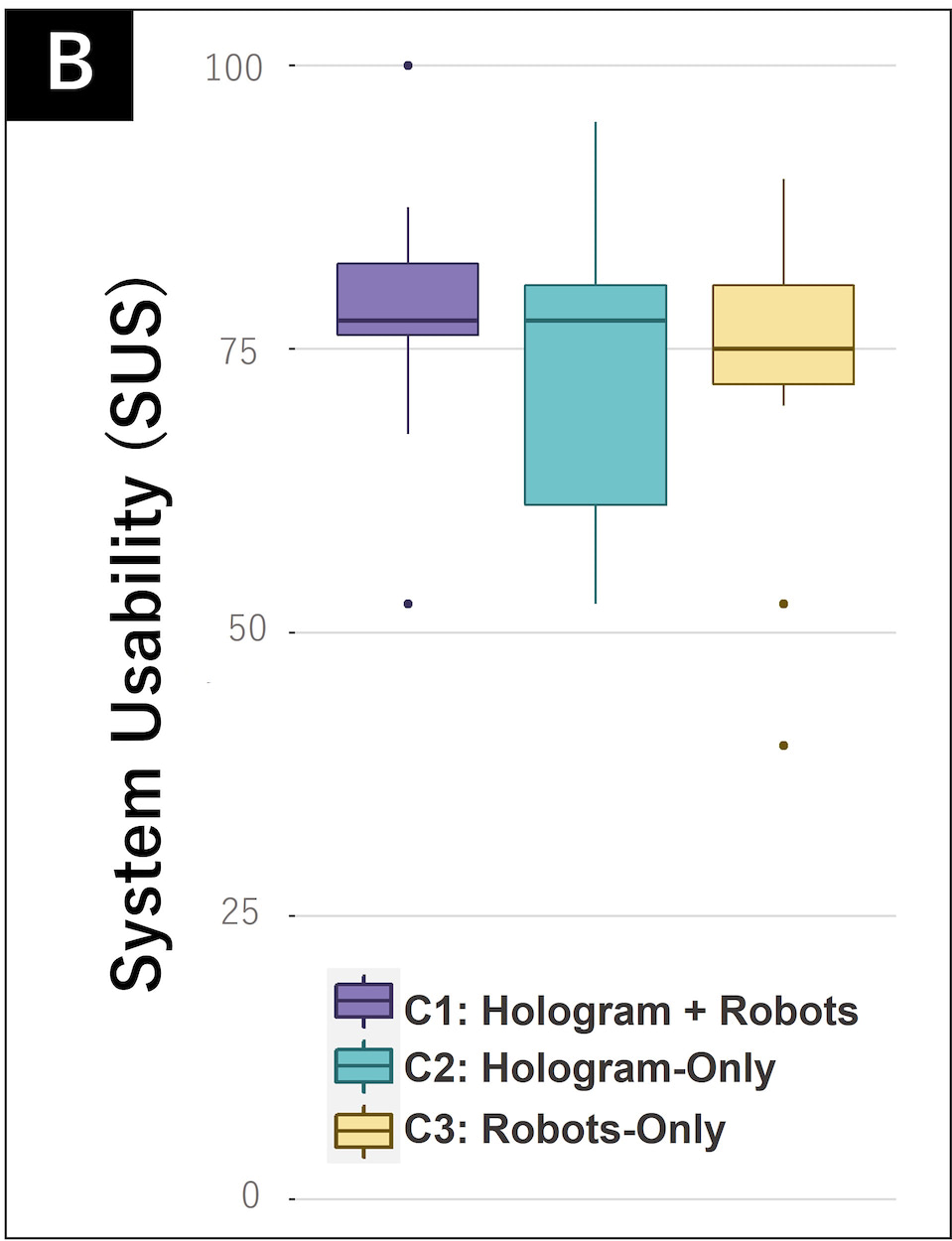}
\hspace{0.035\columnwidth}
\includegraphics[width=0.35\linewidth]{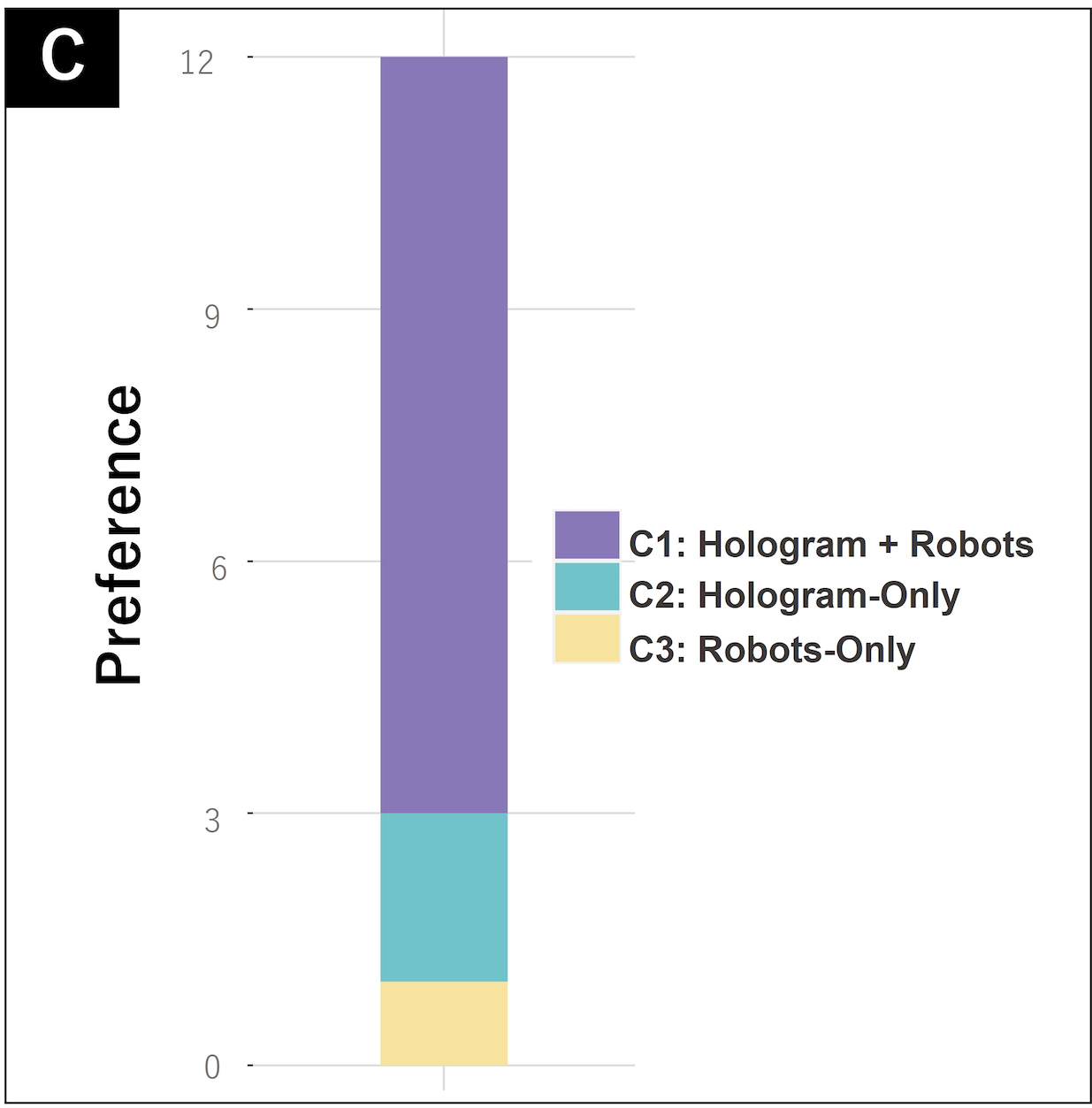}
\caption{A: Cognitive Workload (NASA-TLX), B: System Usability (SUS), C: Preference}
\label{fig:study-sus-nasa}
\end{figure}


\subsubsection{Study Design}
We designed our study with a within-subject design that compares the following three conditions:
\begin{description}
\item[C1. Hologram +  Robots]: Participants interacted with the remote experimenter via hologram and voice chat with using mobile robots.
\item[C2. Hologram-Only]: Participants interacted with the remote experimenter via hologram and voice chat without using mobile robots.
\item[C3. Robots-Only]: Participants interacted with the remote experimenter via voice chat without a hologram with using mobile robots.
\end{description}
To evaluate the difference in these conditions across various interactions, we used four application scenarios that best represent each interaction technique in our design space:
\begin{description}
\item[D1. Object Actuation]: We used the physical storytelling application (Figure~\ref{fig:storytelling}). Participants were instructed to create a short story with the remote experimenter by manipulating virtual or physical dinosaur toys. The remote experimenter could also move the dinosaur toys.
The fundamental elements of the stories shared similarities, including dinosaurs fighting, making up, walking around, and talking to each other. However, participants chose how the dinosaurs fight, where they make up or rest, and which directions they walk. We displayed virtual toys for C2 and used physical toys for C1 and C3.

\item[D2. Shared Tangible UI]: We used virtual image manipulation (Figure~\ref{fig:tangible-ui}). Participants were instructed to adjust the size of a virtual picture until it matched the target size printed on paper, collaborating with the remote experimenter. The virtual cubes or mobile robots were attached to the upper left and bottom right of the virtual picture, and participants and a remote experimenter could adjust the size by moving them.
We used virtual cubes for C2, and used physical cubes for C1 and C3. The virtual picture was displayed in all conditions.
\item[D3. Miniature Body Interaction]: We used the interior and architectural design application (Figure~\ref{fig:interior-design}). The remote experimenter was presented as a miniature body, similar in size to miniature furniture. Participants were instructed to move the furniture and determine furniture placement in discussion with the remote experimenter. The remote experimenter could also physically move the furniture in the conditions with a mobile robot.
\item[D4. Haptic Communication]: We used the haptic notification application (Figure~\ref{fig:notification}). Participants were instructed to read a book and engage in conversation with the remote experimenter when they were contacted. The remote experimenter initiated contact through virtual or physical touch, with mobile robots following the remote  experimenter's fingers to physically touch the participants.
\end{description}

\subsubsection{Measurements}
We measured four different aspects: 1) \textbf{Social Presence} based on the Networked Mind Measure of Social Presence questionnaire \cite{socialpresence}, 2) \textbf{Cognitive Workload} based on NASA Taskload Index (NASA-TLX)~\cite{nasa-tlx}, 3) \textbf{System Usability} based on System Usability Scale (SUS)~\cite{sus}, and 4) \textbf{Preference} based on the questionnaire in which the participants were asked which condition they preferred the most.
In addition to these measurements, we conducted an interview after the study to gather qualitative feedback from participants.

\subsubsection{Procedure}
After participants signed a consent form, we provided them with instructions on how to use the Hololens 2 and Toio robot.
Participants then conducted a task involving 12 sessions (4 applications $\times$ 3 conditions), each lasting 3 minutes.
Participants used four applications in the following order: \UIApp{}, \ObjectApp{}, \BodyApp{}, and \HandApp{}.
Participants conducted each application in all three conditions (C1, C2, and C3). The order of the three conditions was counterbalanced across participants to control for order effects.
After each application, participants answered the social presence questionnaire. 
After each condition, participants answered the SUS and NASA-TLX questionnaire to compare the three conditions.
In total, we asked the participants to complete 12 social presence questionnaires and 3 SUS and NASA TLX questionnaires. 
After the participants finished all of the sessions, we conducted a brief open-ended interview for 10-15 minutes.
The study took approximately 90 minutes in total, and each participant was compensated with 10 USD. 

\subsection{Results}
To analyze the data collected in our study, we employed a Friedman's test for each measurement.
To assess pairwise differences between conditions, we conducted multiple pair-wise comparisons using the Wilcoxon signed-rank test with Bonferroni correction.
We set the significant level at 5 \%.

\subsubsection{Social Presence}
The Social Presence Questionnaire consisted of three sub-scales: Co-Presence (CoP), Attentional Allocation (AA), and Perceived Message Understanding (PU).
Figure~\ref{fig:study-social-presence} shows the result of the social presence questionnaire for a total of 12 sessions (4 applications $\times$ 3 conditions for each).
In addition, we calculated an overall score by averaging the three sub-scales.
We checked the internal consistency with Cronbach’s alpha for each sub-scale:\ $ \alpha_{CoP} = 0.90$, $\alpha_{AA} = 0.78$, $\alpha_{PMU} = 0.93$.

For \ObjectApp{} (D1) and \UIApp{} (D2), Hologram + Robots (C1) condition had significantly higher overall social presence scores than Robots-Only (C3) condition.
For both \ObjectApp{} (D1) and \UIApp{} (D2), pairwise comparisons revealed that Hologram + Robots (C1) condition was significantly higher scores than Robots-Only (C3) condition for CoP (D1: $Z = 3.68$, $p = 0.0007 < 0.001$, D2: $Z = 3.29$, $p = 0.003 < 0.01$), PMU (D1: $Z = 2.46$, $p = 0.042 < 0.05$, D2: $Z = 2.61$, $p = 0.027 < 0.05$), and Overall (D1: $Z = 2.63$, $p = 0.025 < 0.05$, D2: $Z = 2.86$, $p = 0.013 < 0.05$).

In the interviews, participants made comments that suggested that Hologram + Robots (C1) condition resulted in a stronger sense of presence compared to Hologram-Only (C2) condition.
Specifically, one participant noted that \textit{``Hologram + Robots clearly felt the presence of the other party, whereas Hologram alone was less present.'' (P1)}, while another participant mentioned that \textit{``Hologram-only conditions were difficult to react to when the other person was out of sight'' (P2)}.
These comments suggest that combining mobile robots with holographic telepresence could help users better understand the remote user's actions and movements, even when the holographic user is out of sight.
Furthermore, for all four applications, the graph of the data suggested that Hologram + Robots (C1) had the highest scores, followed by Hologram-Only (C2) and Robots-Only (C3).

\subsubsection{Cognitive Workload}
The results for the cognitive workload are shown in Figure~\ref{fig:study-sus-nasa} (A).
A lower score indicates a lower workload.
The average score for each condition was 54.0 ($SD = 19.4$) for Hologram + Robots (C1) conditions, 55.3 ($SD = 18.9$) for Hologram-Only (C2), and 51.6 ($SD = 16.9$) for Robots-Only (C3).
The Friedman test showed no significant difference ($\chi^2(2) = 0.30$, $p = 0.86$).

\subsubsection{System Usability}
The results for the system usability scale are shown in Figure~\ref{fig:study-sus-nasa} (B).
A higher score indicates higher usability.
The average score for each condition was 77.9 ($SD = 11.3$) for Hologram + Robots (C1) conditions, 73.3 ($SD = 14.2$) for Hologram-Only (C2), and 72.9 ($SD = 14.1$) for Robots-Only (C3).
The Friedman test showed no significant difference ($\chi^2(2) = 1.64 $, $p = 0.44$).
During the interviews, participants provided feedback on the usability.
One participant noted that \textit{``Conditions which use Toio were easy to manipulate'' (P3)}, and another participant noted that \textit{``It was easy to adjust the size of the virtual picture using Toio'' (P8)}.
Although Hologram + Robots (C1) had a higher average usability score (77.9) than the average score (68) ~\cite{sauro2011practical}, it was not significantly better than Hologram-Only (C2). 
One participant reported, \textit{``The coupling between the actual movements and the robot was slow and misaligned, which sometimes make it difficult to understand'' (P3)}. This feedback suggests that the low ability of coupling between the hologram and mobile robots may have negatively impacted usability.

\subsubsection{Preference}
The results for the preference are shown in Figure~\ref{fig:study-sus-nasa} (C).
75 \% of the participants preferred Hologram + Robots (C1) as the best, followed by Hologram-Only (C2) (17 \%) and Robots-Only (C3) (8 \%).
Chi-squared goodness of fit test revealed a significant difference from random choice ($\chi^2(2) = 9.5$, $p = 0.009 < 0.01$).
In our study, participants preferred Hologram + Robots (C1) over Hologram-Only (C2) and Robots-Only (C3).
Five out of nine participants mentioned social presence as a key factor in their preference for Hologram + Robots (C1), while the remaining four participants mentioned usability as a determining factor.
Therefore, the high social presence and usability in Hologram + Robots (C1) can enhance the overall user experience.

\subsection{Limitations and Design Implications}

\subsubsection{Precise Coupling between Holographic Users and Robot Movement}
In the applications used in the study, the coupling between the virtual body movements and mobile robots was occasionally slow and misaligned due to the Toio's maximum speed and the calibration error between the avatar and Toios.
Upon testing the start latency, the average latency was 0.483 s, 0.262 s, 0.443 s, and 0.615 s in D1, D2, D3, and D4, respectively.
This issue could potentially impact both the social presence and usability of the Hologram + Robots (C1) condition.
Employing faster mobile robots and implementing a more accurate position calibration method between the avatar and Toios could alleviate this problem.

\subsubsection{Noise of Robot Movement}
Several participants reported that the sound generated by the Toios could be distracting and interfere with their ability to concentrate on the task. 
For example, one participant commented \textit{``Toios sound was sometimes a little loud, and it was difficult to concentrate on the task.'' (P2)}, while another mentioned \textit{``I was distracted by the noise of Toios'' (P3)}.
Upon testing the noise levels generated when moving the Toio 45 cm in 4 seconds, the maximum recorded noise was 64.5 dB, 60.3 dB, 65.0 dB, and 70.0 dB in D1, D2, D3, and D4, respectively.
This issue could potentially impact the user experience and social presence.
To address this problem, we could improve the system to make Toios travel to their destination by the shortest route, reducing travel time and the duration of sound generation.

\subsubsection{Bi-Directional Collaboration between Participants}
Additionally, the collaboration in our study was between a participant and an experimenter.
To gain further insights, it may be beneficial to set up an environment where participants can collaborate with other participants without the presence of an experimenter.
This can provide insights on more realistic collaboration scenarios.

\subsubsection{Group Size}
In our study, collaboration was limited to only two people, one participant and one experimenter.
Using larger groups could potentially increase the number of interactions and affect the social presence and user experience.
However, this could also increase conflicts and misunderstandings.
Therefore, conducting studies with larger groups could help us understand how these factors influence our system.

\subsubsection{Number of Robots}
In our study, we used two Toios, but it is possible to use more.
One participant noted that \textit{``I thought it would be good if the picture application could increase the number of manipulable objects (Toio) and allow more complex UI manipulation'' (P8)}.
This comment suggests that using more Toios for UI manipulation could affect usability and user experience.
Additionally, we could use more Toios for body or hand representation, which could enhance the resolution of the remote user's movements and gestures, which could enhance the social presence.

\subsubsection{Enhancing Holographic Visualization}
In this study, we used a single Kinect camera to capture the remote user's body movements for holographic avatar generation.
Future work could expand this setup by adding more Kinect cameras to capture the user's hologram from multiple angles.
This could improve the remote user's clarity and accuracy via multi-directional coverage.
Through these improvements, the local user would better comprehend the remote user's intentions and interactions with the physical environment and overall body language.


\section{Future Work}

In this paper, our design space exploration is limited to the form factor of \system{}. In this section, we suggest several directions to expand the design space of holographic tangible remote collaboration with different form factors, interaction modality, and user representations.

\begin{figure}[h]
\centering
\includegraphics[width=\linewidth]{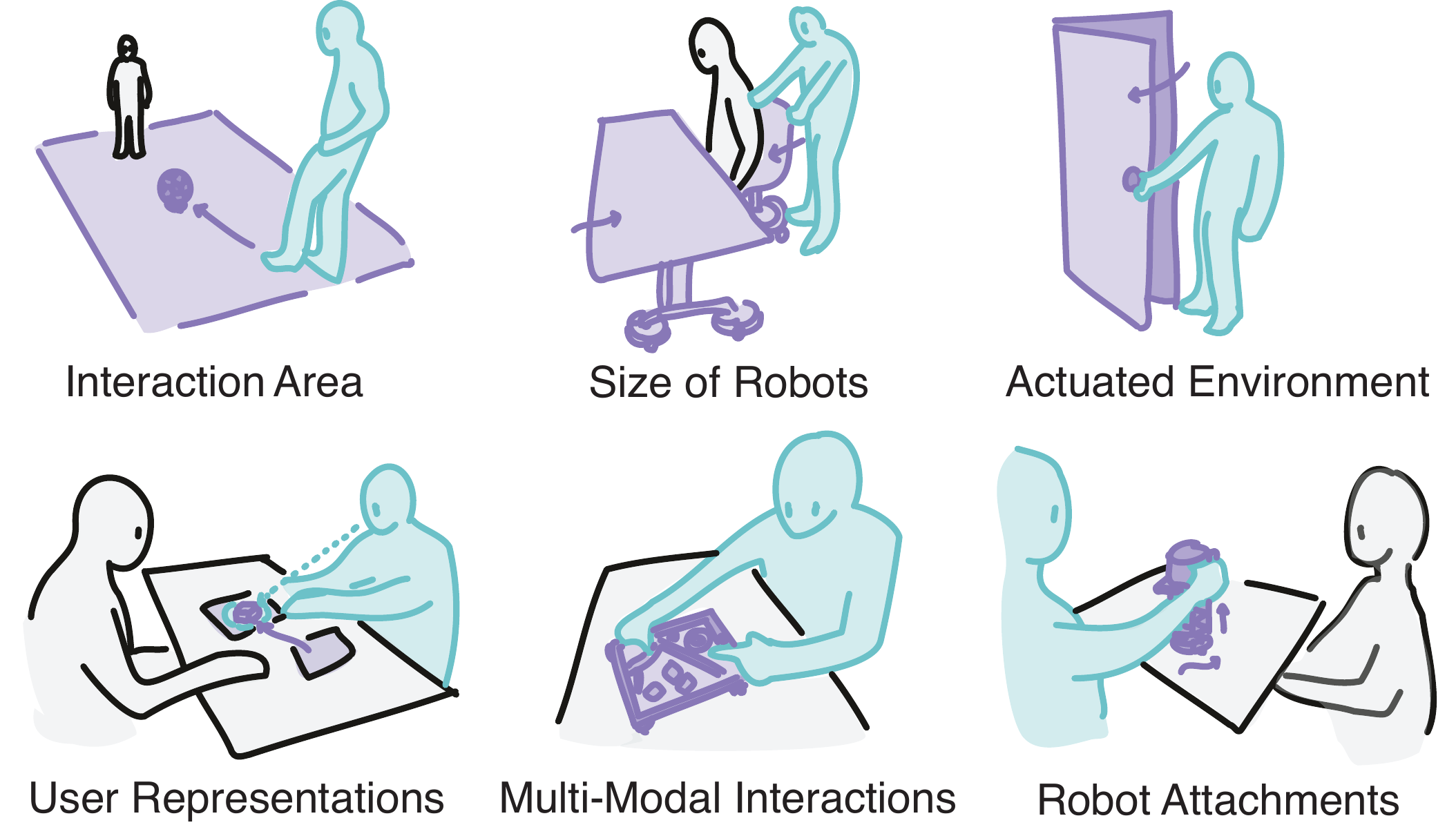}
\caption{Future Work}
\label{fig:future-work}
\end{figure}

\subsubsection*{\textbf{Scalability of Interaction Area}}
In our current implementation, the interaction area was limited to one Toio mat, which restricted users to using the limited space of the table or whiteboard.
However, by aligning multiple Toio mats, we can expand the interaction area to cover larger surfaces, such as larger tables or whiteboards.
This expansion would allow for the use of additional mobile robots and the accommodation of more users.
Furthermore, incorporating fiducial markers and tracking cameras, as seen in \textit{ASTEROIDS}~\cite{li2022asteroids}, could further increase the scalability of the interaction area to encompass an entire table or even a whole room.
This approach is not limited to Toio robots, meaning that various robotic platforms could also be incorporated.

\subsubsection*{\textbf{Different Robotic Form Factors}}
In this study, we focused on the combination of a holographic avatar with tabletop mobile robots.
However, future work could expand on this area by incorporating different types of robots.
For example, larger robots, such as robotic vacuum cleaners, could be utilized.
By synchronizing large robots' movements with the remote user, the holographic user could physically impact the local environment on a larger scale.
This could facilitate physical gaming and sports, and/or enable real-scale interior design by moving actual furniture, similarly to \textit{RoomShift}~\cite{suzuki2020roomshift}.

\subsubsection*{\textbf{IoT and Actuated Environments}}
In future work, we could integrate IoT devices, such as lights, fans, and curtains to enable users to affect the local environment in a different way.
For example, by synchronizing the remote user's movements with a door's movements, the remote user could interact with the door in the local environment.
Also, by using reeling mechanisms like AeroRigUI~\cite{yu2023aerorigui}, remote users can raise and lower the curtains in the local environment.

\subsubsection*{\textbf{Different User Representation}}
In addition to the current finger and miniature body representations, there are other potential ways to embody remote users using Toio robots. We are interested in exploring the following alternative representations in future work.
\textit{Whole Arm:}
Mobile robots could represent the entire arm of the remote user, enabling them to physically move multiple objects simultaneously or to visualize the joints in the user's arm such as their wrist and elbow.
\textit{Gaze and Eye:} Mobile robots could follow the remote user's gaze or eye movement, allowing the local user to physically understand where the remote user is looking, an insightful metric for gauging a user's intent or thought processes.
\textit{Shadows:} Mobile robots could represent the shadow of the remote user, providing additional information about their body position beyond what an avatar can offer.

\subsubsection*{\textbf{Multi-Modal Interaction}} 
Another potential direction for future work would be to explore multi-modal interaction methods that incorporate different modalities of communication.
One such method could involve the use of a tablet device. By connecting mobile robots to the tablet and sharing the screen between the tablets of both remote and local users, the remote user can manipulate the tablet's position and engage with its contents. This interaction enables functionalities such as scaling and rotating maps shown on the tablet, as well as drawing pictures on the tablet.

\remove{
\subsubsection*{\textbf{Supporting More Robots}}
In our current application, we utilized up to three Toio robots, but the system could be expanded by incorporating more robots.
Firstly, incorporating more robots would allow users to work with an increased number of active objects and tangible UIs, enabling complex tasks such as storytelling with multiple characters and collaborative modeling that involves numerous variables.
Secondly, an increased number of robots could enhance user representation, such as by allowing the robots to follow all ten fingers of the remote user.
Moreover, additional robots could represent other body parts and joints of the remote user, such as their elbows and wrists, enabling the remote user to more intuitively manipulate objects and engage in more realistic haptic communication with the local user.
}

\subsubsection*{\textbf{Versatile Robot Attachments}}
In future work, the capabilities of \system{} could be enhanced by incorporating versatile physical attachments, inspired by the \textit{HERMITS}~\cite{nakagaki2020hermits} concept. These on-demand attachments could significantly expand the functionality and flexibility of the tabletop robots and enable them to adapt to a wider range of tasks and interaction scenarios for tangible remote collaboration. Here are some possible directions for versatile physical attachments:
\textit{2.5D Shape Display:}
By adopting the technique presented in \textit{HapticBots}~\cite{suzuki2021hapticbots}, we can enhance our mobile robots to enable height adjustments, allowing them to actuate in 2.5 dimensions.
\textit{Attaching Tangible Controllers:}
By attaching complex tangible controllers, such as joysticks, sliders, or knobs, our system could enable remote users to perform intricate manipulations using the shared tangible UI.
\textit{Attaching Grippers:}
Equipping mobile robots with grippers could allow the remote user to manipulate small objects, thereby making it possible for them to actuate a wider range of object types.
\textit{Force Aggregation:}
Combining multiple robots within a single shell could aggerate their force, enabling the remote user to move heavier objects for object actuation or provide stronger haptic feedback for haptic communication.

\remove{
\subsubsection*{\textbf{Actuation in 2.5 Dimensions}}
In this study, the mobile robots were limited to movement within two dimensions (x and y). However, by adopting the technique presented in \textit{HapticBots}~\cite{suzuki2021hapticbots}, we can enhance our mobile robots to enable height adjustments, allowing them to actuate in 2.5 dimensions.
This improvement could lead to more advanced interaction techniques. For instance, enabling vertical movement would enable the remote user to lift objects, allowing them to stack objects or showcase objects to the local user during object actuation. With shared tangible UIs, the enhanced mobile robots could represent more complex interfaces that utilize push-down or pull-up motions.
Moreover, height adjustments could improve miniature body interaction by allowing for a better representation of the miniature body size, enabling the remote user to physically engage with the local environment based on their size. Additionally, this enhancement would enable more realistic haptic feedback of the remote user's arm to the local user by representing the remote user's arm shape in 2.5D, compared to a 2D representation.
}

\section{Conclusion}
In this paper, we presented \system{}, a novel mixed reality interface that augments holographic telepresence through synchronized tabletop mobile robots. With \system{}, we demonstrated that the remote users can physically engage with local users and the environment, enabling them to touch, grasp, manipulate, and interact with tangible objects as if they were co-located in the same space. This paper expands upon existing physical telepresence by presenting more comprehensive design space and interaction techniques, such as object actuation, virtual hand physicalization, miniature body interaction, shared tangible interfaces, embodied guidance, and haptic communication. We demonstrated various applications for \system{}, such as physical storytelling, remote tangible gaming, and hands-on instruction. A user study with twelve participants revealed that \system{} significantly enhances co-presence and shared experiences in mixed reality remote collaboration, proving its scalability, deployability, and generalizability for a wide range of remote tangible collaboration scenarios.
Additionally, we have outlined several potential avenues for future work that could extend the design space and uncover new opportunities for \system{}.

\begin{acks}
This research was funded part by the Natural Sciences and Engineering Research Council of Canada (NSERC) Discovery Grant RGPIN-2021-02857, NSERC RTI Grant RTI-2023-00582, and Mitacs Globalink Research Award.
We also thank all of the participants for our user study.
\end{acks}

\ifdouble
  \balance
\fi
\bibliographystyle{ACM-Reference-Format}
\bibliography{references}

\end{document}